\documentstyle[11pt,epsfig,fleqn]{article}

\newcommand{\zr}[1]{\mbox{\hspace*{#1em}}}
\newcommand{\RR}{\mbox{\zr{0.1}\rule{0.04em}{1.6ex}\zr{-0.05}{\sf R}}}

\begin{document}

\begin{titlepage}

{\small\noindent
ANL-PHY-8758-TH-97   \hfill   UNITU-THEP-12/1997  \newline
hep-ph/9707327       \hfill   TU-GK-6/97                     }

\vspace{5mm}

\begin{center}
{\Large A Solution to Coupled Dyson--Schwinger Equations\\
for Gluons and Ghosts in Landau Gauge}

\vspace{5mm}

        Lorenz von Smekal$^a $\\
        Physics Division, Argonne National Laboratory,\\
        Argonne, Illinois 60439-4843, USA   
        \\[5mm]
        Andreas Hauck and Reinhard Alkofer \\ 
        Institut f\"ur Theoretische Physik,
        Universit\"at T\"{u}bingen, \\
        Auf der Morgenstelle 14, 72076 T\"ubingen, Germany

\end{center}

\vspace{10mm}

\begin{quote}
  {\bf Abstract}\\
A truncation scheme for the Dyson--Schwinger equations of QCD in Landau gauge
is presented which implements the Slavnov--Taylor identities for the 3--point
vertex functions. Neglecting contributions from 4--point correlations such as
the 4--gluon vertex function and irreducible scattering kernels, a closed
system of equations for the propagators is obtained. For the pure gauge
theory without quarks this system of equations for the propagators of gluons
and ghosts is solved in an approximation which allows for an analytic
discussion of its solutions in the infrared: The gluon propagator is shown
to vanish for small spacelike momenta whereas the ghost propagator is found
to be infrared enhanced. The running coupling of the non--perturbative
subtraction scheme approaches an infrared stable fixed point at a
critical value of the coupling, $\alpha_c \simeq 9.5$. The gluon
propagator is shown to have no Lehmann representation. The results for 
the propagators obtained here compare favorably with recent lattice
calculations.
\end{quote}
\begin{quote}
\end{quote}

\vfill

\hskip 1cm \hrule width 5cm
\vskip .2cm 

{\small \noindent $^a$Address after Nov.\ 1st, 1997: Institut f\"ur
Theoretische Physik III, \\[-4pt]
Universit\"at Erlangen--N\"urnberg, Staudtstr.\ 7, 91058 Erlangen, Germany.} 
\end{titlepage}

\section{Introduction}

The mechanism for confinement of quarks and gluons into colorless hadrons 
still is little understood. Some theoretical insight could be obtained from
disproving the cluster decomposition property for color--nonsinglet
gauge--covariant operators. One idea in this direction is based on the
possible existence of severe infrared divergences, {\it i.e.}, divergences
which cannot be removed from physical cross sections by a suitable summation
over degenerate states by virtue of the Kinoshita--Lee--Nauenberg theorem
\cite{KLN62}. Such divergences can provide damping factors for the emission
of colored states from color--singlet states (see \cite{MP78}). However, the
Kinoshita--Lee--Nauenberg theorem applies to non-Abelian gauge theories in
four dimensions order by order in perturbation theory \cite{Kin76}. 
Therefore, such a description of confinement in terms of perturbation theory
is impossible. In fact, extended to Green's functions, the absence of
unphysical infrared divergences implies that the spectrum of QCD necessarily
includes colored quark and gluon states to every order in perturbation theory
\cite{Pog76}.   

An alternative way to understand the insufficiency of perturbation theory
to account for confinement in four dimensional field theories is that
confinement requires the dynamical generation of a physical mass scale. In
presence of such a mass scale, however, the renormalization group [RG]
equations imply the existence of essential singularities in physical
quantities, such as the $S$--matrix, as functions of the coupling at $g =
0$. This is because the dependence of the RG invariant confinement scale
on the coupling and the renormalization scale $\mu$ near the ultraviolet
fixed point is determined by \cite{GN74}
\begin{equation}
  \Lambda = \mu \exp \left( - \int ^g \frac {dg'}{\beta (g')} \right)
  \stackrel{g\to 0}{\rightarrow } \mu  \exp \left( - \frac 1 {2\beta_0g^2}
\right),    \quad \beta_0>0 .
  \label{Lambda}
\end{equation}
Since all RG invariant masses in massless QCD will exhibit the behavior
(\ref{Lambda}) up to a multiplicative constant the ratios of all bound state
masses are, at least in the chiral limit, determined independent of all
parameters. 

Therefore, to study the infrared behavior of QCD amplitudes non--perturbative
methods are required. In addition, as singularities are anticipated, a
formulation in the continuum is desirable. One promising approach to
non--perturbative phenomena in QCD is provided by studies of truncated systems
of its Dyson--Schwinger equations [DSE], the equations of motion for QCD
Green's functions. Typical truncation schemes resort to additional sources of
information like the Slavnov--Taylor identities, as entailed by gauge
invariance, to express vertex functions and higher n-point functions in terms
of the elementary two--point functions, {\it i.e.}, the quark, ghost and gluon
propagators. In principle, these propagators can then be obtained as
selfconsistent solutions to the non--linear integral equations representing
the closed set of truncated DSEs. 

The underlying conjecture to justify such a truncation of the originally
infinite set of DSEs is that a successive inclusion of higher
$n$--point functions in selfconsistent calculations will not result in 
dramatic changes to previously obtained lower $n$--point functions. To achieve
this it is important to incorporate as much independent information as
possible in constructing those $n$--point functions which close the system. 
Such information, {\it e.g.}, from implications of gauge invariance or
symmetry properties, can be relied on being reproduced from solutions to
subsequent truncation schemes. 

So far, available solutions to truncated Dyson--Schwinger equations of QCD
do not even fully include all contributions of the propagators itself. In
particular, even in absence of quarks, solutions for the gluon propagator in
Landau gauge used to rely on neglecting ghost contributions
\cite{Man79,Atk81,Bro89,Hau96} which, though numerically small in
perturbation theory, are unavoidable in this gauge. While this particular
problem can be avoided by ghost free gauges such as the axial gauge, in
studies of the gluon Dyson--Schwinger equation in the axial gauge 
\cite{BBZ81,Sch82,Cud91}, the possible occurrence of an independent second
term in the tensor structure of the gluon propagator has been disregarded
\cite{Bue95}. In fact, if the complete tensor structure of the gluon propagator
in axial gauge is taken into account properly, one arrives at a coupled system
of equations which is of similar complexity as the ghost--gluon system in the
Landau gauge and which is yet to be solved \cite{Alk--}. 

In addition to providing a better understanding of con\-fine\-ment based on
studies of the behavior of QCD Green's functions in the infrared, DSEs have
proven successful in developing a hadron phenomenology which
interpolates smoothly between the infrared (non--perturbative) and the
ultraviolet (perturbative) regime, for recent reviews see,
{\it e.g.}, \cite{Tan97,Rob94}. In particular, a dynamical description of
the spontaneous breaking of chiral symmetry from studies of the DSE for the
quark propagator is well established in a variety of models for the 
gluonic interactions of quarks~\cite{Miranski}. For a sufficiently large
low--energy quark--quark interaction quark masses are generated dynamically 
in the quark DSE in some analogy to the gap equation in superconductivity. 
This in turn leads naturally to the Goldstone nature of the pion and explains
the smallness of its mass as compared to all other hadrons~\cite{Del79}. 
In this framework a description of the different types of mesons is obtained
from Bethe--Salpeter equations for quark--antiquark bound states~\cite{Mun92}.
Recent progress towards a solution of a fully relativistic three--body
equation extends this consistent framework to baryonic bound states
\cite{Cah89,Ish95,Hel97}.  
 
In this paper we present a truncation scheme for the Dyson--Schwinger
equations of QCD in Landau gauge which implements the Slavnov--Taylor
identities for the 3--point vertex functions. Neglecting contributions from
4--point correlations such as the 4--gluon vertex function and irreducible
scattering kernels, this yields a closed set of non--linear integral
equations for the propagators of gluons, ghosts and, in an obvious extension,
also quarks. We present a solution obtained for the pure gauge theory without
quarks in a combination of numerical and analytical methods. The organization
of the paper is as follows: The general framework is outlined in  section
2. In section 3 we review some earlier results in a scheme
proposed by Mandelstam which is most easily introduced by further simplifying
the general scheme of the preceeding section, and we comment on available
axial gauge studies of the gluon propagator. We concentrate on the couped
system of DSEs for gluons and ghosts from section 4 on, in which we introduce
an approximation to reduce this system to a set of one--dimensional non--linear
integral equations. This approximation is especially designed to preserve
the leading behavior of these propagators in the infrared, and we give an 
analytic discussion of this behavior. In section 5 we introduce the
non--perturbative momentum subtraction scheme and discuss the accordingly
extended definition of the corresponding running coupling. The actual
details of the renormalization, the definition of the renormalization
constants and the renormalized ultraviolet finite equations are given in
section 6. An extended analytic discussion of the solutions to the
renormalized equations in terms of asymptotic expansions, a necessary
prerequisite to a numerical solution, is presented in section 7. The
numerical results and their discussion will be the issue of section 8. The
non--perturbative result for the running coupling, {\it i.e.}, the strong
coupling as function of the renormalization scale, and the corresponding 
Callan--Symanzik $\beta$--function are given in section 9. We compare our
results to recent lattice calculations of the gluon and ghost propagators
in section 10, discuss the implications on confinement for gluons in
section 11, and we wrap things up with our summary and conclusions in
section 12. We wish to point out that one main result, the existence of an
infrared stable fixed point at a critical value of the coupling $\alpha_c
\simeq 9.5$, which could, in principle, be obtained independent of the
details of the renormalization as well as the numerical methods and results,
is found in our previous publication \cite{Sme97}. This result, meaningless in
absence of a complete solution, will be incorporated also here in order to
make the present paper self--contained.

\section{A Truncation Scheme for the Propagators of QCD in Landau Gauge} 

\begin{figure}
  \centerline{ \epsfig{file=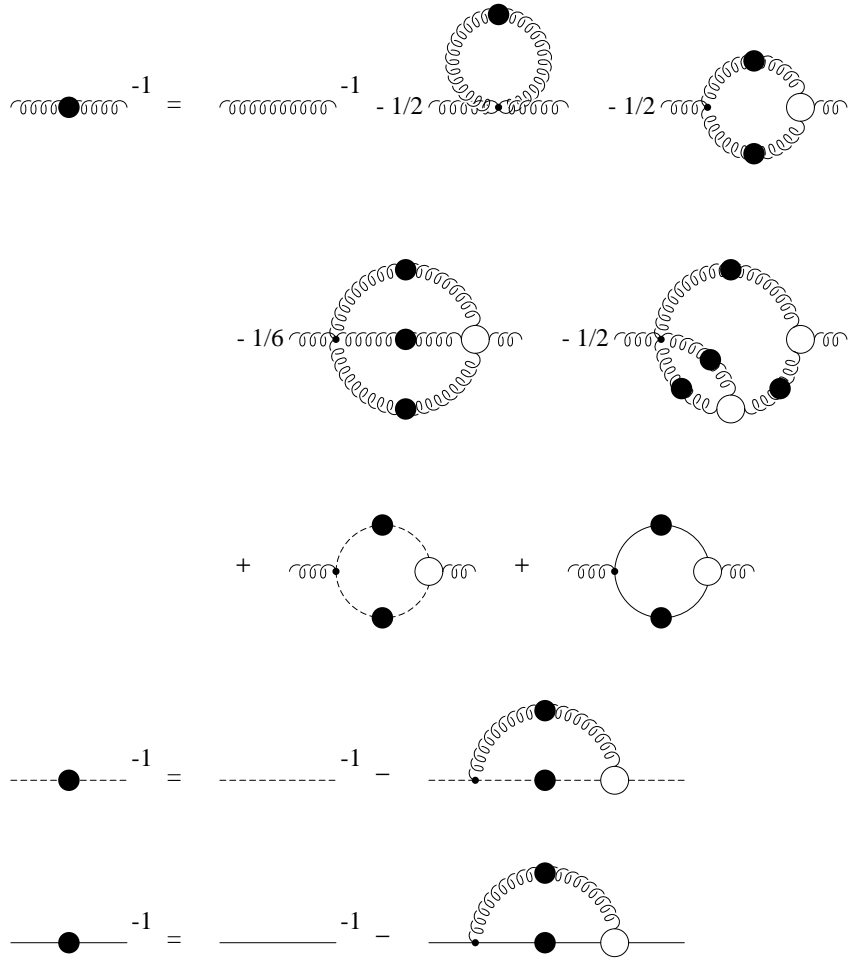,width=0.8\linewidth} }
  \caption{Diagrammatic representation of the gluon, ghost and quark
           Dyson--Schwinger equations of QCD. The wiggly, dashed and solid
           lines represent the propagation of gluons, ghosts and quarks
           respectively. A filled blob represents a full propagator and
           a circle indicates a one--particle irreducible vertex. }
  \label{GlGhQuark}
\end{figure}

Besides all elementary two--point functions, {\it i.e.}, the quark, ghost and
gluon propagators, the Dyson--Schwinger equation for the gluon propagator also
involves the three-- and four--point vertex functions which obey their own
Dyson--Schwinger equations. These equations involve successive higher n--point
functions, see figure~\ref{GlGhQuark}. For simplicity we consider the pure
gauge theory and neglect all quark contributions. As for the truncation
scheme, the extensions necessary to include quarks in future calculations are
straightforward and will be given at the end of this section. 

A first step towards a truncation of the gluon equation is to neglect all terms
with four--gluon vertices. These are the momentum independent tadpole term,
an irrelevant constant which vanishes perturbatively in Landau gauge, and
explicit two--loop contributions to the gluon DSE. The latter are subdominant
in the ultraviolet, and we can therefore expect solutions to have the correct
behavior for asymptotically high momenta without those terms. In the infrared
it has been argued that the singularity structure of the two--loop terms does
not interfere with the one--loop terms \cite{Vac95}. It therefore seems
reasonable to study the non--perturbative behavior of the gluon propagator in
the infrared without explicit contributions from four--gluon vertices, {\it
i.e.}, the two--loop diagrams in figure~\ref{GlGhQuark}. Without quarks the
renormalized equation for the inverse gluon propagator in Euclidean momentum
space with positive definite metric, $g_{\mu\nu} = \delta_{\mu\nu}$, (color
indices suppressed) is thus given by 
\begin{eqnarray} 
  D^{-1}_{\mu\nu}(k)
    &=& Z_3 \, {D^{\hbox{\tiny tl}}}^{-1}_{\mu\nu}(k)\, 
 - \, g^2 N_c \, \widetilde Z_1 \int {d^4q\over (2\pi)^4} \; iq_\mu \,
D_G(p)\, D_G(q)\, G_\nu(q,p)  
\label{glDSE}\\
&+& g^2 N_c\, Z_1  \frac{1}{2} \int {d^4q\over (2\pi)^4} \;
\Gamma^{\hbox{\tiny tl}}_{\mu\rho\alpha}(k,-p,q) 
       \, D_{\alpha\beta}(q) D_{\rho\sigma}(p) \,
                \Gamma_{\beta\sigma\nu}(-q,p,-k)  
  \; ,\nonumber
\end{eqnarray}
where $p = k + q$, $D^{\hbox{\tiny tl}}$ and $\Gamma^{\hbox{\tiny tl}}$ are
the tree--level propagator and three--gluon vertex, $D_G$ is the ghost
propagator and $\Gamma$ and $G$ are the fully dressed 3--point
vertex functions. The DSE for the ghost propagator in Landau gauge QCD,
without any truncations, will remain unchanged in its form when quarks are
included,
\begin{equation}
  D_G^{-1}(k) \, =\,  -\widetilde{Z}_3 \, k^2 \, +\, g^2 N_c\,  \widetilde Z_1
  \int {d^4q\over (2\pi)^4} \; ik_\mu \, D_{\mu\nu}(k-q) \, G_\nu (k,q) \,
D_G(q)   \; .
  \label{ghDSE}
\end{equation}
The coupled set of equations for the gluon and ghost propagator,
eqs. (\ref{glDSE}) and (\ref{ghDSE}), is graphically depicted in
figure~\ref{GluonGhost}. The renormalized propagators for ghosts and gluons,
$D_G$ and $D_{\mu\nu}$, and the renormalized coupling $g$ are defined from the
respective bare quantities by introducing multiplicative renormalization
constants, 
\begin{equation}
  \widetilde{Z}_3 D_G := D^0_G \; , \quad
  Z_3 D_{\mu\nu} := D^0_{\mu\nu} \; , \quad
  Z_g g := g_0 \; .
  \label{Zds}
\end{equation}
Furthermore, $Z_1 = Z_g Z_3^{3/2}$, $\widetilde{Z}_1 = Z_g Z_3^{1/2}
\widetilde{Z}_3$. We use the following notations to separate the
structure constants  $f^{abc}$ of the gauge group $SU(N_c = 3)$ (and the
coupling $g$) from the 3--point vertex functions: 

\begin{figure}
  \centerline{ \epsfig{file=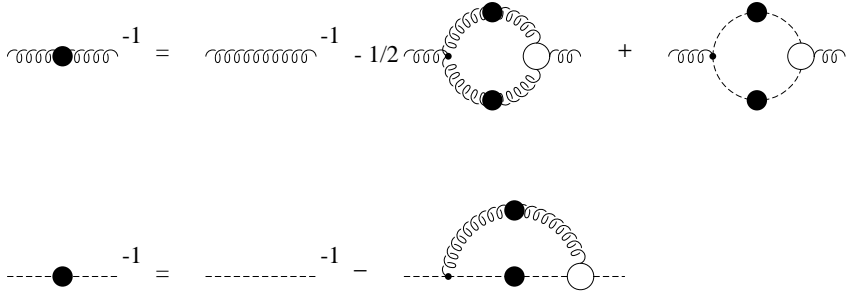,width=0.8\linewidth} }
  \caption{Diagrammatic representation of the gluon and ghost Dyson--Schwinger 
           equations of QCD without quarks. In the gluon DSE terms with
           four--gluon vertices have been dismissed.}
  \label{GluonGhost}
\end{figure}

\begin{equation}
    \Gamma^{abc}_{\mu\nu\rho}(k,p,q) = 
                                   g f^{abc} (2\pi)^4 \delta^4(k+p+q)
\Gamma_{\mu\nu\rho}(k,p,q)     \; .
\quad \hbox{
\begin{minipage}[c]{0.25\linewidth}
  \epsfig{file=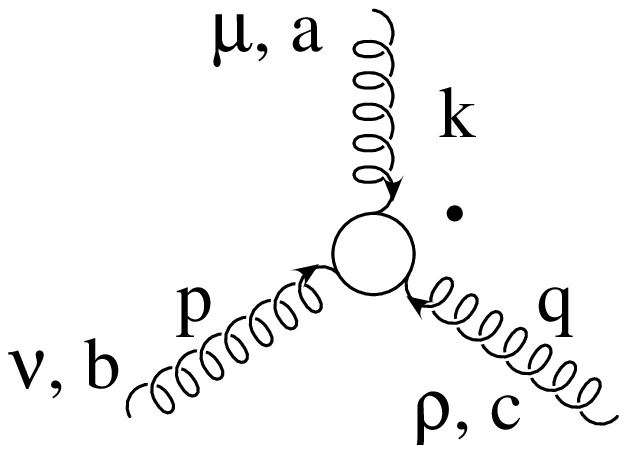,width=0.9\linewidth}
\end{minipage}}
\end{equation}
The arguments of the 3--gluon vertex denote the three incoming gluon momenta
according to its Lorentz indices (counter clockwise starting from the dot). 
With this definition, the tree--level vertex has the form,
\begin{equation}
\Gamma^{\hbox{\tiny tl}}_{\mu\nu\rho}(k,p,q) \, = \,-  i(k-p)_\rho
\delta_{\mu\nu} \, - \, i(p-q)_\mu \delta_{\nu\rho} \, - \, i(q-k)_\nu
\delta_{\mu\rho} \; . 
\end{equation} 
The arguments of the ghost--gluon vertex are the outgoing and incoming ghost 
momenta respectively,

\begin{equation}
     G^{abc}_\mu(p,q) = g f^{abc} G_\mu(p,q)
                      = g f^{abc} iq_\nu \widetilde{G}_{\mu\nu}(p,q) \; ,
\quad \hbox{\hskip .8cm
\begin{minipage}[c]{0.25\linewidth}
  \epsfig{file=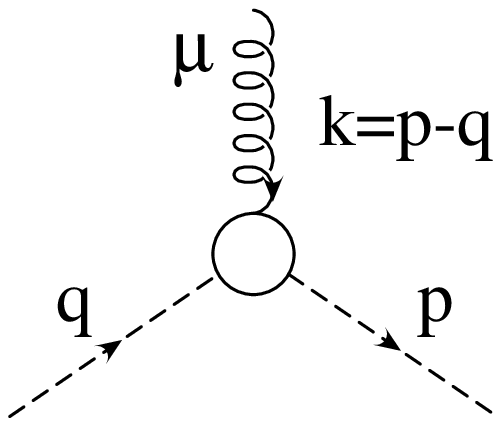,width=0.8\linewidth}
\end{minipage}}
\end{equation}
where the tensor $\widetilde{G}_{\mu\nu}(p,q)$ contains the ghost--gluon
scattering kernel (for its definition see, {\it e.g.}, ref.~\cite{Bar80}). At
tree--level this kernel vanishes which corresponds to
$\widetilde{G}_{\mu\nu}(p,q) = \delta_{\mu\nu}$. Note that the color
structure of all three loop diagrams in fig.~\ref{GluonGhost} is simply given
by $f^{acd} f^{bdc} = - N_c \delta^{ab}$ which was used in
eqs. (\ref{glDSE},\ref{ghDSE}) suppressing the trivial color structure of
the propagators $\sim \delta^{ab}$.

The ghost and gluon propagators in the covariant gauge, introducing the gauge
parameter $\xi$, are parameterized by their respective renormalization 
functions $G$ and $Z$,  
\begin{eqnarray}  
  D_G(k) &=& - {G(k^2)\over k^2}   \qquad \hbox{and} \label{rfD}\\
  D_{\mu\nu}(k) &=&  \left(\delta_{\mu\nu} - {k_\mu k_\nu\over k^2} \right)  \,
  {Z(k^2)\over k^2} + \xi \frac{k_\mu k_\nu}{k^4} \; .
  \nonumber
\end{eqnarray}
In order to arrive at a closed set of equations for the functions $G$ and $Z$,
we still have to specify a form for the ghost--gluon and the 3--gluon vertex
functions. We will now propose a construction of these vertex functions based
on Slavnov--Taylor identities as entailed by BRS invariance. For the 3--gluon
vertex the general procedure in such a construction was outlined in the
literature \cite{Bar80,Bal80,Kim80}. Since this procedure involves unknown
contributions from the ghost--gluon scattering kernel, it cannot be
straightforwardly applied to serve the present purpose, which is to express
the vertex functions also in terms of the functions $G$ and $Z$. To this end
we have to make additional assumptions on the scattering kernel. Since it is
related to the ghost--gluon vertex, we start with the construction of the
latter. 

For our actual solutions later on, we will be using the Landau gauge ($\xi =
0$) in which one has $\widetilde Z_1 = 1$ \cite{Tay71}. Therefore, in
the case of the ghost--gluon vertex the choice of its tree--level form in the
DSEs may seem justified to the extend that it would at least not affect the
leading asymptotic behavior of the solutions in the ultraviolet. This is in
contrast to the  other vertex functions in QCD which have to be dressed at
least in such a way as to account for their anomalous dimensions, if the
solutions to the DSEs in terms of propagators are expected to resemble their
leading perturbative behavior at short distances, {\it i.e.} their correct
anomalous dimensions. 

A good way to achieve this is to construct vertex functions from their
respective Slavnov--Taylor identities. Those identities generally fail,
however, to fully constrain the vertex functions. Methods of refinement of
such constructions from implications of multiplicative renormalization 
have been developed for the fermion--photon vertex in QED \cite{Bro91}. Our
particular focus in the present context is on the 3--gluon vertex which
has a considerably higher symmetry than fermion vertices. This symmetry
alleviates the problem of unconstraint terms and we do not expect our results
to be overly sensitive to such terms in the 3--gluon vertex. 

The Slavnov--Taylor identity for the 3--gluon vertex in momentum space is
given by \cite{Bar80,Bal80,Kim80},
\begin{eqnarray}
  i k_\rho \Gamma_{\mu\nu\rho} (p,q,k) &=&  G(k^2) \, \left\{ \,
  \widetilde{G}_{\mu\sigma}(-k,q) \, {\cal P}_{\sigma\nu} (q) \,
  \frac{q^2}{Z(q^2)}
  \right. \label{glSTI}\\ 
  && \hbox{\hskip 3cm}  - \left.
  \widetilde{G}_{\nu\sigma}(-k,p) \, {\cal P}_{\sigma\mu} 
  (p) \, \frac{p^2}{Z(p^2)} \right\}  \; ,
  \nonumber
\end{eqnarray} 
where ${\cal P}_{\mu\nu}(k) = \delta_{\mu\nu} - k_\mu k_\nu /k^2$ is the
transversal projector. A simple solution to (\ref{glSTI}) is possible if
ghosts are neglected completely, {\it i.e.}, for $G(k^2) = 1$ and
$\widetilde{G}_{\mu\nu} = \delta_{\mu\nu}$ (see the next
section). This is obviously not satisfactory for our present 
study addressing ghost contributions to the gluon propagator in
particular. Given that the anomalous dimension of the ghost--gluon
vertex vanishes in Landau gauge, it may seem appealing to retain its
tree--level form by setting $\widetilde G_{\mu\nu} = \delta_{\mu\nu} $ while
taking the presence of the ghost renormalization function $G(k^2) $ into
account in solving (\ref{glSTI}). Unfortunately, this leads to a
contradiction. The Slavnov--Taylor identity for the 3--gluon vertex can be
shown to have no solution with this Ansatz. 
One possibility to resolve this problem is to add a term corresponding to a
non--vanishing scattering kernel of the form,
\begin{equation}
  \widetilde G_{\mu\nu}(p,q) \, =\, \delta_{\mu\nu} \, + a(p^2,q^2; k^2) \, ( 
  \delta_{\mu\nu} \, pq \, - \,p_\nu q_\mu ) \; .\label{glghkern}
\end{equation}
While the additional term in (\ref{glghkern}) corresponding to a non-trivial
ghost--gluon scattering kernel does not contribute to the ghost--gluon vertex
(being transverse in $q_\nu$), it is possible to make a minimal Ansatz for the
unknown function $a(x,y;z)$ such that (\ref{glSTI}) can be solved. The proof
that (\ref{glSTI}) cannot be solved for $\widetilde G_{\mu\nu} =
\delta_{\mu\nu}$ and the construction of $a(x,y;z)$ are given in appendix
\ref{app:B}. This procedure is certainly not unambiguous, and it is
furthermore against our original intention to neglect explicit contributions 
from irreducible four--point functions as it implies a very specific form for
the ghost--gluon scattering kernel. A truncation in which the ghost--gluon
vertex function is replaced with its tree--level form, $G_\mu (p,q) \to
iq_\mu $, while compatible with the desired short distance behavior of the
solutions, implies that some non-trivial assumptions on the otherwise unknown
ghost--gluon scattering kernel have to be made in order to solve the
Slavnov--Taylor identity for the 3--gluon vertex.  

In light of this, we prefer a second possibility to resolve the present
problem, which is to reconsider the form of the ghost--gluon vertex
to start the truncation scheme with. A suitable source of information is
provided by a Slavnov--Taylor identity which involves the ghost--gluon vertex
directly, similar to the one for the quark--gluon vertex. To our knowledge,
such an identity has not been stated in the literature so far. We will
therefore  outline its derivation from the usual BRS invariance briefly in
the following. Neglecting irreducible ghost--ghost scattering, we will arrive
at an identity, which allows to express the ghost--gluon vertex in terms of
the ghost renormalization function $G(k^2)$. The vertex constructed this way
will have the correct short distance behavior (no anomalous dimension) and
it will constrain $\widetilde G_{\mu\nu}(p,q) $ just enough to admit a simple
solution to the 3-gluon Slavnov--Taylor identity (\ref{glSTI}) without
further assumptions. 
   
In order to construct a non--perturbatively dressed ghost--gluon vertex, we
start from BRS invariance of the generating functional of pure Yang--Mills
theory,
\begin{equation}
  \delta_{\hbox{\tiny BRS}} \, Z[\bar\eta,\eta,J] \,=\, 0
  \qquad \Leftrightarrow
  \label{BRS}
\end{equation}
\begin{displaymath}
  \Delta[\bar\eta,\eta,J] \, := \, \left\langle \int d^4x \left(
  J^a D^{ab} c^b  -
  {g\over 2} f^{abc} \bar\eta^a 
  c^b c^c - \frac{1}{\xi} \partial A^a \eta^a \right)
  \right\rangle_{[\bar\eta,\eta,J]} \, = \, 0 \; ,
\end{displaymath}
where $D^{ab} = \delta^{ab} \partial + g f^{abc} A^c$ is the covariant
derivative. We retained non--zero sources $\{\bar\eta , \eta ,J\}$ for the
ghost and gauge fields $\{c, \bar c, A\}$ in order to perform suitable
derivatives. Fermion fields have been omitted for simplicity. Their 
inclusion would not affect the present derivation. We find that the particular
derivative of (\ref{BRS}) with respect to the sources, suitable to constrain
the ghost--gluon vertex, is given by:
\begin{equation}
  \left. \frac{\delta^3}{\delta\bar\eta^c(z) \delta\eta^b(y) \delta\eta^a(x)}
  \, \Delta[\bar\eta , \eta ,J]  \right|_{\bar\eta = \eta =
  J = 0} \, = \, 0 \; ,
\end{equation}
and we obtain
\begin{eqnarray}
  \frac{1}{\xi}\, \langle\, c^c(z)\, \bar c^b(y) \, \partial A^a(x)\,
  \rangle &-& {1\over \xi} \, \langle \, c^c(z)\, \bar c^a(x) \, \partial
  A^b(y)\, \rangle \label{ghSTI}\\ 
  &&  \quad = \,- {g\over 2} f^{cde} \, \langle\, c^d(z)\, c^e(z)\, \bar
  c^a(x) \, \bar c^b(y) \, \rangle \; .
  \nonumber
\end{eqnarray}
Note that this symbolic notation refers to full reducible correlation
functions. The two terms on the l.h.s.\  of (\ref{ghSTI}) can be decomposed
in the ghost--gluon proper vertex and the respective propagators. The
derivative on the gluon leg thereby projects out the longitudinal part of the
gluon propagator which, by virtue of its own Slavnov--Taylor identity $k_\mu
D_{\mu\nu}(k) = \xi \, k_\nu/k^2$, remains undressed in the covariant gauge
(see eq.~(\ref{rfD})). The r.h.s.\ contains a disconnected part plus terms
due to ghost--ghost scattering. At this point we make the truncating
assumption mentioned previously. We neglect ghost--ghost scattering
contributions to the vertex and retain only the reducible part of the
correlation function on the r.h.s.\ of (\ref{ghSTI}) corresponding to
disconnected ghost propagation, {\it i.e.},
\begin{equation}
  \langle \, c^c \, c^a \, \bar c^b \, \bar c^d \, \rangle \, =\,  \langle
  \, c^a \, \bar c^b \, \rangle \, \langle \, c^c\, \bar c^d\, \rangle \, -
  \langle \, c^a \, \bar c^d \, \rangle \, \langle \, c^c\, \bar c^b\, \rangle
  \, + \; \hbox{connected} \; .
\label{redgh}
\end{equation}
This assumption results in an Abelian Ward--Takahashi type of equation for
the ghost--gluon vertex, relating its longitudinal part to a sum of inverse
ghost propagators. In momentum space with the ghost renormalization function
as introduced in (\ref{rfD}) we obtain (for details see appendix
\ref{app:A}),  
\begin{equation}
  G(p^2) \, ik_\rho G_\rho^{abc}(p,q) \, -\, G(k^2) \, ip_\rho
  G_\rho^{abc}(-k,q) \, = \, g f^{abc} \, { q^2 G(k^2) G(p^2) \over
  G(q^2)}
  \; . \label{ghWTI}
\end{equation}
Note that this equation is valid for all $\xi$ including $\xi=0$
(Landau gauge). Eq. (\ref{ghWTI}) is solved by the following particularly
simple form for the vertex which was also used in ref.~\cite{Elw96},
\begin{equation}
  G_\mu^{abc}(p,q) = g f^{abc} iq_\mu  \, {G(k^2)\over G(q^2)} \;
  . \label{ghv1}
\end{equation}
It is important to note, however, that the solution (\ref{ghv1}) to
eq.~(\ref{ghWTI}) violates the symmetry properties of the vertex
explicitly. In order to check, whether a symmetric solution to (\ref{ghWTI})
exists, we make the general Ansatz,
\begin{equation}
  G_\mu(p,q) \, = \,i (p+q)_\mu \,  a(k^2; p^2,q^2) \, - \,
  i(p-q)_\mu  \, b(k^2; p^2,q^2) \; .
  \label{ans}
\end{equation}
The reason for this particular choice is that in general covariant gauges one
has 
\begin{equation}
  a(x;y,z) = a(x;z,y) \; ,
  \label{asy}
\end{equation}
whereas the symmetry of $b(x;y,z)$ is undetermined. In Landau gauge, however,
the mixed symmetry term proportional to $b$ does not contribute to observables
directly, since it is purely longitudinal in the gluon momentum. With the
Ansatz (\ref{ans}) we find:
\begin{equation}
  G(x) \, \biggl( (z-x) \, a(y;x,z) + y \, b(y;x,z) \biggr) \label{aux1}
\end{equation}
\[\hbox to 1 true cm {\hfill }   + \, G(y) \, \biggl( (z-y)\, a(x;y,z) + x \,
b(x;y,z) \biggr) \, = \, {z G(x) G(y)  \over G(z) } \; . \]
From all terms proportional to $z$ it follows that 
\begin{displaymath}
  a(x;y,z) \, = \, {1\over 2} \, {G(x)\over G(z)} + a_1(x,y;z)
  \frac{1}{G(y)} \; , \quad a_1(x,y;z) \, = \, - a_1(y,x;z)
  \; .
\end{displaymath}
Here the undetermined antisymmetric function $a_1(x,y;z)$ has to guarantee
the symmetry of $a(x;y,z)$ {\it c.f.} (\ref{asy}). This can be achieved
most easily by setting
\begin{displaymath}
  a_1(x,y;z)\, =\, {1\over 2} \, \biggl( G(x) \, -\, G(y) \biggr) \,
  \Rightarrow \; 
  a(x;y,z) \, =\,  {1\over 2} \, \biggl( {G(x)\over G(z)} \,+ \,
  \frac{G(x)}{G(y)} \, - 1 \biggr)   .
\end{displaymath}
The mixed symmetry term follows readily from terms proportional to $x$
or $y$ in (\ref{aux1}),
\begin{equation}
  b(x;y,z) = \frac{G(x)}{G(y)} \, a(y;x,z)
           = \frac{1}{2} \, \biggl( {G(x)\over G(z)}
                \, - \, {G(x)\over G(y)} \, +  1 \biggr)  \; .
\end{equation}
Collecting the above terms, from (\ref{ans}) the ghost--gluon vertex can be
written,
\begin{eqnarray}  
G_\mu(p,q) &=&  iq_\mu \, {G(k^2) \over G(q^2)} \, +\, ip_\mu
\, \biggl( {G(k^2)\over G(p^2)} \, - 1 \biggr) \label{fvs}\\ 
&=& iq_\mu \,  \biggl({G(k^2) \over G(q^2)} \, + \, {G(k^2)\over
G(p^2)} \, - 1 \biggr) \, + \quad \hbox{longitudinal terms} \; .\label{lgvs}
\end{eqnarray} 
In the last line we rearranged the terms and displayed explicitly only the
ones relevant for the vertex in Landau gauge. While terms longitudinal in the
gluon momentum $k = p - q$ are irrelevant in the vertex, they contribute to
the 3--gluon vertex Slavnov--Taylor identity (\ref{glSTI}).
The form of $\widetilde G_{\mu\nu}(p,q) $ necessary to generate the complete
ghost--gluon vertex (\ref{fvs}), is given by
\begin{equation}  
\widetilde G_{\mu\nu}(p,q) \, =\, {G(k^2)\over G(q^2)} \, \delta_{\mu\nu} \,
+ \, \biggl( {G(k^2)\over G(p^2) } \, - 1 \biggr) \, {p_\mu q_\nu\over q^2}
\; . \label{tG} 
\end{equation}
Here, the last term does not contribute to (\ref{glSTI}). Therefore, while
the relevant term to the ghost--gluon vertex in Landau gauge is given by
the one explicitly stated in (\ref{lgvs}), the term of the tensor
(\ref{tG}) to be used in (\ref{glSTI}) is $\widetilde G_{\mu\nu}(p,q) =
(G(k^2)/G(q^2))\, \delta_{\mu\nu}$ and not what would follow from
(\ref{lgvs}) with the longitudinal terms neglected before constructing
$\widetilde G$, {\it i.e.}, $(G(k^2)/G(q^2) + G(k^2)/G(p^2) - 1 )\,
\delta_{\mu\nu}$. Additional contributions to $\widetilde G_{\mu\nu}(p,q)$
which are purely transverse in $q_\nu$ can arise from the ghost--gluon
scattering kernel. Such terms cannot be constraint from the form of the
ghost--gluon vertex. In contrast to the scheme with the tree--level
ghost--gluon vertex, where precisely these terms are necessary to solve
the 3--gluon Slavnov--Taylor identity, using the dressed vertex
(\ref{fvs}) and (\ref{tG}), unknown contributions from the ghost--gluon
scattering kernel can be neglected in (\ref{glSTI}). Thus using (\ref{tG})
in (\ref{glSTI}) we obtain
\begin{equation}
 i k_\rho\Gamma_{\mu\nu\rho}(p,q,k) =
    G(k^2) \left( {\cal P}_{\mu\nu}(q) {q^2 G(p^2)\over G(q^2) Z(q^2)}
    \, -\, {\cal P}_{\mu\nu}(p) {p^2 G(q^2)\over G(p^2)Z(p^2) } \right) ,
  \label{glWTI}
\end{equation}
which is only slightly generalized as compared to
the {\it Abelian} like identity obtained neglecting ghosts
completely. Using the symmetry of the vertex the solution to (\ref{glWTI})
fixes the vertex up to completely transverse parts. It can be derived
straightforwardly along the lines of the general procedure outlined in
\cite{Bal80,Kim80}. In the present case, the solution is
\[  \Gamma_{\mu\nu\rho}(p,q,k) 
       \, =\, - A_+(p^2,q^2;k^2)\,  \delta_{\mu\nu}\,  i(p-q)_\rho\,  
          - \,  A_-(p^2,q^2;k^2)\,  \delta_{\mu\nu} i(p+q)_\rho  \]
\begin{equation}
   \hskip 6mm - \, 2\frac{A_-(p^2,q^2;k^2)}{p^2-q^2}
( \delta_{\mu\nu} pq \, -\,  p_\nu
q_\mu) \, i(p-q)_\rho\,  + \; \hbox{cyclic permutations} \; ,
  \label{3gv}
\end{equation}
with
\begin{equation}  \label{3gva}
  A_\pm (p^2,q^2;k^2)
    = G(k^2) \, \frac{1}{2} \, \left( \frac{G(q^2)}{G(p^2)Z(p^2)} \pm
      \frac{G(p^2)}{G(q^2)Z(q^2)} \right) \quad .
\end{equation}
At this point, additional unconstrained transverse terms in the
vertex are ignored. A counter example for this to be justified is the
fermion--photon vertex in QED, for which it has been shown that the
transverse part is crucial for multiplicative renormalizability (see,
refs.\ \cite{Bro91}). A similar construction to fix transverse pieces of
the vertices in QCD is still lacking. The most important difference, however,
which tends to alleviate the problem of undetermined transverse terms in the
case of the 3-gluon vertex as constructed from its Slavnov-Taylor identity is
its high symmetry. In contrast to the QED fermion vertex, in which 8 of the
12 independent tensor terms are transverse in the photon momentum, in the
case of the 3--gluon vertex only 4 out of 14 terms (involving 2 unknown
functions) are transverse in all three gluon momenta and thus
unconstraint by the Slavnov--Taylor identity (see \cite{Bal80}). 
Therefore, 
its Slavnov-Taylor identity together with its full Bose (exchange) symmetry
puts much tighter constraints on the 3--gluon vertex than the
Ward--Takahashi/Slavnov--Taylor identities do on the fermion--photon/gluon
vertices. Note also that there are no undetermined transverse terms
in the ghost-gluon vertex of the present truncation scheme. 

A further important difference between fermion vertices and the 3--gluon
vertex function is that transverse terms in the vertices of (electrons) quarks 
are of particular importance due to their coupling to transverse
(photons) gluons in the Landau gauge. In contrast, for the present purpose of
the 3--gluon vertex function, to be used in truncated gluon DSEs, it is well
known that the longitudinal contribution to the 3--gluon loop is the one of
particular importance \cite{Bro88}. This implies that even though the two
gluons within the loop are transverse in Landau gauge, the third (external)
leg of the 3--gluon vertex must not be connected to a transverse projector
(see below). This makes the important difference, because only the
terms that are transverse with respect to all three gluon momenta are
unconstraint in the vertex.

Now, we have set up a closed system of equations for the renormalization
functions $G(k^2)$ and $Z(k^2) $ of ghosts and gluons, consisting of their
respective DSEs (\ref{glDSE}) and (\ref{ghDSE}) with the vertex functions
given by (\ref{lgvs}) and (\ref{3gv}/\ref{3gva}). Thereby we neglected
explicit 4--gluon vertices (in the gluon DSE (\ref{glDSE})), irreducible
4--ghost correlations (in the identity for the ghost--gluon vertex
(\ref{ghWTI})) and contributions from the ghost--gluon scattering kernel (to
the Slavnov--Taylor identity (\ref{glSTI})). This is the most important
guideline for our truncation scheme according to the general idea of
successively taking higher n--point functions into account. 

It is straightforward to extend the present scheme to selfconsistently
include the quark propagator which has the general structure, 
\begin{equation}
  S(p) \, = \, \frac{1}{-ip\hskip -5pt \slash  A(p^2) + B(p^2)} \; . 
\end{equation}
For the quark--gluon vertex $\Gamma^a_\mu (p,q) $ with momentum arguments $p$
and $q$ for outgoing and incoming quarks respectively, the Slavnov--Taylor
identity is \cite{Ei74},
\begin{eqnarray}
  G^{-1} (k^2)  \,  ik_\mu \Gamma^a_\mu(p,q)    &=& \Bigl( g t^a -
  B^a(k,q)\Bigr) \, iS^{-1}(p) \label{qkSTI}\\  
  && \hskip 1.5cm - \, iS^{-1}(q) \, \Bigl( g t^a - B^a(k,q)\Bigr) \; , 
  \quad k = p -q \; . \nonumber 
\end{eqnarray}
Here,  $t^a$ is the $SU(3)$ generator in the fundamental representation, and
$B^a(k,q)$ is the ghost--quark scattering kernel which again represents 
irreducible 4--point correlations. In the present truncation scheme
in which such 4--point correlations are consistently neglected, the solution
to this Slavnov--Taylor identity is obtained from a particularly simple
extension to the construction of Ball and Chiu for the solution to the
analogous Ward--Takahashi identity of Abelian gauge theory \cite{BC80}.
For $B^a(k,q) = 0$ the Slavnov--Taylor identity for the quark--gluon
vertex (\ref{qkSTI}) is solved by
\begin{displaymath}
  \Gamma^a_\mu(p,q) \, =\, - g t^a  \, G(k^2) \, \Bigg\{ \frac{1}{2} \left(
  A(p^2) + A(q^2)\right) i\gamma_\mu + \frac{p_\mu + q_\mu}{p^2-q^2}
  \biggl( \Bigl( A(p^2) 
\end{displaymath}
\begin{equation}
  \hskip 2cm - A(q^2)\Bigr)
  \frac{ip\hskip -5pt \slash  + iq\hskip -5pt \slash}{2} \,
- \, \left(B(p^2) - B(q^2)\right) \biggr) \Bigg\} \, + \;
  \hbox{transverse terms} \; . \label{qkSTIsol}
\end{equation}
This is justified for a study with emphasis on the infrared
behavior of the propagators even beyond the present truncation scheme,
because it has been shown that $B^a(k,q) \to 0$ for $k \to 0$ in Landau
gauge \cite{Tay71}. Note that the only difference to the Abelian case
considered in \cite{BC80} at this point is the presence of the ghost
renormalization function appearing in the quark--gluon vertex in Landau
gauge. This being a minor modification to the structure of the vertex, the
ghost renormalization function in (\ref{qkSTIsol}) is, however, crucial for
the vertex to resemble its correct anomalous dimension in the perturbative
limit. Furthermore, our present solution demonstrates that the ghost
renormalization function $G(k^2)$, being infrared enhanced as we will see
below, can give an essential contribution to the effective interaction of
quarks as part of the quark--gluon vertex also in the infrared.

With the form (\ref{qkSTIsol}) for the vertex function in the quark DSE,
possibly improved by additional transverse terms as discussed for quenched
QED in \cite{Bro91}, the present truncation scheme is extended to a closed
set of equations for all propagators of Landau gauge QCD as parameterized by
the four functions $Z(k^2), G(k^2), A(k^2)$ and $B(k^2)$. Its solution would
represent for the first time a systematic and complete solution to the DSEs
of QCD at the level of propagators. In the following, we will present our
solution to the subset of equations restricted to the pure gauge theory
without quarks. The selfconsistent inclusion of the quark DSE will be subject
to further investigations.

\goodbreak

\section{The Infrared Behavior of the Gluon Propagator in Previous DSE Studies}

Before we turn to the discussion of the coupled system of DSEs for gluons and 
ghosts and its simultaneous solution in an one--dimensional approximation
in the following sections, we briefly review the current status in DSE
studies of the gluon propagator in Landau gauge which is of particular
relevance to our present scheme going beyond an approximation by Mandelstam 
that was previously employed in these studies. For comparison we also
summarize available axial gauge studies of the gluon propagator. Their review
is necessarily incomplete and we refer the reader to the quoted literature
for further information.  

A first approximation scheme for the gluon Dyson--Schwinger equation in
Landau gauge was originally proposed by Mandelstam \cite{Man79}. Compared to
the scheme outlined above, it consists of a further truncating assumption
which is, even though working in Landau gauge, to neglect all ghost
contributions to the gluon DSE in pure QCD (without quarks). As a
justification for this, it was usually referred to perturbative calculations
which yield numerically small ghost contributions to the gluon
self--energy. Even though there was never any doubt about the importance of
ghosts for fundamental reasons such as transversality of the gluon propagator
and unitarity, it was asserted that their quantitive contributions to many
hadronic observables might remain negligible even beyond perturbation
theory. We will see in the following sections that our present solutions to
the coupled system of gluon {\sl and} ghost DSEs, in the systematic
truncation scheme outlined above, yield qualitatively quite different results
as compared to the Mandelstam approximation, and are thus counter examples to
this assertion.   

\begin{figure}
  \centerline{
 \epsfig{file=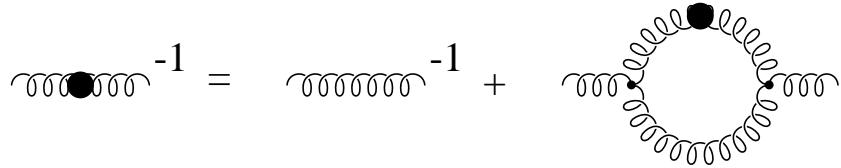,width=0.8\linewidth}}
  \caption{Diagrammatic representation of the gluon Dyson--Schwinger 
           equation in Mandelstam's approximation.}
  \label{Mandelstam}
\end{figure}

Without ghosts, the solution to the Slavnov--Taylor identity for the 3--gluon
vertex is obtained from eq.~(\ref{glWTI}) by setting  $G = 1$, {\it i.e.},
eq.~(\ref{3gv}) with 
\begin{equation}   \label{3gvb}
  A_\pm (p^2,q^2;k^2)  =   A_\pm (p^2,q^2)  = 
      \frac{1}{2} \, \left( \frac{1}{Z(p^2)} \pm
      \frac{1}{Z(q^2)} \right) \quad .
\end{equation}
Assuming that $Z(p^2)$ is a slowly varying function one may then approximate 
the corresponding solution (\ref{3gv}) by
\begin{equation}
  \Gamma_{\mu\nu\rho}(p,q,k)
    = A_+(p^2,q^2) \Gamma^{\hbox{\tiny tl}}_{\mu\nu\rho}(p,q,k)  \;
    . \label{eq:MaAn} 
\end{equation}
While this form for the full three--gluon vertex simplifies the 3--gluon loop
in the gluon DSE even more than the use of another bare vertex, {\it i.e.},
$\Gamma =\Gamma^{\hbox{\tiny tl}}$, it was observed by Mandelstam to be
superior to the latter since it accounts for some of the dressing of the
vertex as it results from the corresponding Slavnov--Taylor identity. The
nature of this dressing is such that it cancels the dressing of one of the
gluon propagators in the 3--gluon loop, and without ghost contributions the
gluon DSE~(\ref{glDSE}) in the Mandelstam approximation thus simplifies to  
\begin{eqnarray} \label{glMA}
  D^{-1}_{\mu\nu}(k)
    &=&  Z_3 \, {D^{\hbox{\tiny tl}}}^{-1}_{\mu\nu}(k) \\ 
&& \hskip -1cm +  g^2 N_c\, \frac{1}{2} \int {d^4q\over (2\pi)^4}
    \; \Gamma^{\hbox{\tiny tl}}_{\mu\rho\alpha}(k,-p,q) 
       \, D_{\alpha\beta}(q) D^{\hbox{\tiny tl}}_{\rho\sigma}(p) \,
                \Gamma^{\hbox{\tiny tl}}_{\beta\sigma\nu}(-q,p,-k)  
  \; ,\nonumber
\end{eqnarray}
where $p=k+q$. This equation, the Mandelstam equation, is schematically
depicted in fig.~\ref{Mandelstam}. It was already pointed out by Mandelstam
that in order to solve this equation selfconsistently it is necessary to
implement an additional constraint: Without ghosts the
Slavnov--Taylor identity~(\ref{glWTI}) with $G = 1$ and its
solution~(\ref{3gv}/\ref{3gvb}) for $p^2 \to 0$ entail that 
\begin{equation}
  \lim_{k^2 \to 0} \frac{k^2}{Z(k^2)} = 0,
  \qquad \hbox{or} \qquad
  \lim_{k^2 \to 0} D_{\!\mu\nu}^{-1}(k^2) = 0 \quad . \label{massless.MA}
\end{equation}
While imposing this additional condition seems consistent with the other
assumptions in the Mandelstam approximation, it has to be emphasized that
concluding (\ref{massless.MA}) as a result of (\ref{glWTI}) relies solely on
neglecting all ghost contributions in covariant gauges (see also the
discussion at the end of the next section). 

The Mandelstam equation in its original form is obtained from
eq.~(\ref{glMA}) upon contraction with the transversal projector
${\cal P}_{\mu\nu}(k) = \delta_{\mu\nu} - k_\mu k_\nu /k^2$ and 
integration of the angular variables,
\begin{eqnarray}  \label{eq:Mandelstam}
  \frac{1}{Z(k^2)}
     &=& Z_3 \,+\, \frac{g^2}{16\pi^2} \int_0^{k^2} \frac{dq^2}{k^2}
              \left(   \frac{7}{8}\frac{q^4}{k^4}
                     - \frac{25}{4}\frac{q^2}{k^2}
                     - \frac{9}{2} \right) Z(q^2) \nonumber \\
         && \hskip -1cm +\, \frac{g^2}{16\pi^2}
\int_{k^2}^{\Lambda^2} \frac{dq^2}{k^2}
              \left(   \frac{7}{8}\frac{k^4}{q^4}
                     - \frac{25}{4}\frac{k^2}{q^2}
                     - \frac{9}{2} \right) Z(q^2) \; . 
\end{eqnarray}
The quadratic ultraviolet divergence in this equation is absorbed by a
suitably added counter term introduced in order to account for the masslessness
condition~(\ref{massless.MA}), see refs.~\cite{Man79,Atk81}. The solution to
this equation proposed by Mandelstam's infrared analyses proceeds briefly as
follows: Assume $Z(k^2) \sim 1/k^2 $. On the r.h.s of
eq.~(\ref{eq:Mandelstam}) this exclusively yields contributions which violate
the masslessness condition~(\ref{massless.MA}). Such terms have to be
subtracted. Since the kernel on the r.h.s of eq.~(\ref{eq:Mandelstam}) is
linear in $Z$, this is achieved by simply subtracting the corresponding
contribution from $Z$ in the integrand. Thus defining  
\begin{equation} \label{MAsol}
        Z(k^2) =  \frac{b}{k^2} + C(k^2) \; , \quad  b = \hbox{const.} 
\end{equation}
and retaining only the infrared subleading second term in the integrals, 
eq.~(\ref{eq:Mandelstam}) leads to a non--linear integral equation for the
function $ C(k^2) $ which, as a solution to this equation, can be shown to
vanish in the infrared by some non--integer exponent of the
momentum \cite{Man79},  
\begin{equation} 
  C(k^2) \sim  (k^2)^{\gamma_0} \; , \quad
  \gamma_0 = \sqrt{\frac{31}{6} - 1} \simeq 1.273 \; ,
  \quad \hbox{for} \; k^2 \to 0 \; .
\end{equation}
Subsequently, an existence proof, a discussion of the singularity structure
and an asymptotic expansion of the solution generalizing Mandelstam's
discussion of the leading behavior of $C(k^2)$ in the infrared was given by
Atkinson et al. \cite{Atk81}. 

It was later observed
by Brown and Pennington that it is superior for several quite general
reasons, to be discussed in more detail in the next section, to contract
eq.~(\ref{glMA}) with the tensor ${\cal R}_{\mu\nu}(k) = \delta_{\mu\nu} - 4
k_\mu k_\nu /k^2$. This led to a somewhat modified equation for the gluon
renormalization function, in particular, without quadratically ultraviolet
divergent terms~\cite{Bro89}, 
\begin{eqnarray}  \label{eq:Brown}
  \frac{1}{Z(k^2)}
     &=& Z_3 \,+\, \frac{g^2}{16\pi^2}\int_0^{k^2} \frac{dq^2}{k^2} 
              \left(   \frac{7}{2}\frac{q^4}{k^4}
                     - \frac{17}{2}\frac{q^2}{k^2}
                     - \frac{9}{8} \right) Z(q^2) \nonumber \\
         && +\, \frac{g^2}{16\pi^2} \int_{k^2}^{\Lambda^2}\frac{dq^2}{k^2}
              \left(   \frac{7}{8}\frac{k^4}{q^4}
                     - 7\frac{k^2}{q^2} \right) Z(q^2)  . 
\end{eqnarray}
In a previous publication \cite{Hau96} we demonstrated that the solution to this equation
has an infrared behavior quite similar to the solution of Mandelstam's
original equation~(\ref{eq:Mandelstam}). In particular, explicitly separating
the leading infrared contribution according to (\ref{MAsol}) we obtained
a unique solution of the form 
\begin{equation} 
  C(k^2) \sim  (k^2)^{\gamma_0} \, , \;
  \gamma_0 = \frac{2}{9} \sqrt{229} \cos\left(\frac{1}{3}
  \arccos\left(-\frac{1099}{229\sqrt{229}}\right)\right) - \frac{13}{9}
  \simeq 1.271 \; ,
\end{equation}
for $k^2 \to 0$. We solved both equations, eq.~(\ref{eq:Mandelstam}) as well as
eq.~(\ref{eq:Brown}), using a combination of numerical and analytic methods. In
the infrared, we applied the asymptotic expansion technique of ref.\
\cite{Atk81} and calculated successive terms recursively. The asymptotic
expansions obtained this way were then matched to the iterative numerical
solution. The results proved independent of the matching point for a
sufficiently wide range of values, see \cite{Hau96} for details. 

Logarithmic ultraviolet divergences are absorbed in the gluon
renormalization constant $Z_3$ which can be shown to obey the identity 
$Z_g Z_3 = 1$ in Mandelstam approximation \cite{Hau96}. This entails that the
product of the coupling and the gluon propagator, $g D_{\mu\nu}(k)$, does not
acquire multiplicative renormalization in this approximation scheme. Using a
non--perturbative momentum subtraction scheme corresponding to the
renormalization condition 
\begin{equation} 
  Z(k^2 = \mu^2) = 1
\end{equation} 
for some arbitrary renormalization point $\mu^2 > 0 $, the resulting equation
can in both cases be cast in a renormalization group invariant form
determining the renormalization group invariant product $g Z(k^2)$ which
is equivalent to the running coupling $\bar g(t,g)$ of the scheme,
\begin{equation} \label{rc.MA}
  g Z(k^2) = \bar g(t_k,g) \; , \quad t_k = \frac{1}{2} \ln k^2/\mu^2 \; . 
\end{equation} 
The scaling behavior of the solution near the
ultraviolet fixed point is determined by the coefficients $\beta_0 = 25/2$
and $\gamma_A^0 = 25/4 $ for eq.~(\ref{eq:Mandelstam}) vs. $\beta_0 = 14$
and $\gamma_A^0 = 7$ for~(\ref{eq:Brown}) which are reasonably close to the
perturbative values for $N_f = 0$, {\it i.e.},  $\beta_0 = 11$ and
$\gamma_A^0 = 13/2$, the difference being attributed to neglected ghost
contributions.  

\begin{figure}[t]
 \centerline{ \epsfig{file=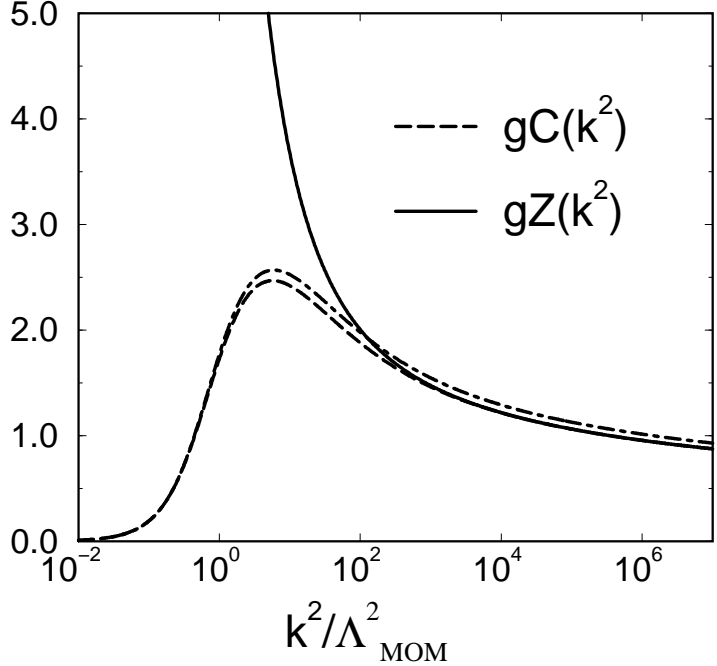,width=0.8\linewidth} }
 \caption{The gluon renormalisation function $g Z(k^2) = 8\pi\sigma /k^2
          + gC(k^2)$ for eq.~(\protect\ref{eq:Brown}). The dashed lines
          show the corresponding function $g C(k^2)$ for eqs.\
          (\protect\ref{eq:Brown}) and (\protect\ref{eq:Mandelstam}).} 
 \label{fig:gluon}
\end{figure}

The full non--perturbative solutions to both equations resulting from
Mandelstam's approximation scheme,
eq.~(\ref{eq:Mandelstam}) and eq.~(\ref{eq:Brown}), are compared in figure
\ref{fig:gluon}. The most important feature of these results to be noted in
the present context is their implication towards an infrared enhanced quark
interaction from the (when ghosts are neglected) renormalization group
invariant product   
\begin{equation} 
g D_{\mu\nu}(k)  \, = \, {\cal P}_{\mu\nu}(k) \,\left( \,
 \frac{8\pi\sigma}{k^4}\,
 + \, \frac{gC(k^2)}{k^2} \, \right)  \label{ir_enh}
\end{equation} 
which would allow to identify the string tension $\sigma $ and relate it to
the scale $\Lambda_{\hbox{\tiny MOM}}$ of the subtraction scheme \cite{Hau96}.
The initial question that led to our present study was whether the presence
of ghosts in Landau gauge could affect these conclusions.  

It is worth mentioning that this result of the Mandelstam approximation, the
infrared enhanced gluon propagator (\ref{ir_enh}), was in
some contradiction with implications from other studies ranging from lattice
calculations \cite{Ber94,Mar95} to implications from complete gauge fixings
\cite{Zwa94}. In addition, an alternative approach to studying the 
Dyson--Schwinger equations of QCD (in Landau gauge) exists which also leads to
dissentive conlusions~\cite{Hae90,Sti95}. This approach is based on the
observation that an exact vertex function and its perturbative estimate differ
by expressions which, in general, contain essential singularities in the
coupling of the same structure as in the spontaneous mass scale, see eq.\
(\ref{Lambda}). The idea therefore is to improve the perturbative expansion in
approximating the Euclidean proper vertices $\Gamma$ by a double series in the
coupling as well as in the degree of a rational approximation  with respect to
the spontaneous mass scale $\Lambda_{\rm QCD}$ being non--analytic in $g^2$,
{\it i.e.}, the proper vertices are approximated by a truncated double series
in functions $\Gamma ^{(r,l)}$ where $r$ denotes the degree of the rational
approximation while $l$ represents the loop order of perturbative corrections
in $g^2$ calculated from $\Gamma ^{(r,0)}$ instead of $\Gamma ^{(0)}_{pert.}$,
see ref.~\cite{Sti95}. The requirement that these non--perturbative terms
reproduce themselves in the Dyson--Schwinger equations is then used to further
restrict the rational Ans\"atze leaving only a few possible solutions. This
method preserves the perturbative renormalizibility. However, due to the 
complexity of the resulting equations the approach has to be restricted to only
a few proper vertex functions. These are then taken as a subset of the seven
primitively divergent proper vertex functions of QCD: the (inverse) gluon,
ghost and quark propagators and the 3--gluon, 4--gluon,  gluon--ghost and
gluon--quark vertex functions. Preliminary studies within this approach,
restricted to $r=1$ and $l=0$ for pure QCD in Landau gauge and assuming a
perturbative ghost propagator, find an infrared vanishing gluon propagator
$\sim k^2$ for small spacelike momenta~\cite{Hae90}. A possible problem of this
approach might be the existence of  light--cone singularities in the 3--gluon
vertex function which may be attributed to the low order of the employed
approximation. A non--trivial ghost selfenergy becomes unavoidable also in this
approach when perturbative corrections are taken into account via Operator
Product Expansion techniques \cite{Ahl92}. In the meantime several solutions to
the order $r=1$ and $l=1$ have been found. However, even though at the present
level the results might not be regarded as absolutely conclusive, neither of
these solutions gives  rise to an infrared enhancement of the gluon propagator
\cite{Sti97}.

One might think that the particular problem with ghosts in the Landau gauge can
be avoided using gauges such as the axial gauge, in which there are no ghosts
in the first place. As for studies of the gluon Dyson--Schwinger equation in
the axial gauge \cite{BBZ81,Sch82,Cud91} it is important to note that, so far,
these rely on an assumption on the tensor structure of the gluon propagator. In
particular, an independent additional term in the structure of the gluon
propagator has not been included in presently available studies of the gluon
DSE in axial gauge~\cite{Bue95}. This term vanishes in perturbation theory. If,
however,  the complete tensor structure of the gluon propagator in axial gauge
is taken into account properly, one arrives at a coupled system of equations
which is of considerably higher complexity than the ghost--gluon system in the
Landau gauge~\cite{Alk--}. Some preliminary progress in this direction has so
far only been obtained in light--cone gauge~\cite{Vac94}. Available studies of
the gluon propagator in the axial gauge can therefore not be regarded any more
conclusive than those in Landau gauge. In fact, the present situation in axial
gauge DSE studies can be summarized to be quite comparable to the Mandelstam
approximation in Landau gauge: Early truncation and approximation schemes to
the gluon DSE in axial gauge, due to the simplified tensor structur, lead to
equations quite similar to Mandelstam's equation~\cite{BBZ81,Sch82}.
Furthermore, these studies led to an infrared enhanced gluon propagator, $D(k)
\propto \sigma/k^4 $ for $ k^2 \to 0$, analogous to the solution to
Mandelstam's equation~\cite{Man79,Atk81,Bro89,Hau96}. Subsequent studies came
to a somewhat dissentive conclusion obtaining an infrared vanishing  gluon
propagator in the identical scheme in axial gauge~\cite{Cud91}. This was,
however, later attributed to a sign error in the resulting
equation~\cite{Bue95}. Therefore, previous DSE studies of the gluon propagator
in Landau gauge as well as in the conventional axial gauge seemed to agree in
indicating an infrared enhanced gluon propagator $\propto \sigma/k^4 $. On one
hand, such a solution is known to lead to an area law in the Wilson
loop~\cite{Wes82}, on the other hand it was also argued from positivity
constraints that the full axial gauge gluon propagator should not be more
singular than $1/k^2$ in the infrared \cite{Wes83}. While this apparent puzzle
might be resolved by using suitable principle value prescriptions for the axial
gauge singularities \cite{Procs}, it is to be expected that the complete tensor
structure of the gluon propagator in the axial gauge will have to be fully
taken into account before a more conclusive picture can emerge. As for the
Landau gauge, we will find that, instead of the gluon propagator, the
previously neglected ghost propagator assumes an infrared enhancement similar
to what was then obtained for the gluon.   

Progress is  desirable in axial gauge as well, of course. A further
prerequisite necessary for this, however, will be a proper treatment of the
spurious infrared divergences which are well known to be present in axial gauge
due to the zero modes of the covariant derivative \cite{Procs}. This can be
achieved by either introducing redundant  degrees of freedom, {\it i.e.}
ghosts, also in this gauge \cite{Lav89} (by which it obviously looses its
particular advantage) or by using a modified axial gauge \cite{Len94}, which is
specially designed to account for those zero modes. Ultimately, progress in
more than one gauge will be the only reliable way to asses the influence of
spurious gauge dependencies.

\section{Infrared Dominance of Ghost Contributions} 

From now on we concentrate on the coupled system of integral equations
derived in section 2 for the propagators of gluons {\sl and} ghosts. Instead
of attempting its direct numerical solution, we will solve this
system simultaneously in an one--dimensional approximation which allows for a
thorough analytic discussion of the asymptotic infrared behavior of its
solutions. This is necessary in order to obtain stable numerical results from
a matching procedure which generalizes the techniques successfully applied to
the gluon DSE in Mandelstam approximation \cite{Hau96}. The approximation we
use to arrive at the one--dimensional system of equations is thereby designed
to preserve the leading order of the integrands in the infrared limit of
integration momenta. At the same time, it will account for the correct short
distance behavior of the solutions (behavior at high integration
momenta). Because of the particular simplicity of the ghost DSE, we start
with this equation to introduce and motivate a modified angle
approximation. From (\ref{ghDSE}) with the vertex (\ref{lgvs}) we obtain the
following equation for the ghost renormalization function $G(k^2)$,
\begin{equation}
  \frac{1}{G(k^2)} \, =\,  \widetilde{Z}_3 -  g^2 N_c \int {d^4q\over
  (2\pi)^4}  \, \biggl( k {\cal P}(p) q \biggr) \,
  \frac{Z(p^2) G(q^2)}{k^2\, p^2\, q^2} \,
  \label{ghDSE1}
\end{equation}
\[ \hskip 50mm
  \times \left( \frac{G(p^2)}{G(q^2)} + \frac{G(p^2)}{G(k^2)} - 1 \right) \; ,
  \quad p = k - q\; .
\]
Here and in the following we use that in Landau gauge $\widetilde Z_1 = 1$
\cite{Tay71}. In order to be able to perform the integration over the
4--dimensional angular variables analytically, we make the following
approximations:

For $q^2 < k^2$ we use the angle approximation for the arguments of the
functions $Z$ and $G$, {\it i.e.}, $G(p^2) = G((k-q)^2) \to G(k^2)$ and
$Z(p^2) \to Z(k^2) $. This obviously preserves the limit $q^2 \to 0$ of the
integrand. 
 
For $q^2 > k^2 $ we make a slightly different assumption, which is that the
functions $Z$ and $G$ are slowly varying with their arguments, and we are
thus allowed to replace $G(p^2) \simeq G(k^2) \to G(q^2)$. This assumption
ensures the correct leading ultraviolet behavior of the equation according
to the resummed perturbative result at one--loop level. For all momenta
being large, {\it i.e.} in the perturbative limit, this approximation is well
justified by the slow logarithmic momentum dependence of the perturbative
renormalization functions for ghosts and gluons. Our solutions will resemble
this behavior, justifying the validity of the approximation in this limit.  

With this approximation, we obtain from (\ref{ghDSE1}) upon angular
integration,
\begin{eqnarray}
 \frac{1}{G(k^2)} &=&  \widetilde{Z}_3 - \frac{g^2}{16\pi^2} \frac{3 N_c}{4} \left\{
\int_0^{k^2} \, \frac{dq^2}{k^2} \frac{q^2}{k^2} \, Z(k^2) G(k^2)
\, + \,\int_{k^2}^{\Lambda^2} \frac{dq^2}{q^2} \,  Z(q^2) G(q^2) \right\} \nonumber \\
&=&  \widetilde Z_3 -  \frac{g^2}{16\pi^2} \frac{3 N_c}{4}  \left( \frac{1}{2} \,
Z(k^2) G(k^2) \, + \,\int_{k^2}^{\Lambda^2} \frac{dq^2}{q^2} \,  Z(q^2) G(q^2)
\right) \; ,
  \label{odGDSE}
\end{eqnarray}
where we introduced an $O(4)$--invariant momentum cutoff $\Lambda$ to account
for the logarithmic ultraviolet divergence, which will have to be absorbed by
the renormalization constant. 
 
A preliminary discussion of the implications of eq.~(\ref{odGDSE}) on the
infrared behavior of the renormalization functions reveals the following:

Making the Ansatz that for $x := k^2 \to 0$ the product of $G$ and $Z$
behaves as $Z(x)G(x) \sim x^\kappa$ for $\kappa \not= 0$, it follows readily
that 
\begin{equation}
  G(x) \sim x^{-\kappa} \quad \hbox{and} \quad  Z(x) \sim x^{2\kappa}
  \quad \hbox{as} \quad  x \to 0 \; .
  \label{irGZ}
\end{equation}
Furthermore, in order to obtain a positive definite function $G(x)$ for
positive $x$ from a positive definite $Z(x)$, as $x\to 0$, we find the
necessary condition $1/\kappa - 1/2 > 0$ which is equivalent to 
\begin{equation}
  0 < \kappa < 2 \; . \label{0lkl2}
\end{equation}
The special case $\kappa = 0$ leads to a logarithmic singularity in
eq.~(\ref{odGDSE}) for $x\to 0$. In particular, assuming that $ZG = c$
with some constant $c > 0$ and $x < x_0$ for a sufficiently small $x_0$,
we obtain $G^{-1}(x) \to c \,(3 N_c g^2/64\pi^2) \,\ln (x/x_0) + \hbox{const}$
and thus $G(x) \to 0^-$ for $x \to 0$, showing that no positive definite
solution can be found in this case either. 

The gluon DSE (\ref{glDSE}) is more complicated. In a manifestly gauge
invariant formulation the gluon Dyson--Schwinger equation in the covariant
gauge would be transverse without further adjustments. For the following
decomposition of all its contributions to the inverse gluon propagator: 
\begin{equation}
  D^{-1}_{\mu\nu}(k) \, =
  \,  A(k^2) \, k^2 \,\delta_{\mu\nu}  \, -\, B(k^2)\, k_\mu k_\nu \, +
  \,\frac{Z_3}{\xi_0} \, k_\mu k_\nu \; ,
\end{equation}
this implies that $A(k^2) = B(k^2) = Z^{-1}(k^2)$. The longitudinal part of the
gluon propagator does not acquire dressing and cancels with the one of the
tree--level propagator on the r.h.s. of the gluon DSE (\ref{glDSE}) \ ($\xi =
Z_3^{-1} \xi_0$). Although we can do our best to use implications of gauge
invariance in constructing vertex functions, we cannot expect to arrive at an
exactly gauge covariant truncation scheme. This fact is reflected in $A \not=
B$ as obtained from truncated gluon DSEs. An additional source of spurious
longitudinal terms in the gluon DSE is the regularization by an
$O(4)$--invariant Euclidean cutoff $\Lambda $ which violates the residual
local invariance, {\it i.e.}, the invariance under transformations generated
by harmonic gauge functions ($\partial^2 \Lambda(x) =0$). The straightforward
elimination of spurious longitudinal terms by contracting eq.~(\ref{glDSE})
with the transversal projector ${\cal P}_{\mu\nu}(k)$ is known to result in
quadratically ultraviolet divergent contributions which are of course
artifacts of the regularization not being gauge invariant. As observed by
Brown and Pennington \cite{Bro88}, in general, quadratic ultraviolet
divergences can occur only in $A(k^2)$. Therefore, this part cannot be
unambiguously determined, it depends on the momentum routing. An unambiguous
procedure is to isolate $B(k^2)$ by  contracting the truncated gluon DSE with
the projector 
\begin{equation}
  {\cal R}_{\mu\nu}(k) = \delta_{\mu\nu} - 4 \, \frac{k_\mu k_\nu}{k^2}
  \quad , \label{proR}  
\end{equation}
and to set $Z(k^2) = 1/B(k^2)$. Note that an additional advantage of this
prescription is that the constant contribution of the tadpole term in the
full gluon DSE does not enter in $B(k^2)$ either. Since the
non--perturbative tadpole does not necessarily vanish, it is in fact an
example of a contribution which, if neglected, can lead to $ A \not= B$. From
these various reasons it is clear that one should concentrate on $B$ rather
than $A$ in truncated gluon DSEs \cite{Bro88,Bro89}.

We gave the detailed justification for this, because it implies that the usual
argument of the irrelevance of longitudinal terms in the ghost--gluon vertex in
Landau gauge does not apply to the ghost--loop in the gluon DSE (\ref{glDSE}).
We are interested in just the unambiguous term $B$ proportional to $k_\mu
k_\nu$. Therefore, contracting (\ref{glDSE}) with ${\cal R}(k)$, we see that
we have to use the full form of the vertex given in eq.~(\ref{fvs}) in the
gluon DSE. With this and the 3--gluon vertex as given in
eqs.~(\ref{3gv}/\ref{3gva}) we obtain,
\begin{eqnarray}
  \frac{1}{Z(k^2)} &=&
     Z_3 \,-\, Z_1\frac{ g^2 N_c}{6} \int \frac{d^4q}{(2\pi)^4} \, \left\{
       N_1(p^2,q^2;k^2)  \, \frac{Z(p^2)G(p^2) Z(q^2)G(q^2)}{Z(k^2)G^2(k^2)} 
     \right. \nonumber \\
     &&\hskip -1.5cm \left. + \,
        N_2(p^2,q^2;k^2) \, \frac{Z(p^2)G(p^2)}{G(q^2)} \, +
        N_2(q^2,p^2;k^2) \, \frac{Z(q^2)G(q^2)}{G(p^2)}\,
\right\} \,    \frac{G(k^2)}{k^2\, p^2\, q^2}  \nonumber \\
    && \hskip -1cm +  \frac{ g^2N_c}{3} \int \frac{d^4q}{(2\pi)^4} 
       \Bigg\{ \Bigl(q {\cal R}(k) q\Bigr) 
\Bigl( G(k^2) G(p^2) - G(q^2) G(p^2) \Bigr)     
 \nonumber \\
    &&\hskip 2cm          
                 - \Bigl( q {\cal R}(k) p \Bigr)  G(k^2) G(q^2) \Bigg\} 
                \frac{1}{k^2\, p^2\, q^2}  \; .  \label{ZDSE}
\end{eqnarray}
The functions $N_1(x,y;z) = N_1(y,x;z)$ and $N_2(x,y;z)$ are given in 
appendix \ref{app:C} for completeness. These functions have singularities
for coinciding arguments canceling only in their sum, $N_1(x,y;z) +
N_2(x,y;z) + N_2(y,x;z)$, which determines the 3--gluon loop contribution in
the Mandelstam approximation \cite{Man79,Atk81,Bro89,Hau96}. In the present
case, the modified angle approximation allows us to combine the three terms
of the 3--gluon loop for $q^2 > k^2$ in the same way as in Mandelstam
approximation. For $q^2 < k^2$, however, the second term in the
3--gluon loop in (\ref{ZDSE}) can only be combined with the other two by
encountering an additional error of the form
\begin{equation}
   -  Z_1 g^2 \frac{N_c}{6} \int \frac{d^4q}{(2\pi)^4}\, N_2(p^2,q^2;k^2)
       \left( \frac{Z(p^2)G(p^2)}{G(q^2)} - \frac{Z(q^2)G(q^2)}{G(p^2)}
\right) \frac{G(k^2)}{k^2\, p^2\, q^2} \; . \label{ochfott}
\end{equation}
In this contribution the singularity in $N_2(x,y;z) \sim 1/({x-y})$ is
cancelled by the terms in brackets which vanish for $q^2 \to p^2$. 
However, employing an angle approximation on these terms, the cancellation of
the singularity is destroyed and an artificial singular contribution would
arise. This demonstrates that approximations have to be used with some care
in order to avoid spurious divergences (and imaginary parts). The angle
approximation cannot be reasonably applied to the contribution
(\ref{ochfott}). We will omit this contribution to the gluon DSE for $q^2 <
k^2$ in addition to the modified angle approximation, since its
inclusion would preclude a one-dimensional reduction of the equations. 

Furthermore, this additional approximation has no influence on the
motivation for the angle approximation for momenta $q^2 < k^2$, {\it i.e.},
to preserve the infrared limit of the integrands, because all contributions
of the 3--gluon loop to the r.h.s in eq. (\ref{ZDSE}) are subleading in the
infrared as compared to the contributions of the ghost loop, as our
preliminary discussion of the infrared behavior of the solutions reveals (see
below). We will therefore put particular emphasis on the ghost contributions
to the coupled system of Dyson--Schwinger equations. In order to isolate
their effect, it seems reasonable to treat the 3--gluon loop as analogous as
possible to previous studies of the gluon DSE in the Mandelstam
approximation. This is achieved by the modified angle approximation to
(\ref{ZDSE}), if in addition, for $q^2 < k^2$, the contribution
(\ref{ochfott}) in the 3--gluon loop is omitted. This yields from
(\ref{ZDSE}) upon angular integration, 
\begin{eqnarray}
  \frac{1}{Z(k^2)}
     &=& Z_3 + Z_1 \frac{g^2}{16\pi^2} \frac{N_c}{3}
         \left\{ \int_{0}^{k^2} \frac{dq^2}{k^2}
         \left( \frac{7}{2}\frac{q^4}{k^4}
         - \frac{17}{2}\frac{q^2}{k^2}
         - \frac{9}{8} \right) Z(q^2) G(q^2) \right. \nonumber \\
     && + \, \left. \int_{k^2}^{\Lambda^2} \frac{dq^2}{q^2} \left(
            \frac{7}{8} \frac{k^2}{q^2} - 7 \right) Z(q^2) G(q^2)
          \right\} \label{odZDSE} \\
     && \hskip -2cm 
+ \, \frac{g^2}{16\pi^2} \frac{N_c}{3} \left\{ \int_{0}^{k^2}
         \frac{dq^2}{k^2}
         \frac{3}{2} \frac{q^2}{k^2} G(k^2) G(q^2) - \frac{1}{3} G^2(k^2)
         + \frac{1}{2} \int_{k^2}^{\Lambda^2} \frac{dq^2}{q^2}
         G^2(q^2) \right\} \; . \nonumber
\end{eqnarray}
The only difference in the 3--gluon loop as obtained here versus the
Mandelstam approximation is that the gluon renormalization function $Z$ is
replaced by the product $ZG$, see eq.~(\ref{eq:Brown}) in sec.~3. This is one
consequence the presence of the ghost renormalization function has on the
dressing of the 3--gluon vertex. The system of equations (\ref{odGDSE}) and
(\ref{odZDSE}) is a direct extension to the gluon DSE in the Mandelstam
approximation. As such it is very well suited to investigate the influence of
ghosts on conclusions of previous studies based on this approximation. 

\begin{figure}
  \centerline{ \epsfig{file=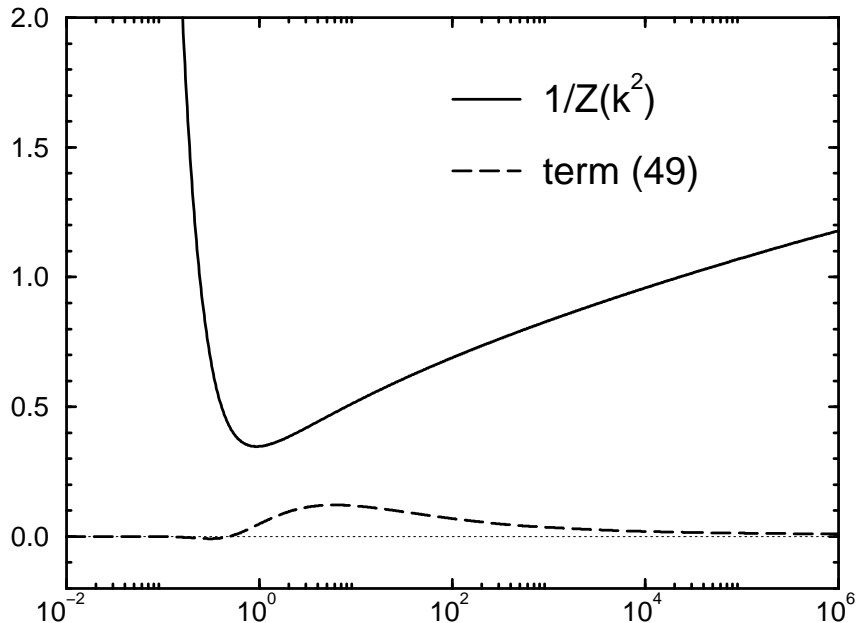,width=0.8\linewidth} }
  \caption{The dismissed contribution (\protect\ref{ochfott}) (dashed line)
  compared to the contributions retained on the r.h.s of
  (\protect\ref{odZDSE}), {\it i.e.}, $1/Z(k^2)$ (solid line) as a function
  of $x = k^2/\sigma$ with $\sigma\simeq (350\hbox{MeV})^2$, {\it c.f.},
  sec. 8.} 
  \label{dismissed}
\end{figure}

To asses the relative importance of the neglected term 
(\ref{ochfott}) we calculated this contribution without angle
approximation using, however, the selfconsistent solutions for $Z(k^2)$ and
$G(k^2)$ as obtained from the one--dimensional set of equations
(\protect\ref{odGDSE}) and (\ref{odZDSE}), {\it c.f.}, the numerical method
and results given in section 8 below. In figure
\ref{dismissed} we plot the resulting contribution in comparison with the
terms retained on the r.h.s. of eq. (\ref{odZDSE}) which, for selfconsistent
solutions, are equivalent to $1/Z(k^2)$. Though even small terms can, in
principle, have a considerable effect on the non--linear selfconsistency
problem, the fact that the additional contribution to the gluon DSE due to
(\ref{ochfott}) is comparably small for all momenta supports this additional
approximation. In particular, neither the infrared nor the ultraviolet
asymptotic regimes are affected by this contribution at all.

Before we turn to the renormalization of equations (\ref{odGDSE}) and 
(\ref{odZDSE}) in the next sections, we conclude this section with an
extension of our preliminary infrared discussion. From (\ref{odGDSE}) we find
that $ZG \to c x^\kappa $ for $x \to 0$ implies,
\begin{eqnarray}
  G(x) &\to & \left( g^2\gamma_0^G \left(\frac{1}{\kappa} - \frac{1}{2}
  \right) \right)^{-1}  c^{-1} x^{-\kappa} \; , \quad \gamma_0^G =
  \frac{1}{16\pi^2}  \; \frac{3 N_c}{4}  \; ,
  \label{loirG}\\ 
  Z(x) &\to &  \left( g^2\gamma_0^G \left(\frac{1}{\kappa} - \frac{1}{2}
  \right) \right) \,  c^{2} x^{2\kappa}   \; , \quad 0<\kappa < 2  \; ,
  \label{loirZ1}
\end{eqnarray} 
where $\gamma_0^G$ is the leading order perturbative coefficient of the
anomalous dimension of the ghost field. Accordingly, the ghost--loop
gives infrared singular contributions $\sim x^{-2\kappa}$ to the gluon
equation (\ref{odZDSE}) while the 3--gluon loop yields terms proportional to
$ x^\kappa $ as $x\to 0$, which are thus subleading contributions to
the gluon equation in the infrared. With eq. (\ref{loirG}) the leading
asymptotic behavior of eq. (\ref{odZDSE}) for $x \to 0$ shows that
\begin{equation} Z(x) \,\to \,  g^2\gamma_0^G \, \frac{9}{4} \left(\frac{1}{\kappa} -
\frac{1}{2}   \right)^2 \left( \frac{3}{2}\, \frac{1}{2-\kappa} - \frac{1}{3}
+ \frac{1}{4\kappa} \right)^{-1} \, c^2 x^{2\kappa} \quad \Rightarrow
\label{loirZ2}\end{equation}  
\begin{equation}
  \left( \frac{3}{2}\, \frac{1}{2-\kappa} - \frac{1}{3} +
  \frac{1}{4\kappa} \right) \, \stackrel{!}{=} \, \frac{9}{4}
  \left(\frac{1}{\kappa} - \frac{1}{2} \right)  \; \Rightarrow \;
  \kappa \, = \, \frac{61 \stackrel{(+)}{-} \sqrt{1897}}{19} \,
         \simeq \, 0.92  \; .
  \label{kappa}
\end{equation}
In the last line we compared (\ref{loirZ2}) to (\ref{loirZ1}) and used the
restriction (\ref{0lkl2}) obtained from the ghost equation on the exponent
$\kappa$ (ruling out the other sign in (\ref{kappa})). Note that $\kappa$
is independent of the number of colors $N_c$.

This leading behavior of the gluon and ghost renormalization functions and
thus their propagators is entirely due to ghost contributions. The details of
the treatment of the 3--gluon loop have no influence on above
considerations. This is in remarkable contrast to the Mandelstam
approximation, in which the 3--gluon loop alone determines the infrared
behavior of the gluon propagator and the running coupling in Landau gauge
\cite{Man79,Atk81,Bro89,Hau96}. As a result of this, the running coupling as
obtained from the Mandelstam approximation is singular in the infrared
\cite{Hau96}. In contrast, as we will show in the next section, the infrared
behavior derived from the present truncation scheme implies an infrared
stable fixed point. This is certainly a counter example to the frequently
quoted assertion that the presence of ghosts in Landau gauge may have
negligible influence on physical observables at hadronic energy scales.

With the infrared behavior of gluon and ghost propagators obtained
above, a comment on its implications for the vertex functions is in order. 
Starting with the ghost--gluon vertex (\ref{fvs}) we realize that the 
limit of vanishing ghost momenta is regular: 
\begin{eqnarray}
  G_\mu(p,q) \to & -\, ik_\mu & \qquad\mbox{for}\quad p \to 0  \\
  G_\mu(p,q) \to & 0     & \qquad\mbox{for}\quad q \to 0
  \quad ,
\end{eqnarray}
where we used that $G(k^2) \sim (k^2)^{-\kappa}$ for $ k^2 \to 0$. On the
other hand, for vanishing gluon momentum $k \to 0$ the vertex diverges as
\begin{equation}
  G_\mu(p,q) \to 2 ip_\mu \, \frac{G(k^2)}{G(p^2)}\,
             \sim \, \frac{2 ip_\mu }{G(p^2)}\,  \frac{1}{ (k^2)^{\kappa}}
    \qquad\mbox{for}\quad k \to 0
  \quad .
\end{equation}
Since the exponent $\kappa$ is a (positive) irrational number, see
eq.~(\ref{kappa}), the corresponding divergence cannot be interpreted to
reflect the presence of a massless excitation. This divergence is in fact
weaker than a massless particle pole ($\kappa < 1$) and presumably lacking a
physical interpretation.

Similarly, the 3--gluon vertex as given in equations
(\ref{3gv}/\ref{3gva}) shows analogous infrared divergences resulting from 
the infrared enhanced ghost renormalization function,
\begin{eqnarray}
\Gamma_{\mu\nu\rho}(p,q,k) &\to&   G(k^2) \, \Bigg\{ \Bigl( ip_\mu
\delta_{\nu\rho} + ip_\nu \delta_{\mu\rho} - 2 ip_\rho \delta_{\mu\nu} \Bigr)
\frac{1}{Z(p^2)} \label{diff3gl} \\
&&  +\, 2 ip_\rho \,  p^2  \, {\cal P}_{\mu\nu}(p)\, \biggl(
\frac{2G'(p^2)}{G(p^2)Z(p^2)} + \frac{Z'(p^2)}{Z^2(p^2)} \biggr) \Bigg\}
    \qquad\mbox{for}\quad k \to 0 \; .
 \nonumber
\end{eqnarray}
This is the correct result to satisfy the differential Slavnov-Taylor
identity obtained from eq. (\ref{glWTI}) for $k\to 0$. Note that in the
previous studies of the gluon DSE in Mandelstam approximation
\cite{Man79,Atk81,Bro89,Hau96}, the requirement for the 3--gluon vertex
function to obey the differential Slavnov--Taylor identity led to the
so--called masslessness condition, eq. ~(\ref{massless.MA}), $p^2/Z(p^2) \to
0$ for $p^2 \to 0$. In absence of ghost contributions, the gluon DSE had to
be supplemented by this as an additional constraint. This original condition
is violated by the infrared behavior of the gluon propagator found here, $
Z(x) \to x^{2\kappa}$. The correct replacement of this condition for the
present case using (\ref{glWTI}) with the solution (\ref{3gv}/\ref{3gva}) is,
however, 
\begin{equation} 
p^2 G(p^2) \to 0 \qquad \hbox{and} \quad \frac{p^2}{G(p^2)Z(p^2)} \to 0
\qquad \hbox{for} \quad p^2 \to 0 \; .
\label{massless}
\end{equation}
These two necessary conditions are obeyed by the infrared behavior obtained
here, {\it i.e.}, $G(x) \sim 1/(G(x)Z(x)) \sim x^{-\kappa}$  with $\kappa
\simeq 0.92$ for $ x \to 0$, without further adjustments. 
We see, however, that generally the original bounds on $\kappa $ obtained
from the consistency of the ghost DSE, when combined with conditions
(\ref{massless}), have to be restricted further to $ 0 < \kappa < 1$. The
possibility that the masslessness condition in its original form might not
hold for non--Abelian gauge theories has been pointed out previously
\cite{Kug94,Hat83}. Our present results support this conjecture.

Finally, since we do not expect quarks to screen the divergence in the ghost
renormalization function completely, its presence in the quark--gluon vertex
function, see eq. (\ref{qkSTIsol}), will have a similar effect there, too. We 
will demonstrate on the example of the running coupling in the next section,
how these apparently unphysical divergences in the elementary correlation
functions of gluons, ghosts and quarks can, in principle, nevertheless cancel
in physical quantities. The lack of a physical interpretation of these
divergences (other than maybe reflecting confinement) should, however, not be
too surprising for a Euclidean field theory violating reflection positivity
(see sec. 11). Note also that in all skeletons of the kernels in relativistic
bound state equations for hadrons, the combination of dressed quark--gluon
vertices with the non--perturbative gluon propagator will give contributions
of the form ({\it c.f.}, eqs. (\ref{qkSTI},\ref{qkSTIsol})), 
\begin{equation}
\sim   g^2 \,   G^2(k^2)\, D_{\mu\nu} (k)  \;  \Gamma_\nu^{\hbox{\tiny
BC}}(p,p-k)  \otimes \Gamma_\mu^{\hbox{\tiny BC}}(q-k,q)  \; ,
\end{equation} 
where $\Gamma^{\hbox{\tiny BC}}_\mu(p,q)$ stands for a 
vertex function with non--perturbative dressing of similar structure to what
occurs in an Abelian theory, {\it e.g.}, a Ball--Chiu construction from its
Ward--Takahashi identity. This combination of the ghost renormalization
functions with the gluon propagator, $g^2 G^2(k^2) D(k)$, in the effective
quark interaction is free of the unphysical infrared divergences for $k \to
0$. It is furthermore independent of the renormalization scale in Landau
gauge and suited to define a non--perturbative running coupling as we discuss
next.  

\section{Subtraction Scheme and Non--Perturbative Running Coupling}

Before we discuss the renormalization of the Dyson--Schwinger equations for
gluons and ghosts in the next section, some introductory remarks on the choice
of the non--perturbative subtraction scheme and its relation to the definition
of the running coupling are necessary. In particular, we will see that the
preliminary infrared discussion of the last section already yields an
important first result: it implies the existence of an infrared fixed point.

We begin with the identity for the renormalization constants (as
introduced in eq.~(\ref{Zds})), 
\begin{equation}
  \widetilde{Z}_1 \, = \, Z_g Z_3^{1/2} \widetilde{Z}_3 \, = \, 1 \; ,
  \label{wtZ1}
\end{equation}
which has been shown to hold in the Landau gauge \cite{Tay71}. In
fact, our Slavnov--Taylor identity (\ref{ghWTI}) proves that it remains valid
in general covariant gauges to the extend that irreducible 4--ghost
correlations are neglected. From eq.~(\ref{wtZ1}) it follows that the product
$g^2 Z(k^2) G^2(k^2)$ is renormalization group invariant. In absence of any
dimensionful parameter this (dimensionless) product is therefore a function
of the running coupling $\bar g$,
\begin{equation}
  g^2 Z(k^2) G^2(k^2) = f( \bar{g}^2(t_k, g)) \; , \quad
  t_k = \frac{1}{2} \ln k^2/\mu^2 \; .
  \label{gbar}
\end{equation}
Here, the running coupling $\bar g(t,g)$ is the solution of
$d/dt \, \bar g(t,g) = \beta(\bar g) $ with $\bar g(0,g) = g$ and the
Callan--Symanzik $\beta$--function $\beta (g) = - \beta_0 g^3 + {\cal
O}(g^5)$. The perturbative momentum subtraction scheme is asymptotically
defined by $f(x) \to x$ for $ x\to 0$. This is realized by independently
setting
\begin{equation}
  Z(\mu^2) = 1 \quad \mbox{and} \quad G(\mu^2) = 1
  \label{persub}
\end{equation}
for some asymptotically large subtraction point $k^2 = \mu^2$.
If the invariant product $g^2 Z(k^2) G^2(k^2)$ is to have a physical
meaning, {\it e.g.}, in terms of a potential between static color sources, it
should be independent under changes $(g,\mu) \to (g',\mu')$ according to
the renormalization group for arbitrary scales $\mu'$. Therefore,
\begin{equation} 
  g^2 Z(\mu'^2) G^2(\mu'^2) \, \stackrel{!}{=} \,  g'^2 = \bar g^2(\ln
(\mu'/\mu) , g) \;
, \end{equation} 
and, $f(x) \equiv x $, $ \forall x$. We adopt this as a physically sensible
definition of a non--perturbative running coupling in the Landau gauge.

This definition is an extension to the one we used in the Mandelstam
approximation, eq.~(\ref{rc.MA}). In ref.~\cite{Hau96} we proved in this
approximation without ghosts the identity $Z_g Z_3 = 1 $ implying that $g
Z(k^2)$ is the renormalization group invariant product in this case. The
according identification of this product with the running coupling $ g Z(k^2)
= \bar g(t_k,g)$ is equivalent to the non--perturbative renormalization
condition $Z(\mu^2) = 1, \forall \mu$ in Mandelstam approximation (in which
there is no ghost renormalization function). 

In the present case, it is not possible to realize $f(x) \equiv x$ by simply 
extending the perturbative subtraction scheme (\ref{persub}) to arbitrary
values of the scale $\mu$, as this would imply a relation between the functions
$Z$ and $G$ which is inconsistent with the leading infrared behavior 
of the solutions discussed in the last section. For two independent
functions the condition (\ref{persub}) is in general too restrictive to be
used for arbitrary subtraction points. Rather, in extending the perturbative
subtraction scheme, one is allowed to introduce functions of the coupling
such that 
\begin{equation}
  Z(\mu^2) \, = \, f_A(g) \quad \hbox{and} \quad G(\mu^2) \, = \, f_G(g)
  \quad \hbox{with} \quad f_G^2 f_A \, = \, 1 \; ,
  \label{npsub}
\end{equation}
and the limits $ f_{A,\, G} \to 1 \, , \; g \to 0 $. Using this it is
straightforward to see that for $k^2 \not= \mu^2$ one has ($t_k = (\ln
k^2/\mu^2)/2$),
\begin{eqnarray}
  Z(k^2) &=& \exp\bigg\{ -2 \int_g^{\bar g(t_k, g)} dl \,
  \frac{\gamma_A(l)}{\beta(l)} \bigg\} \, f_A(\bar g(t_k, g)) \; ,
  \label{RGsol} \\ 
   G(k^2) &=& \exp\bigg\{ -2 \int_g^{\bar g(t_k, g)} dl \, 
  \frac{\gamma_G(l)}{\beta(l)} \bigg\} \, f_G(\bar g(t_k, g)) \; .
  \nonumber
\end{eqnarray}
Here $\gamma_A(g)$ and $\gamma_G(g)$ are the anomalous dimensions
of gluons and ghosts respectively, and $\beta(g)$ is the Callan--Symanzik 
$\beta$--function. Eq. (\ref{wtZ1}) corresponds to the following identity 
for these scaling functions in Landau gauge:
\begin{equation}
  2 \gamma_G(g) \, +\, \gamma_A(g)  \, = \, -\frac{1}{g} \, \beta(g)  \; .
  \label{andim}
\end{equation}
Therefore, we verify that the product $g^2 Z G^2$ indeed gives the running
coupling ({\it i.e.}, eq. (\ref{gbar}) with $f(x) \equiv x$). Perturbatively,
at one--loop level eq. (\ref{andim}) is realized separately, {\it i.e.},
$\gamma_G(g) = - \delta \, \beta(g) /g$ and $ \gamma_A(g) = - (1-2\delta)\,
\beta(g)/g$ with $\delta = 9/44$ for $N_f=0$ and arbitrary
$N_c$. Non--perturbatively we can separate these contributions from the
anomalous dimensions by introducing an unknown function $\epsilon(g)$,
\begin{equation}
  \gamma_G(g) \, =:\, - (\delta + \epsilon (g) ) \,
  \frac{\beta(g)}{g}
  \; \Rightarrow \quad  \gamma_A(g) \, =\, - (1 - 2\delta - 2\epsilon (g)
  ) \, \frac{\beta(g)}{g}  \;  . \label{RGgam}
\end{equation}
This allows us to rewrite (\ref{RGsol}) as follows:
\begin{eqnarray}
  Z(k^2) &=& \biggl( \frac{\bar g^2(t_k, g)}{g^2} \biggr)^{1-2\delta} 
  \exp\bigg\{ - 4\int_g^{\bar g(t_k, g)} dl \, \frac{\epsilon(l)}{l} \bigg\}
  \, f_A(\bar g(t_k, g)) \; , \label{RGsol1} \\ 
  G(k^2) &=& \biggl( \frac{\bar g^2(t_k, g)}{g^2}  \biggr)^{\delta}  \,
  \exp\bigg\{ 2 \int_g^{\bar g(t_k, g)} dl \, \frac{\epsilon(l)}{l} \bigg\}
  \, f_G(\bar g(t_k, g)) \; .
  \nonumber
\end{eqnarray}
This is generally possible, also in the presence of quarks, in which case 
$\delta = \gamma_0^G/\beta_0 = 9N_c/(44 N_c - 8 N_f)$ for $N_f$ flavors in
Landau gauge. The above representation of the renormalization functions
expresses clearly that regardless of possible contributions from the unknown
function $\epsilon(g)$, the resulting exponentials cancel in the product $G^2
Z$. For a parameterization of the renormalization functions, these
exponentials can of course be absorbed by a redefinition of the functions
$f_{A,\, G}$. The only effect of such a redefinition is that the originally
scale independent functions $f_{A,\, G} (\bar g(t_k, g))$ will acquire a
scale dependence by this, if $\epsilon \not= 0$. We therefore use the form
(\ref{RGsol1}) to motivate the following parameterization for $G$ and $Z$,
\begin{eqnarray}
  Z(k^2) &=& \left( \frac{ F(x) }{F(s)} \right)^{1-2\delta}  \, R^2(x)
  \; , \label{parZG}\\
  G(k^2) &=& \left( \frac{ F(x) }{F(s)} \right)^\delta  \, \frac{1}{R(x)} \quad
  \hbox{with}  \quad x := k^2/\sigma \quad \hbox{and} \quad  s :=
  \mu^2/\sigma  \; ,
  \nonumber
\end{eqnarray}
where $\sigma$ is some currently unfixed (RG invariant) scale parameter. This
obviously implements the renormalization condition $G^2 Z\vert_{k^2 = \mu^2}
= 1$ and from the definition of the running coupling (\ref{gbar}) we find
that $\bar g^2(t_k,g) \sim F(x)$. We fix the constant of proportionality for
later convenience by setting (with $\beta_0 = 11 N_c/(48\pi^2)$ for $N_f=0$
quark flavor), 
\begin{equation}
  \beta_0 \, \bar g^2(t_k,g) \, = \,  F(x)
  \quad \hbox{and} \quad
  \alpha_S(\mu) \, = \, \frac{g^2}{4\pi}
                \, = \, \frac{1}{4\pi\beta_0} \, F(s) \; ,
  \label{rcpl}
\end{equation}
accordingly. As stated above, in general, a non--trivial contribution from
$\epsilon(g)$ will result in the function $R$ being scale dependent. For the
present truncation scheme it will be possible, however, to obtain explicitly
scale independent equations to determine the running coupling $\sim F$ as well
as the function $R$ thus showing that the solutions for the renormalization
functions $G$ and $Z$ obey one--loop scaling at all scales. In particular, it
implies that the products $g^{2\delta} G$ and $g^{2(1-2\delta)} Z$ are
separately renormalization group invariants in the present scheme (as they are
at one-loop level). As for the renormalization scale dependence, the
non--perturbative nature of the result is therefore buried entirely in the
result for the running coupling. 

We are now in a position to discuss the implications of the preliminary
results for the infrared behavior of the solutions $G$ and $Z$ at the end
of the last section without actually solving equations
(\ref{odZDSE},\ref{odGDSE}). From eqs.~(\ref{loirG}) and (\ref{loirZ1}) we
find for $k^2 \to 0$, 
\begin{equation}
   g^2 Z(k^2) G^2(k^2) \, = \, \bar g^2(t_k,g) \, \stackrel{t_k \to
  -\infty}{\longrightarrow} \, \left(
  \gamma_0^G \left( \frac{1}{\kappa} - \frac{1}{2} \right) \right)^{-1} 
  =:\, g_c^2  \; .
  \label{gcrit}
\end{equation}
The critical coupling scales with the number of colors as $g_c^2 \sim 1/N_c$
implying that in our approach $g_c^2 N_c$ is constant. This agrees with the
general considerations of the large $N_c$--limit in which $g^2 N_c$ is 
kept fixed as $N_c$ becomes large \cite{tHo74}. For $N_c=3$ and with
eq.~(\ref{kappa}) for $\kappa$ we obtain $g_c^2 \simeq 119.1$ which corresponds
to a critical coupling $\alpha_c = g^2_c/(4\pi) \simeq 9.48$. This is a
remarkable result in its own, if compared to the running coupling as it was
analogously obtained from the Mandelstam approximation \cite{Hau96}. The
dynamical inclusion of ghosts changes the infrared singular coupling of the
Mandelstam approximation to an infrared finite one implying the existence of an
infrared stable fixed point.

\section{Renormalization}

We discuss the actual renormalization of equations (\ref{odGDSE}) and
(\ref{odZDSE}) once again beginning with the simpler and thus more evident
case of the ghost DSE. To see the effect of the renormalization condition we
use the parameterization (\ref{parZG}) of the last section in the ghost DSE
(\ref{odGDSE}) and set $k^2 = \mu^2$ ($\Leftrightarrow x = s$) with $\beta_0
g^2 = F(s)$ and obtain,
\begin{equation}
R(s) \, =\, \widetilde Z_3 - \delta \left( \frac{1}{2} R(s)F(s) \, +\, F^\delta
(s)\, \int_s^L \frac{dy}{y} \, R(y) F^{1-\delta}(y) \right) \; ,
\label{wtZ3}\end{equation} 
where $L := \Lambda^2/\sigma $ is the ultraviolet cutoff  and $\delta =
\gamma_0^G/\beta_0$ ( = 9/44 for $N_f = 0$). This determines the
renormalization constant $\widetilde Z_3$ for arbitrary scales $s$. We can use
(\ref{wtZ3}) to eliminate $\widetilde Z_3$ in (\ref{odGDSE}) and obtain analogously
for general $x \not= s$,
\begin{equation}
\frac{R(x)}{F^\delta(x)} \, - \, \frac{R(s)}{F^\delta(s)}  \, =
\label{renG1}\end{equation}  
\[ \hskip 1cm \delta \left(
\frac{1}{2} \left( R(s)F^{1-\delta}(s) -  R(x)F^{1-\delta}(x) \right)\, +\,
\, \int_s^x \frac{dy}{y} \, R(y) F^{1-\delta}(y) \right) \; . \]
This equation is now ultraviolet as well as infrared finite. Note that the
preliminary infrared analyses corresponds to
\begin{equation} F(x) \to a := \beta_0 \,g_c^2 \simeq 8.3 \quad \hbox{and} \quad R(x) \to b
\, x^\kappa \; , \quad x\to 0 \; , \label{IRFR} \end{equation}
where $b$ is some currently unfixed constant depending on the choice of
the scale parameter $\sigma $ which we did not specify so far. Equation
(\ref{renG1}) is valid for all scales $s$, in particular, we may let $s\to 0$
to obtain, 
\begin{equation}
  \frac{R(x)}{F^\delta(x)} \, = \, \delta \left( \int_0^{\;\; x}
\frac{dy}{y} \,
R(y) F^{1-\delta}(y) \,  - \, \frac{1}{2} \, R(x)F^{1-\delta}(x) \right)
\; .
  \label{renG2}
\end{equation}
We can use this equation in (\ref{wtZ3}) exchanging $x \leftrightarrow s$
to write, 
\begin{equation}
  \widetilde{Z}_3
    = F^\delta(s) \left( \delta \int_0^L \frac{dy}{y} \,
                          R(y) F^{1-\delta}(y) \right) \; .
  \label{wtZ3a}
\end{equation} 
If we use this form for $\widetilde Z_3$ in (\ref{odGDSE}), we arrive
directly at (\ref{renG2}) which shows that it remains valid also for
non--zero scales $s$ (we have thus shown that all $x$--dependent terms cancel
separately from the $s$--dependent ones in eq. (\ref{renG1}) with the
integration range $(s,x)$ split into $(0,s)$ and $(0,x)$). It is instructive
to replace the integral in (\ref{wtZ3a}) by again using eq. (\ref{renG2}) now
with $x \leftrightarrow L$, 
\begin{equation}  
\widetilde Z_3 \, =\, \frac{R(L)}{F^\delta (L)} \, \left( 1 \, +\,
\frac{\delta}{2} \, F(L) \right) \, F^\delta(s) \, \stackrel{L\to
\infty}{\longrightarrow} \, \frac{F^\delta(s)}{F^\delta(L)} \, =\,
\left(\frac{g^2}{g_0^2}\right)^\delta \; , \label{wtZ3b}
\end{equation}
for the perturbative limits of large $x$, $F(x) \to 1/\ln x$ and $R(x) \to 1$,
showing nicely the scaling limit, in particular, the possible interpretation
of the bare coupling as the running coupling at the cutoff scale in the
present scheme. Furthermore, the renormalization constant in the scaling
limit (large $L$) becomes identical to the multiplicative factor for
finite renormalization group transformations of the ghost propagator,
$\widetilde Z_3(\mu^2 ,\mu'^2)$ for $\mu \to \mu'$, at the cutoff scale,
$\widetilde Z_3 =  \widetilde Z_3(\mu^2 ,L)$. This detailed study of the
scaling behavior of the ghost DSE justifies the use of the improved angle
approximation for integration momenta $q^2 > k^2$ in this case. We now turn
to the gluon DSE using the discussion above as a guideline for the
anticipated scaling behavior of the gluon equation (\ref{odZDSE}).  
 
The gluon DSE has the additional problem of the presence of the
renormalization constant for the 3--gluon vertex $Z_1$ in the gluon loop. We
will address this issue in the following. The first step is to rewrite
(\ref{odZDSE}) with (\ref{parZG}) and set $ x = s$,
\begin{eqnarray}
\frac{1}{R^2(s)} &=& Z_3 \, + \, Z_1 \,\frac{F^\delta(s)}{11} \, \Bigg\{
\int_{0}^{s} \frac{dy}{s}
\, \left(  \frac{7}{2}\frac{y^2}{s^2}
                     - \frac{17}{2}\frac{y}{s}
                     - \frac{9}{8} + 7 \frac{s}{y} \right) R(y)
F^{1-\delta}(y)  \nonumber\\
&& + s \frac{7}{8} \, \int_s^\infty \frac{dy}{y^2} \, R(y)
F^{1-\delta}(y) \,  - \, \frac{7}{\delta}
\, \frac{\widetilde Z_3}{F^\delta(s)} \Bigg\} \, + \,
\frac{F^{1-2\delta}(s)}{11} \, 
\Bigg\{ \frac{3}{2} \, \frac{F^\delta(s)}{R(s)} \nonumber\\
&& \times \int_0^s  \frac{dy}{s} \frac{y}{s} \, \frac{F^\delta(y)}{R(y)} \,
-\, \frac{1}{3} \frac{F^{2\delta}(s)}{R^2(s)}  \, +\, \frac{1}{2} \int_s^L
\frac{dy}{y} \, \frac{F^{2\delta}(y)}{R^2(y)} \Bigg\} \; ,
\label{eq:15} 
\end{eqnarray}
where we have used (\ref{wtZ3a}) to express the logarithmically divergent
term in the 3--gluon loop in terms of the constant $\widetilde{Z}_3$.
Furthermore, we used that the integral $\int_x^L (dy/y^2) RF^{1-\delta}
\to \int_x^\infty (dy/y^2) RF^{1-\delta}$ is ultraviolet finite for
functions $R$ and $F$ resembling the perturbative behavior at high
momenta. We now have to specify $Z_1$. We will discuss three possibilities
in the following. The first and probably most obvious will lead to a
contradiction. The numerical results of the other two will not show major
qualitative differences except at short distances. Comparing to the known
perturbative behavior at short distances, we suggest that the third method
of treating $Z_1$ is suited to restore some of the effects of the
truncation which are at the root of the present problem.

\bigskip
{\bf a)} $Z_1 = Z_3/ \widetilde{Z}_3$
\smallskip

The most rigorous way would be to use the Slavnov--Taylor identity $Z_1 = 
Z_3/\widetilde Z_3$ which follows from $\widetilde{Z}_1 = 1$ in Landau
gauge. We could then attempt to eliminate $Z_3$ from (\ref{odZDSE}) in
analogy to (\ref{renG1}) for $x\not= s$ writing, 
\begin{eqnarray}
\frac{1}{R^2(x) F^{1-2\delta}(x)} &-& \frac{1}{R^2(s) F^{1-2\delta}(s)}
\, = \label{eq:16}\\
&& \hskip -2cm 
\frac{Z_3 \, F^\delta(L)}{F^{1-2\delta}(s)}  \, \frac{1}{11} \, \Bigg\{
\int_{0}^{x} \frac{dy}{x}
\, \left(  \frac{7}{2}\frac{y^2}{x^2}
                     - \frac{17}{2}\frac{y}{x}
                     - \frac{9}{8} + 7 \frac{x}{y} \right) R(y)
F^{1-\delta}(y) \nonumber\\
&& 
+ \, x \frac{7}{8} \, \int_x^\infty \frac{dy}{y^2} \, R(y)
F^{1-\delta}(y)   \, - \, ( x\leftrightarrow s ) \Bigg\}  \nonumber\\
&& \hskip -2cm + \, \frac{1}{11} \,
\Bigg\{ \frac{3}{2} \left( \frac{F^\delta(x)}{R(x)}
\int_0^x  \frac{dy}{x} \frac{y}{x} \, \frac{F^\delta(y)}{R(y)} \, -\,
\frac{F^\delta(s)}{R(s)}
\int_0^s  \frac{dy}{s} \frac{y}{s} \, \frac{F^\delta(y)}{R(y)} \right)
\nonumber\\ 
&& 
- \frac{1}{3} \left( \frac{F^{2\delta}(x)}{R^2(x)}  \, - \,
\frac{F^{2\delta}(s)}{R^2(s)} \right) \,  - \, \frac{1}{2} \int_s^x
\frac{dy}{y} \, \frac{F^{2\delta}(y)}{R^2(y)} \Bigg\} \; ,
  \nonumber
\end{eqnarray}
where we used $F^\delta(s) Z_1 = Z_3 F^\delta(L)$ from (\ref{wtZ3b}) for
large $L$. This factor $Z_3 F^\delta(L)$ is the only cutoff dependence left
in (\ref{eq:16}). Expecting a scaling limit as in the ghost case, {\it i.e.},
$Z_3 \to (F(s)/F(L))^{1-2\delta}$, we see that the prefactor of the 3--gluon
loop contribution to (\ref{eq:16}) behaves as $\sim 1/F^{1-3\delta}(L)$ which
is singular in the limit $L\to \infty $, if $1-3\delta > 0$. Perturbatively we
expect this to be the case since $3\delta = 3\gamma_0^G/\beta_0 < 1
\Leftrightarrow  \beta_0 > 27/4$ which is true for $N_f < 7$. There is no
term left in the otherwise ultraviolet finite eq. (\ref{eq:16}) to absorb the
singularity in this case. 

It is interesting to note that perturbatively for $N_f \geq 7$ the leading
anomalous dimensions are such that $Z_1 = Z_3/\widetilde{Z}_3 \to 0 $ in
the scaling limit in Landau gauge. This implies that the contribution of
the 3--gluon loop to the gluon DSE vanishes completely for $N_f \geq 7$.

In order to reproduce the leading logarithmic behavior from perturbation
theory for high momenta, being determined by $ 1 - 3\delta  = 17/44 $ for $N_f
= 0$, we find that it is not possible to renormalize the DSEs for gluons
and ghosts in the present truncation scheme while retaining the identity $Z_1
= Z_3/\widetilde Z_3$. Could it alternatively be possible to find a consistent
solution for the special case $1-3\delta = 0$? Obviously, eq.~(\ref{eq:16})
would have a finite scaling limit in this case. However, 
a detailed study of the asymptotic behavior of eqs.~(\ref{renG2}) and
(\ref{eq:16}) shows that this leads to a contradiction. From
eq.~(\ref{renG2}) for large $x$ with $R\to 1$ we obtain a differential
equation for $F$ asymptotically, which has a solution $F(x) = 1/(const. +\ln
x)$ with $ \delta = 9/(4\beta_0)$. Therefore, in order to have $\delta = 1/3$
we need $\beta_0 = 27/4$. The same sort of solution is possible asymptotically 
from (\ref{eq:16}), however, this time we find $\delta = 13/(2\beta_0) =
1/3$ which gives the contradiction. The coefficients of the ultraviolet
dominant terms of eqs.~(\ref{renG2}) and (\ref{eq:16}) are consistent only
with the perturbative values for $\gamma_0^G$, $\gamma_0^A$ and $\beta_0$.
We therefore conclude that it is necessary to stay with the perturbative
short distance character of the equations and abandon the identity $Z_1 =
Z_3/\widetilde{Z}_3$ instead. These considerations show that our
truncation scheme is unable to obey the implications of gauge invariance
exactly which may not be surprising after all. However, we will see in c)
below that the hope to obey all Slavnov--Taylor identities for the
renormalization constants was only slightly too optimistic. 

\bigskip
{\bf b)} $Z_1 = 1$
\smallskip

The rigorous case above being inconsistent for the present truncation scheme,
it may seem natural to try the easiest next, setting $Z_1 = 1$ since this
constant is not necessary to absorb any ultraviolet divergences at the present
level anyway. We will see below that $Z_1 =1 $ is essentially
equivalent to setting it to an arbitrary (finite) constant $c_{Z_1}$. As in
the case of the ghost DSE, we look at the limit of eq. (\ref{eq:15}) for $s
\to 0$ in which this equation is singular in contrast to the ghost
eq.~(\ref{wtZ3a}). Using the limits (\ref{IRFR}) the leading behavior of
eq.~(\ref{eq:15}) up to contributions vanishing for $s \to 0$ is given by, 
\begin{eqnarray}
\frac{1}{b^2 s^{2\kappa}} &=& Z_3 \, - \, c_{Z_1} \, \frac{28}{9}\,
\widetilde Z_3 
\, + \, \frac{a}{11} \,\left\{ \frac{3}{2}\,
\frac{1}{2-\kappa} \, -\, \frac{1}{3} \right\}  \, \frac{1}{b^2 s^{2\kappa}}
\\
&& \hskip 5cm  +\, \frac{a^{1-2\delta}}{22}
\, \int_s^L \frac{dy}{y} \, \frac{F^{2\delta}(y)}{R^2(y)} \; . \nonumber
\end{eqnarray}  
From the definitions of $\kappa$ and $a = \beta_0 g_c^2$ , eqs. (\ref{kappa})
and (\ref{gcrit}), we find  
\[  \frac{a}{11} \left(  \frac{3}{2}\,
\frac{1}{2-\kappa} \, -\, \frac{1}{3} \right) \, = \, 1 \, - \, \frac{a}{11}
\, \frac{1}{4\kappa} \quad \hbox{and with}  \quad \frac{s^{-2\kappa}}{4\kappa}
\, =\, \frac{1}{2} \int_s^\infty \frac{dy}{y^{1+2\kappa}} \; ,\]
we obtain
\begin{equation} Z_3 \, = \,  c_{Z_1} \, \frac{28}{9}\, \widetilde Z_3
\, -  \, \frac{a^{1-2\delta}}{22} 
\, \int_0^L \frac{dy}{y} \, \left( \frac{F^{2\delta}(y)}{R^2(y)} \, - \,
\frac{a^{2\delta}}{b^2 y^{2\kappa}} \right) \; .\label{eq:17}\end{equation}
We assumed that $Z_3 = Z_3(s)$ is finite in the limit $s\to 0$ which is
necessary from eq. (\ref{wtZ1}), since due to the infrared fixed point
$Z_g$ and $\widetilde Z_3$ (see eq. (\ref{wtZ3a})) are infrared
finite. Furthermore, the above integral exists only if $ F^{2\delta}/R^{2}
\to  a^{2\delta}/(b^2 y^{2\kappa}) + {\cal O}(y^{\tau}) \, ,\; \tau > 0 $,
which we will verify when we come to discuss the infrared expansion of the
solutions for $R$ and $F$. We are now  able to eliminate all ultraviolet
divergences from the gluon DSE which can be written as
\begin{eqnarray}
\frac{1}{R^2(x)F^{1-2\delta}(x)} &=& \frac{c_{Z_1}}{a^{1-3\delta}} \,
\frac{1}{11} \, \Bigg\{
\int_{0}^{x} \frac{dy}{x}
\, \left(  \frac{7}{2}\frac{y^2}{x^2}
                     - \frac{17}{2}\frac{y}{x}
                     - \frac{9}{8} + 7 \frac{x}{y} \right) R(y)
F^{1-\delta}(y)  \nonumber\\
&& \hskip -3cm + x \frac{7}{8} \, \int_x^\infty \frac{dy}{y^2} \, R(y)
F^{1-\delta}(y) \Bigg\} \, + \, \frac{1}{11} \,
\Bigg\{ \frac{3}{2} \, \frac{F^\delta(x)}{R(x)} \int_0^x  \frac{dy}{x}
\frac{y}{x} \, \frac{F^\delta(y)}{R(y)}  \label{eq:18}  \\
&& \hskip -3cm  -\,
\frac{1}{3} \frac{F^{2\delta}(x)}{R^2(x)}  \, - \, \frac{1}{2} \int_0^x
\frac{dy}{y} \,\left( \frac{F^{2\delta}(y)}{R^2(y)} \, - \,
\, \frac{a^{2\delta}}{b^2 y^{2\kappa}} \right) \, + \,
\frac{1}{4\kappa} \, \frac{a^{2\delta}}{b^2 x^{2\kappa}} \Bigg\} \quad . 
\nonumber \end{eqnarray}
We see from this equation that with a redefinition $R' := c_{Z1}^{1/3} R$ and
accordingly $b' =  c_{Z1}^{1/3} b$ we can eliminate the finite constant
$c_{Z_1}$ from (\ref{eq:18}). The ghost DSE (\ref{renG2}) is linear in R and
thus in its form unchanged by this redefinition. It implies a scaling of the
propagators by a multiplicative constant which does not appear in physical
quantities like the running coupling and is thus unessential. We will
therefore solve the coupled system (\ref{renG2}) and (\ref{eq:18}) for $Z_1 =
c_{Z_1} = 1$. It is easy to see from those terms in eq.~(\ref{eq:18})
which are dominant for large $x$, that the resulting gluon propagator cannot
be expected to have the correct perturbative ultraviolet behavior (its
perturbative anomalous dimension): The leading ultraviolet contribution from
the ghost loop has the wrong sign and should be compensated by the according
contribution from the 3--gluon loop. However, in eq. (\ref{eq:18}) the
latter gives a contribution qualitatively equal to the asymptotic behavior
of the ghost DSE (\ref{renG2}) which is logarithmically subleading as
compared to the ghost loop (since $\widetilde Z_3/Z_3 \to 0$ for $N_f <
7$). It is clear that we cannot expect the solutions to resemble the
perturbative behavior at asymptotically high momenta for this reason. 

Evidently, the insufficiencies of the truncation scheme appear here in the
approximations used to simplify the 3--gluon loop. We reiterate, however,
that the 3--gluon loop does not affect the leading behavior of the
ghost--gluon DSEs in the infrared. We suggest a further manipulation in the
following which is motivated by the desired short distance behavior of the
solutions and which is suited for its restoration. We will compare our
numerical results obtained for $Z_1 = 1$ to those from the procedure outlined
in the following. In particular, this will verify that the additional
manipulation improves the short distance (high momentum) behavior of the
solutions, and that this is indeed all it does (not affecting conclusions on
the infrared behavior).  

\bigskip
{\bf c)} $Z_1 = (F(s)/F(y))^{1-3\delta}$
\smallskip

As a first remark, we note that we found $\widetilde{Z}_3 =
(F(s)/F(L))^\delta$ in the scaling limit. Analogously, we would like to
obtain $Z_3 = (F(s)/F(L))^{1-2\delta}$ and $Z_1 = (F(s)/F(L))^{1-3\delta}$
from the corresponding Slavnov--Taylor identity. We saw above that the
latter, {\it i.e.}, $Z_1 = Z_3/\widetilde{Z}_3$, was not consistently
possible at the present level of truncations. Therefore, we suggest to set
$Z_1 = (F(s)/F(y))^{1-3\delta}$ instead, where $y = q^2/\sigma$ denotes
the loop momentum. We repeat the same steps as before to eliminate the
renormalization constants and, in place of eq.~(\ref{eq:16}), we arrive at
\begin{eqnarray}
\frac{1}{R^2(x) F^{1-2\delta}(x)} &-& \frac{1}{R^2(s) F^{1-2\delta}(s)}
\, = \label{eq:19}\\
&& \hskip -3cm \frac{1}{11} \, \Bigg\{
\int_{0}^{x} \frac{dy}{x}
\, \left(  \frac{7}{2}\frac{y^2}{x^2}
                     - \frac{17}{2}\frac{y}{x}
                     - \frac{9}{8} + 7 \frac{x}{y} \right) R(y)
F^{2\delta}(y) \, + \, x \frac{7}{8} \, \int_x^\infty \frac{dy}{y^2} \, R(y)
F^{2\delta}(y)  \nonumber\\
&&  \hskip -1cm - \, ( x\leftrightarrow s ) \Bigg\}  \, + \quad \hbox{ghost
loop contributions} \; ,
  \nonumber
\end{eqnarray}
where we did not explicitly repeat the unchanged contributions from the ghost
loop again (see eq.~(\ref{eq:16})). If we now repeat the
exact same steps as in the case of the ghost DSE, the limit $s \to 0$ now
leads to
\begin{eqnarray}
  \frac{11}{R^2(x)F^{1-2\delta}(x)} &=& 
  \int_0^x \frac{dy}{x}
  \, \left(  \frac{7}{2}\frac{y^2}{x^2}
                     - \frac{17}{2}\frac{y}{x}
                     - \frac{9}{8} + 7 \frac{x}{y} \right) R(y)
F^{2\delta}(y)  \nonumber\\
&& \hskip -3cm + x \frac{7}{8} \, \int_x^\infty \frac{dy}{y^2} \, R(y)
F^{2\delta}(y) \, + \,  \frac{3}{2} \, \frac{F^\delta(x)}{R(x)} \int_0^x
\frac{dy}{x} \frac{y}{x} \, \frac{F^\delta(y)}{R(y)} \,  -\,
\frac{1}{3} \frac{F^{2\delta}(x)}{R^2(x)}   \nonumber  \\
&& \hskip -3cm  - \, \frac{1}{2} \int_0^x
\frac{dy}{y} \,\left( \frac{F^{2\delta}(y)}{R^2(y)} \, - \,
\, \frac{a^{2\delta}}{b^2 y^{2\kappa}} \right) \, + \,
\frac{1}{4\kappa} \, \frac{a^{2\delta}}{b^2 x^{2\kappa}} \quad ,
\label{eq:20} \end{eqnarray}
which can again be use to obtain the gluon renormalization constant,
exchanging variables $x \leftrightarrow s$, with the result,
\begin{equation}
  Z_3 =  F^{1-\delta}(s) \left( \frac{7}{11} \int_0^L \frac{dy}{y} \,
  R(y) F^{2\delta}(y) \, - \, \frac{1}{22}  \int_0^L \frac{dy}{y} \left(
  \frac{F^{2\delta}(y)}{R^2(y)} -  \frac{a^{2\delta}}{b^2 y^{2\kappa}}
  \right) \right)  .
\end{equation}
This is analogous to eq. (\ref{wtZ3b}) and valid for all $s$. Using
eq.~(\ref{eq:20}) with $x \leftrightarrow L$ for large $L$ we furthermore
obtain the desired scaling limit,
\begin{equation}
  Z_3 \, \stackrel{L\to\infty}{\longrightarrow} \, \left( \frac{F(s)}{F(L)}
  \right)^{1-2\delta} =\, \left( \frac{g^2}{g_0^2} \right)^{1-2\delta} \; .
\end{equation}
While $Z_1 = (F(s)/F(y))^{1-3\delta}$ violates the Slavnov--Taylor identity
which instead would demand $Z_1 = (F(s)/F(L))^{1-3\delta}$, it thus reproduces
the correct scaling limit for the gluon propagator. This will be reflected in
the numerical solutions, in particular, in the leading logarithmic behavior of
the gluon propagator.

Note that eqs. (\ref{eq:18}) and (\ref{eq:20}) are equivalent to order $g^2$
in perturbation theory. One--loop scaling requires, however, that a certain
class of diagrams of the perturbative series is subsummed, which includes
contributions of orders $g^4$ and higher. It should therefore not be
surprising that a truncation scheme neglecting 4--gluon correlations (which
appear at order $g^4$) does not automatically reproduce the perturbative
anomalous dimensions. We have seen that this problem is also related to some
violation of the implications of gauge invariance.

We will proceed mainly with discussing the solutions to (\ref{renG2}) and
(\ref{eq:20}) analytically as well as the numerical results, and use the
alternative version (\ref{eq:18}) only for comparison and in order to
demonstrate that our main conclusions such as the existence of the
infrared fixed point do not depend on the particular choice for $Z_1$ in the
3--gluon loop. We will verify that the sole effect of this choice is to fix 
the behavior of the solutions, in particular, the gluon propagator, at short
distances where their anticipated asymptotic perturbative form was used 
as the guiding principle. 

\section{Asymptotic Expansion and Scale Invariance}

The set of equations (\ref{renG2}) and (\ref{eq:20}) does not depend on the
scale $s$ showing that its solutions represent gluon and ghost
renormalization functions obeying one--loop scaling at all scales, {\it 
i.e.}, $\epsilon (g) = 0$, see eqs.~(\ref{RGsol1}) and (\ref{parZG}). We have
thus rewritten the problem in terms of the renormalization group invariant
functions $F(x)$ and $R(x)$. In particular, the scaling behavior of the
propagators follows trivially from the solution for the non--perturbative
running coupling $F/\beta_0$.

Similar to our previous solution to the gluon DSE in Mandelstam
approximation, a detailed analyses of the solutions in the infrared in terms
of asymptotic series is necessary in order to obtain numerically stable
iterative solutions. Due to the nature of the coupled set of equations a
recursive calculation of the respective coefficients of the asymptotic series
for $F$ and $R$ is considerably more difficult than in Mandelstam
approximation in which simple recursion relations allowed to calculate these 
coefficients to any desired order \cite{Atk81,Hau96}. Fortunately,
calculating the leading corrections to the asymptotic infrared behavior of
$R$ and $F$, as given above, proved sufficient to obtain numerically stable
results. Thereby, for $x < x_0$  with some infrared matching
point $x_0$, the asymptotic series to at least second order, in a sense to be
explained below, is used in obtaining iterative solutions for  $ x >
x_0$. The matching point $x_0$ has to be sufficiently small for the
asymptotic series to provide the desired accuracy. However, limited by
numerical stability, it cannot be chosen arbitrarily small either. This leads
to a certain range of values of $x_0$ for which stable solutions are obtained
with no matching point dependence to fixed accuracy. We verified that the
additional inclusion of third order contributions in the asymptotic series
has no effect other than increasing the allowed range for the  matching
point.

The discussion of the solutions in the infrared is alleviated by the
observation that one of the equations, eq.~(\ref{renG2}),
\begin{equation}
  \frac{R(x)}{F^\delta(x)} \, = \, \delta \left( \int_0^{\;\; x} \frac{dy}{y} \,
  R(y) F^{1-\delta}(y) \,  - \, \frac{1}{2} \, R(x)F^{1-\delta}(x) \right)
  \; ,
  \label{eq:13}
\end{equation}
can be converted in a first order homogeneous linear differential equation
for $R(x)$. Differentiating eq. (\ref{eq:13}) with respect to $x$ one obtains,
\begin{equation}
R'(x) \,=\, T(x) R(x) \;, \qquad  
T(x) \, := \, \frac{\delta}{1 + \frac{\delta}{2} F} \,  \left( \frac{F}{x} +
\frac{F'}{F} - \frac{1-\delta}{2} F' \right) \; . \end{equation}
The second equation to solve, eq. (\ref{eq:20}), can be written,
\begin{eqnarray}
\frac{11}{R^2(x)F^{1-2\delta}(x)} &=& 
\int_{0}^{x} \frac{dy}{x}
\,\Bigg\{ \left(  \frac{7}{2}\frac{y^2}{x^2}
                     - \frac{17}{2}\frac{y}{x}
                     - \frac{9}{8} + 7 \frac{x}{y} - \frac{7}{8}
\frac{x^2}{y^2} \right) R(y)F^{2\delta}(y)  \nonumber\\
&& \hskip -3cm + \frac{7}{8} \frac{x^2}{y^2} \, b a^{2\delta}\, y^\kappa
\Bigg\} \, + \,   \frac{7}{8} \frac{ b\, a^{2\delta}}{1-\kappa}\, x^\kappa \,
+ \, A\, x \, 
+ \, \frac{3}{2} \, \frac{F^\delta(x)}{R(x)} \int_0^x
\frac{dy}{x} \frac{y}{x} \, \frac{F^\delta(y)}{R(y)}  \nonumber  \\
&& \hskip -3cm  -\,
\frac{1}{3} \frac{F^{2\delta}(x)}{R^2(x)}  \, - \, \frac{1}{2} \int_0^x
\frac{dy}{y} \,\left( \frac{F^{2\delta}(y)}{R^2(y)} \, - \, 
\, \frac{a^{2\delta}}{b^2 y^{2\kappa}} \right) \, + \,
\frac{1}{4\kappa} \, \frac{a^{2\delta}}{b^2 x^{2\kappa}}    \; , 
\label{eq:23} \end{eqnarray}
where we have used that
\begin{eqnarray}  
  x \frac{7}{8} \, \int_x^\infty \frac{dy}{y^2} \, R(y) F^{2\delta}(y)
  &=& - x \frac{7}{8} \, \int_0^x \frac{dy}{y^2} \left( RF^{2\delta}
      -  b a^{2\delta} y^\kappa \right)
\, + \,   \frac{7}{8} \frac{ b\, a^{2\delta}}{1-\kappa}\,  x^\kappa \, + \,
A\,  x \nonumber \\
\hbox{with} \hskip 2.2cm  A&=&   \frac{7}{8} \, \int_0^\infty  \frac{dy}{y^2}
\left(  RF^{2\delta}    -  b\, a^{2\delta} y^\kappa \right) \; . 
  \label{defA}
\end{eqnarray}
From the leading infrared behavior, {\it i.e.}, $F \to a$ and $R \to b
x^\kappa$ for $x \to 0$, we see from eq.~(\ref{eq:13}) that an asymptotic
infrared expansion of $1/(R^2F^{1-2\delta})$ has to contain powers of
$x^\kappa $ as well as powers of $x$ in subsequent subleading terms. This
motivates the following Ansatz,
\begin{eqnarray}
R(x) &=&  b \, x^\kappa \sum_{l,m,n = 0}^{\Sigma =N} C_{lmn} \; x^{m\nu +
3n\kappa + l(1+2\kappa)}\label{eq:24a} \\
F(x) &=&  a \, \sum_{l,m,n = 0}^{\Sigma =N} D_{lmn} \; x^{m\nu +
3n\kappa + l(1+2\kappa)}  \; , \quad \hbox{with} \quad  \Sigma := l+m+n \;
\label{eq:24b}\end{eqnarray}
and $C_{000} = D_{000} = 1$. The additional fractional power $x^\nu$ in
these expansions was introduced to find the most important subleading behavior 
possible from the consistency in the infrared. We will determine it to be
$\nu \simeq 2.05$. With $2 < 3\kappa < 1 + 2 \kappa
\stackrel{<}{\approx} 3$ this shows that for not too large orders $N$ powers of
different orders in this expansions do not mix in their successive importance
at small $x$. Furthermore, we have repeatedly subtracted leading infrared
contributions explicitly from integrals such as the one in 
(\ref{eq:17}) above, assuming that the remaining contributions are integrable
for $x \to 0$. For subleading contributions to $R$ and $F$ suppressed by
powers of $x^\nu$ with $\nu \simeq 2.05$ this is justified {\it a
posteriori}. 

With the series (\ref{eq:24b}) we can calculate $T(x)$ in eq. (\ref{eq:13}) to
the same order $N$,
\begin{equation}
  T(x)  \, =\, \frac{\kappa}{x}  \sum_{l,m,n = 0}^{\Sigma=N} E_{lmn} \;
x^{\tau_{lmn}} \; , \; \hbox{with} \quad \tau_{lmn} := {m\nu + 3n\kappa + 
l(1+2\kappa)} \; ,
\end{equation}
where the coefficients $E_{lmn}$ can be straightforwardly calculated from
the coefficients $D_{lmn}$ of $F$ (with $E_{000} = 1$). The solution of
eq.~(\ref{eq:13}) for $R$ with the integration constant set to $b$ is then
given by  
\begin{equation}
  R(x) = b \, x^\kappa \exp\left\{
    \kappa \, \sum_{\Sigma =1}^N \frac{E_{lmn}}{\tau_{lmn}} \, x^{\tau_{lmn}}
  \right\} \; .
\end{equation}
Expanding this series to an appropriate order allows to relate the
coefficients $C_{lmn}$ to $E_{lmn}$ and thus to $D_{lmn}$. For $N = 1$ 
the result is,
\begin{eqnarray}  
   C_{100} &=& \frac{\kappa}{1+2\kappa} E_{100} = \left( \frac{\kappa
  (1-3\kappa ) }{2(1+2\kappa)} + \delta \right) D_{100} \nonumber \\
  C_{010} &=& \frac{\kappa}{\nu}  E_{010} = \left( \frac{\kappa}{\nu} -
  \frac{\kappa}{2} - \frac{\kappa^2}{2\nu}  + \delta \right) D_{010} 
  \label{eq:27}\\
  C_{001} &=& \frac{1}{3} E_{001} = \left( \frac{1}
  {3}- \frac{2}{3} \kappa + \delta \right) D_{001}  \; .\nonumber 
\end{eqnarray}
At higher orders in $N$ this procedure recursively yields relations that
uniquely determine the coefficients $C$ in terms of the coefficients $D$.
Analogous relations are obtained from eq. (\ref{eq:23}) by expanding all
ratios of $R$ and $F$ which occur with dependence on $x$ and $y$, and by
comparison of the respective orders, ${\cal O}(x^{\tau_{lmn}-2\kappa})$, on
both sides. We verify from eq. (\ref{eq:23}) that to leading order,
${\cal O}(x^{-2\kappa})$:
\begin{equation}
  \frac{11}{b^2\, a^{1-2\delta}} \, = \, \left( \frac{3}{2}\,
  \frac{1}{2-\kappa} - \frac{1}{3} + \frac{1}{4\kappa} \right)  \,
  \frac{a^{2\delta}}{b^2} \; ,
\end{equation}
which was used to determine $\kappa$ with $a =\beta_0 g_c^2 = ((9/44) 
(1/\kappa - 1/2))^{-1}$ (see eqs.~(\ref{gcrit}) and (\ref{kappa})). At
order $N = 1$ we obtain,
\begin{eqnarray}
  && \hskip -1cm {\cal O}(x^{\nu - 2\kappa}) : \hskip 1cm
  \frac{11}{a} \left( D_{010} + 2 (C_{010} - \delta D_{010}) \right) =
  \nonumber\\
  && \left( \frac{3}{2} \left( \frac{1}{2+\nu -\kappa} +
  \frac{1}{2-\kappa}\right)
-\frac{2}{3} - \frac{1}{\nu - 2\kappa} \right)  (C_{010} - \delta D_{010}) 
\label{eq:28a}\\
 && \hskip -1cm {\cal O}(x^{\kappa}) : \hskip 1.5cm
\frac{11}{a} \left( D_{001} + 2 (C_{001} - \delta D_{001}) \right) =  - b^3
f(\kappa)  + \nonumber\\&& 
\left( \frac{3}{2} \left( \frac{1}{2+2\kappa} + \frac{1}{2-\kappa}\right)
-\frac{2}{3} - \frac{1}{\kappa} \right)  (C_{001} - \delta D_{001})  
\label{eq:28b}\\
 && \hskip -1cm {\cal O}(x) :  \hskip 1.7cm
\frac{11}{a} \left( D_{100} + 2 (C_{100} - \delta D_{100}) \right) = 
- \frac{b^2}{a^{2\delta}} \, A \, + \nonumber\\&& 
\left( \frac{3}{2} \left( \frac{1}{3+\kappa} + \frac{1}{2-\kappa}\right)
-\frac{5}{3} \right)  (C_{100} - \delta D_{100}) \; ,
\label{eq:28c}\end{eqnarray}
with $f(\kappa) := 7/(2(3+\kappa)) - 17/(2(2+\kappa)) -9/(8(1+\kappa)) +
7/\kappa + 7/(8(1-\kappa )) $.

These equations together with eqs. (\ref{eq:27}) determine the coefficients $D$
and $C$ to lowest non--trivial order. In particular, we have 3 decoupled sets
of 2 equations each for 2 of the constants, respectively. For $(l,m,n) =
(1,0,0)$, {\it c.f.}, eq. (\ref{eq:28c}), we obtain,
\begin{equation}
  C_{100} \, \simeq \, 0.05554 \, b^2 A \; , \quad \hbox{and} \quad D_{100}
  \, \simeq -0.6992 \, b^2 A \; .
\end{equation}
The set of equations for $(l,m,n) = (0,1,0)$ is homogeneous, see
eq.~(\ref{eq:28a}). The determinant of its 2--dimensional coefficient matrix
is zero for 
\begin{equation}
\nu \, = \,{{-6 - \kappa  - 3\,{{\kappa }^2} \pm
           {\sqrt{-4\,\left( -26 - 23\,\kappa  \right) \,{{\kappa }^2}\,
                \left( 3 + 2\,\kappa  \right)  + 
               {{\left( 6 + \kappa  + 3\,{{\kappa }^2} \right) }^2}}}}\over 
         {2\,\left( 3 + 2\,\kappa  \right) }} \; ,
\end{equation}
with one positive root for the plus sign which determines the positive
exponent $\nu $. The coefficients $C_{010}$, $D_{010}$ follow up to a common
factor. With $\kappa = (61 - \sqrt{1897})/19$ we obtain,
\begin{equation}
  \nu \simeq 2.051
  \quad \hbox{and} \quad
  C_{010} = - 0.0124 D_{010} \; .
\end{equation}
For $(l,m,n) = (0,0,1)$ the scale of the coefficients is set by the
inhomogeneity, $b^3 f(\kappa)$, in eq.~(\ref{eq:28b}) and we obtain,
\begin{equation}
  C_{001} \, \simeq \, 1.969\,  b^3 \; , \quad \hbox{and} \quad D_{001} \,
  \simeq \, - 26.52 \, b^3 \; .
\end{equation}
Using these results, higher orders, though increasingly tedious, can be
obtained recursively by analogous sets of equations. The general pattern is
such that the order $N=1$ above fixes the scales for higher order
coefficients. This allows us to define scale independent coefficients
$\widetilde{C}$ and $\widetilde{D}$ by extracting their respective scales
according to the exponent $\tau_{lmn} = {m\nu + 3n\kappa + l(1+2\kappa)} $ of
$x$ for a given set $(l,m,n)$, 
\begin{equation} 
C_{lmn}   \, =: \, \widetilde C_{lmn} \, b^{3 n + 2 l}\, t^m \, A^l \; ,
\quad \hbox{and}\quad  D_{lmn}   \, =: \, \widetilde D_{lmn} \, b^{3 n + 2
l}\, t^m \, A^l \; , 
\end{equation} 
where the scale of the powers of $x^\nu$ is set by $t$ which we fix 
for convenience to
\begin{equation}
  t := - D_{010}
\end{equation}
{\it i.e.}, $\widetilde{C}_{010} \simeq 0.0124$ and $\widetilde{D}_{010} = -1$.
We summarize the values of the coefficients $\widetilde{C}$ and $\widetilde{D}$
for $N = 2$ in table \ref{coefftab}.

\begin{table}[h,t]
\begin{center}
\renewcommand{\arraystretch}{1.4}
\begin{tabular}{l|rrrrrr}
$(l,m,n)$ &  
$(2,0,0)$ &  $(1,1,0)$ & $(1,0,1)$ &  $(0,2,0)$ &  $(0,1,1)$ & $(0,0,2)$  \\ 
\hline
$\widetilde C$  &
 -0.1042  & -0.3034    & -7.933    & -0.2160    &   -11.55   & -151.0 \\
$\widetilde D$  & 
 0.5246   & 1.590      &  40.10    & 1.226      &  60.98     & 766.8 
\end{tabular}
\caption{Coefficients of the asymptotic expansion for $N=2$.} 
\label{coefftab} 
\end{center}
\end{table}

The constant $A$ as given in (\ref{defA}) is determined numerically in the
iterative process. Of the remaining two parameters $b$ and $t$ in the
asymptotic forms, one can be related to the overall momentum scale so that we
are left with one independent parameter. Note that we neither specified the
momentum scale $\sigma $ (in $ x = k^2/\sigma$) nor the infrared constant $b$
so far. The problem is scale invariant, however, {\it i.e.}, a change in the
scale $\sigma $ according to $\sigma \to \sigma ' = \sigma /\lambda $ or,
equivalently, $ x \to x' = \lambda x $ can be compensated by
\begin{equation} b \, \to\,  b' \, = \, b/\lambda^\kappa \; , \quad \hbox{and} \quad t \,
\to \, t' \, = \, t/\lambda^\nu \; . \label{scale}\end{equation}
We can thus choose the scale without loss of generality such that the
positive number $b = 1$. The parameter $t$ can in principle be any real number
including zero. We can find numerically stable iterative solutions for not too
large absolute values of $t$ (see below). Furthermore, it can be verified
numerically, that a solution for a value of $b \not= 1$ for fixed $t$ is
identical to a solution for $b=1$ and $t' = t\, b^{\nu/\kappa} $, if $x$ is
substituted by $x' = x/b^{1/\kappa} $. This is the numerical manifestation of
the scale invariance mentioned above (for $ \lambda =  1/b^{1/\kappa}$). Note
that under scale transformations (\ref{scale}) the constant $A$ trivially
transforms according to its dimension, $A \to A' = A/\lambda $, without any
adjustments from the way it is calculated, since
\begin{equation} A' \, = \, A/\lambda \, = \, \lim_{x_0' \to 0} \, \frac{7}{8}
\left( \, \int_{x_0'}^\infty  \, \frac{dy'}{y'^2} \,R(y') F^{2\delta}(y') \, - 
\, b' \, a^{2\delta}\,  \frac{(x_0')^{\kappa -1}}{1-\kappa} \right) \; . \end{equation}
\begin{figure}
  \centering{\
        \epsfig{file=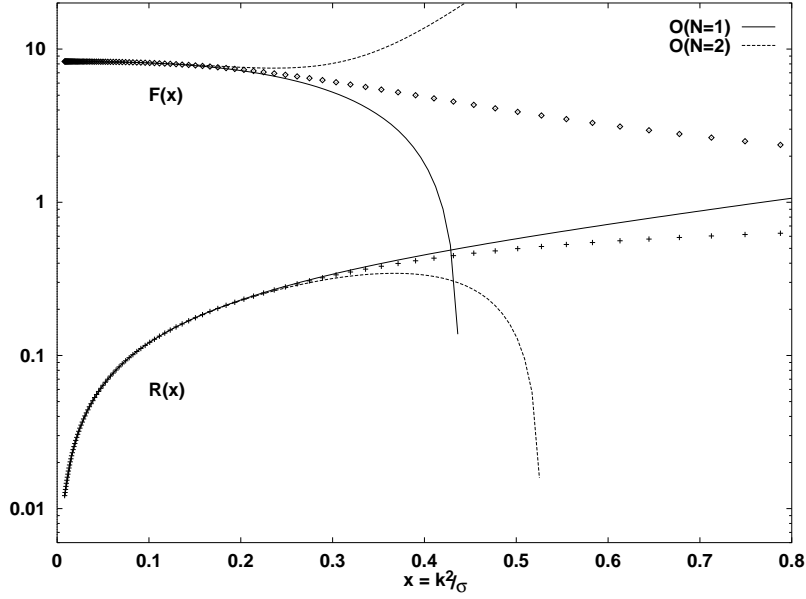,width=0.8\linewidth} }
  \caption{The numerical solutions of $F(x)$ and $R(x)$ for $t=0$ and $b=1$
together with their asymptotic expansions to order $N=1$ as well as $N=2$ at
small $x$.}
  \label{FRir}
\end{figure}

In figure \ref{FRir} we plotted the numerically obtained solutions for $F(x)$
and $R(x)$ for $b=1$ and $t=0$ at small $x$ together with their respective
asymptotic forms to order $N=1$ and $N=2$. The contributions of the order
$N=2$ in the asymptotic expansion become comparable in size to the lower
order at about $x \simeq 0.2$ and dominant for increasingly higher values of
$x$. The error in the asymptotic series being of the order of the first terms
neglected, this is the usual indication for the range of $x$ in which the
asymptotic expansion to the given order can yield reliable results. In the
particular calculation we used a value of about $x_0 = 0.008$ for the
matching point of the iterative process to the asymptotic result. This is
obviously well below the estimated range of the validity of the asymptotic
expansion.

\section{Numerical Results and Perturbative Limit}

The numerical results were obtained with the order $N=1$ in the asymptotic
expansion for which no dependence on $x_0$ was observed for $x_0 < 0.1$. 
Higher values for the matching point are possible at higher orders. This
shows that the expansion to next to leading order in the infrared ($N=1$)
is sufficient to find numerically stable results for a satisfactory regime
of matching points. Similar results follow for $t\not= 0$. We calculated
$F$ and $R$ for several values in the range $-5 \le t \le 16$. At lower
negative values the procedure became numerically unstable due to a
developing (tachyonic) pole in $F(x)$. The fact that the integral
equations for $R$ and $F$ possess a one--parameter family of solutions
characterized by $t$ is in fact the reason for the necessity of the infrared
expansion up to next to leading order, since no stable solution can be
found numerically without fixing the leading $x$--dependence of $F(x)$ at
small $x$ by fixing the parameter $t$. This is a boundary condition to be
imposed on the solutions from physical arguments. 

\begin{figure}
  \centering{\
        \epsfig{file=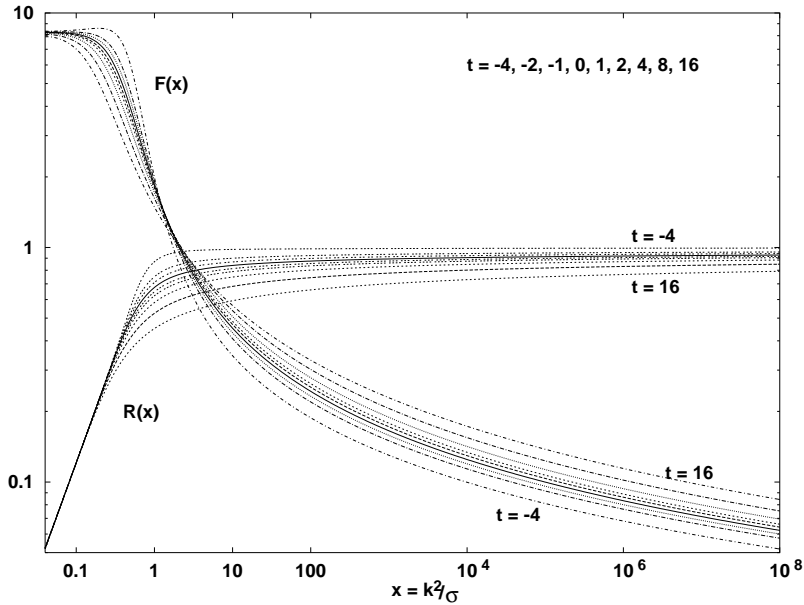,width=0.8\linewidth} }
  \caption{The numerical solutions of $F(x)$ and $R(x)$ with $b= 1$ for
different values of the parameter $t = \{ -4, -2, -1, 0 ,1, 2, 4, 8, 16\} $
(solid lines represent $t = 0$ solutions).}
  \label{FRdt}
\end{figure}

In figure \ref{FRdt} the numerical results are plotted for different values of
the parameter $t$ (all with $b = 1$). Perturbatively, we expect $R(x)$ to
approach a constant value and $F(x) \to 1/\ln (\lambda x)$ for $x \to \infty$.
The reason we introduced the constant $\lambda$ in the one--loop running
coupling $F$ is that we fixed the momentum scale in our calculations by
arbitrarily setting $b = 1$. The relation between the scale of perturbative
QCD $\Lambda_{\hbox{\tiny QCD}} $ and $\sigma$ cannot be determined this
way. Therefore, we set $\Lambda^2_{\hbox{\tiny QCD}} = \sigma/\lambda$ for
some scale parameter $\lambda $.  Qualitatively, all 
solutions display a similar behavior at high momenta. The solutions for
positive values of $t$ seem to have more residual momentum dependence in $R$ at
high momenta than those for $t \le 0$. A very good fit to the perturbative form
can be obtained for $t = -4.2$. For negative values of $t$, however, the
running coupling, $\alpha_S(\mu) = F(s)/(4\pi\beta_0)$,  has a maximum,
$\alpha_{\hbox{\tiny max}} > \alpha_c$, at a finite value of the
renormalization scale $s = \mu^2/\sigma$. This is because the dominant
subleading term of the running coupling in the infrared is determined by $t$,
\begin{equation}
  F(x) \, \to \, a \, ( 1 - t\, x^\nu + D_{001} \, x^{3\kappa} + D_{100} \,
  x^{1+2\kappa} ) \; , \quad x \to 0  \; .
\end{equation}
With $\nu \simeq 2.05 < 3\kappa < 1 + 2\kappa $ and $D_{001} < 0 $, it is
clear that for $t<0 $ the running coupling increases for smallest scales
close to $\mu = 0$ before higher order terms dominate. There necessarily has to
be a maximum  $\alpha_{\hbox{\tiny max}} > \alpha_c $ at some finite scale
$\mu $ for any solution with $t<0$. Such a maximum at finite scale implies a
zero of the beta function which cannot be interpreted as a fixed point (since
it appears at a finite scale $\mu_0$). It is easy to see that the slope of
the beta function $\beta(g)$ at such a point $g_{\hbox{\tiny max}}$ is
infinite. This precludes its possible classification as being infrared or
ultraviolet stable. The present case ($t<0$) corresponds to a double valued
$\beta (g)$ with a positive branch and a negative branch coexisting for $g_c
<  g < g_{\hbox{\tiny max}}$. In lack of an understanding of this scenario,
we therefore disregard solutions for $t < 0$ as likely to be unphysical. 

In contrast, for $t \ge 0 $, $\alpha_c = \alpha(\mu = 0)$ is the only
maximum of the running coupling for all real values of the renormalization
scale, and $\alpha_c $ is thus a true infrared stable fixed point.
Comparing the behavior of the resulting gluon and ghost renormalization
functions in the ultraviolet we observe that, for the $t \ge 0$ solutions,
the case $t = 0$ yields the best resemblance of their one--loop anomalous
dimensions. We therefore restrict the discussion of our results to the case
$t=0$ from now on, interpreting the existence of solutions for $t \not= 0$ as
a mathematical peculiarity. 

\begin{figure}
  \centering{\
        \epsfig{file=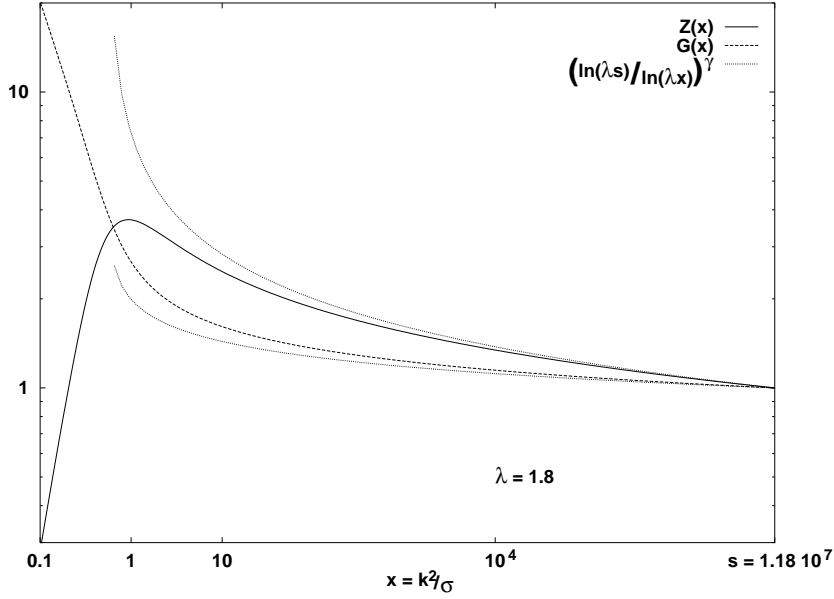,width=0.8\linewidth} }
  \caption{The renormalization functions $Z(x)$ and $G(x)$ obtained for
           $t = 0$ and their leading logarithmic forms
           (with $\gamma \to 1 - 2\delta$ for the gluon, and
           $\gamma \to \delta$ for the ghost).} 
  \label{ZG}
\end{figure}

In order to obtain an estimate of the factor $\lambda$ to relate the infrared
scale $\sigma $ to the perturbative QCD scale $\Lambda_{\hbox{\tiny QCD}}$,
we compare the results for the renormalization functions $Z(x)$ and $G(x)$
(with $x = k^2/\sigma$) to their respective one--loop forms,
\begin{eqnarray}
  Z_{\hbox{\tiny pt}} &=&
    \left(
      \frac{\ln(\mu^2/\Lambda_{\hbox{\tiny
QCD}}^2)}{\ln(k^2/\Lambda_{\hbox{\tiny QCD}}^2)} 
    \right)^{1-2\delta}
  \, =  \,
    \left(
      \frac{\ln(\lambda s)}{\ln (\lambda x)}
    \right)^{1-2\delta} \; , \quad \hbox{and} \\
  G_{\hbox{\tiny pt}}  &=&
    \left( \frac{\ln (\mu^2/\Lambda^2_{\hbox{\tiny
QCD}})}{\ln(k^2/\Lambda^2_{\hbox{\tiny QCD}})} \right)^{\delta}
  \, =  \, \left( \frac{\ln (\lambda s)}{\ln (\lambda x)} \right)^{\delta} \; .
\end{eqnarray}
In figure \ref{ZG} we plot the functions $Z(x)$ and $G(x)$ as obtained
from $F(x)$ and $R(x)$ (for $t=0$) according to the parameterization
(\ref{parZG}) along with their respective one--loop forms using $\lambda
= 1.8$. The renormalization point $s = \mu^2/\sigma = 1.18 \cdot 10^7$
(with $Z(s) = G(s) = 1$) was chosen to be the numerical ultraviolet cutoff
of the particular calculation for aesthetic reasons. A different choice
would result in no more than a constant vertical shift of the curves on
the logarithmic plot, and the choice of the renormalization point is thus
absolutely inessential for the present consideration. It is possible to
match one of the renormalization functions $Z$ and $G$ with its
perturbative form even closer than in this figure by varying the value of
$\lambda $ in the perturbative logarithms.  This will, however, at the
same time affect the perturbative form of the other renormalization
function in the opposite direction, {\it e.g.}, for $\lambda$
approximately $4 \sim 5$ the perturbative logarithm of the gluon is hardly
distinguishable from $Z(x)$ for $x > 10 $ on the scales as used in
figure~\ref{ZG}, while a similarly good fit of the one--loop form to
$G(x)$ is obtained for $0.2\sim 0.3$. To fit the perturbative logarithms
at high momenta as close as possible to both renormalization functions
with a unique value of $\lambda $, we can restrict its value approximately
to the range $1.5 \le \lambda \le 2$. 

\begin{figure}
  \centering{\epsfig{file=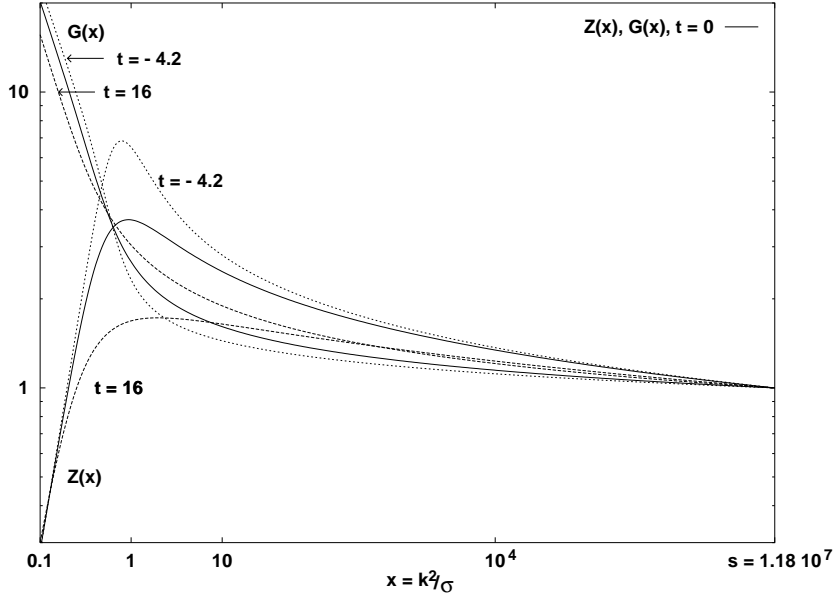,width=0.8\linewidth}}
  \caption{The renormalization functions $Z(x)$ and $G(x)$ for
           $t = \{ -4.2, 0, 16 \}$. Their leading logarithmic form
           at large $x$ is not given (see fig.~\protect\ref{ZG}), it
           is indistinguishable from the $t = -4.2$ results for $x > 10$
           (and with $\lambda$ in 1.5 --- 2). } 
  \label{ZGvgl}
\end{figure}

For a scale factor in the same range, $1.5 \le \lambda \le 2$, the
resemblance of the leading logarithms at high momenta can be optimized 
by an {\sl unphysical} choice of the parameter $t$ of approximately $t = -4.2$.
The according solutions approach their leading logarithmic behavior at momenta
as low as $k^2 \simeq 10 \sigma$ and are practically identical to the latter
for all momenta higher than that. In contrast, as mentioned above, increasing
positive values of $t$ lead to solutions fit by the leading perturbative
logarithms decreasingly well (see figure \ref{ZGvgl}).

To summarize the discussion of our results at high momenta, we conclude
that we obtain quite good agreement with the leading logarithmic behavior
(as known from perturbation theory) for the renormalization functions $Z$
and $G$ at high momenta. The perturbative QCD scale parameter
$\Lambda_{\hbox{\tiny QCD}}$ is thereby related to the infrared scale
$\sigma $ by $\Lambda^2_{\hbox{\tiny QCD}} = \sigma /\lambda$. The factor
$\lambda$ can be numerically estimated to be approximately 1.5 --- 2. 

Note that even though we introduced a numerical ultraviolet cutoff, for
sufficiently high values, our solutions to the ultraviolet finite equations
for $F(x)$ and $R(x)$ show no dependence on this cutoff. There are no
ultraviolet boundary conditions imposed on the solutions. Fixing the infrared
boundary condition by choosing a value for the parameter $t$ (and the scale
by setting $b = 1$) we have no further influence on the high momentum
behavior of the solutions which selfconsistently results from the iterative
process alone. Deviations from leading logarithmic behavior
can have several origins:

\begin{itemize}
\item Even though the gluon and ghost renormalization functions obey
one--loop scaling, the non--perturbative running coupling, $\alpha_S (\mu) =
F(s)/(4\pi\beta_0)$, contains two-- and higher loop contributions (if
certainly not all of them). Therefore, the fact that the results for $t = 0$
approach the leading logarithmic behavior at higher momenta than some of the 
apparently unphysical results for negative values of $t$ is not unreasonable.

\item Depending on $t$ also, the function $R(x)$ can have residual
momentum dependence even at very high momenta. In figure \ref{FRdt} the 
solution for, {\it e.g.}, $t = -4$ approaches the value one at momenta as low
as $k^2 \simeq 10 \sigma$, whereas for positive $t$ considerable
logarithmic momentum dependence can be left even at  $k^2 \simeq 10^8 \sigma
$ (see fig \ref{FRdt} for $t = 16$). While this has of course no influence on
the running coupling, it does affect the asymptotic behavior of the
renormalization functions.

\item Since $\ln (\lambda x) = \ln \lambda + \ln x$, all different asymptotic
forms used above yield an equivalent leading logarithmic contribution (in the
limit $x\to \infty$). The relevant question is the one for the momentum scale
above which the results should follow their leading logarithms in
phenomenological applications.  
\end{itemize}

Fixing the momentum scale in our calculations from the rough estimate $\sigma
= \lambda \, \Lambda^2_{\hbox{\tiny QCD}}$ given above is not satisfactory. 
Not only that the parameter $\lambda$ is poorly determined, also the value of
$ \Lambda_{\hbox{\tiny QCD}}$, appropriate for the present subtraction scheme
(and $N_f = 0$), is not a phenomenologically well known quantity. We discuss
in more detail in the next section, how to avoid this problem by fixing
$\sigma $ directly, independent of the precise values for $\lambda$ and thus
$\Lambda_{\hbox{\tiny QCD}}$. As a result of this, we find that $k^2 \simeq
10 \sigma$ corresponds to a momentum scale of 1GeV, and we conclude that 
the high momentum behavior of the solutions for $t=0$ is in reasonable
agreement with the perturbative result. 

\begin{figure}
  \centering{\
        \epsfig{file=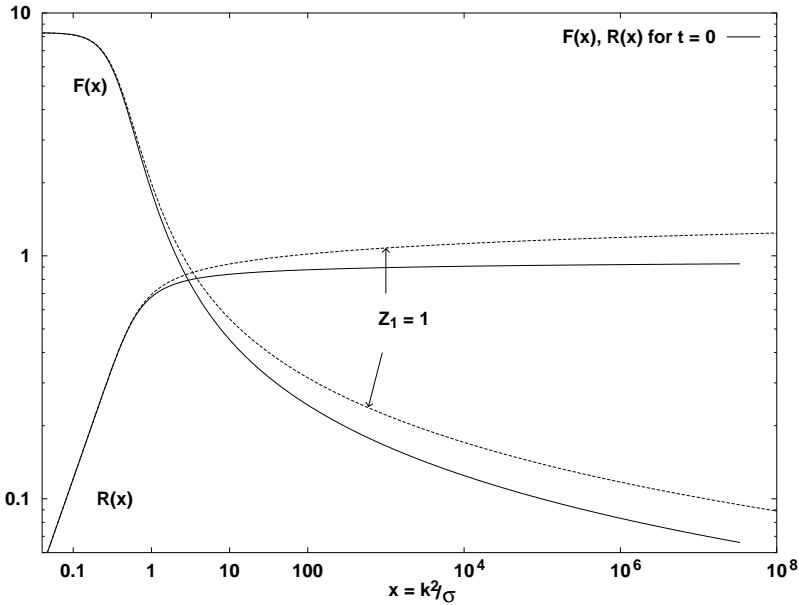,width=0.8\linewidth} }
  \caption{The functions $F(x)$ and $R(x)$ for $t = 0$ compared to the
respective results obtained from setting $Z_1 = 1$.} 
  \label{FRZ1}
\end{figure}

In connection with the renormalization procedure discussed in the previous
section we motivated the substitution of the cutoff dependence in the
constant $Z_1$ by a dependence on the integration momentum to improve the 
ultraviolet behavior of the solutions. In figure~\ref{FRZ1} we compare the
$t = 0$ results for $F(x)$ and $R(x)$ obtained this way (from
eq. (\ref{eq:20}) for the running coupling $F$) with those of an analogous
calculation using (\ref{eq:18}) with $Z_1 = c_{Z_1} = 1$ instead. It can be 
seen clearly that both solutions represent identical results in the infrared.
This shows that the existence of the solutions and, in particular, of the
infrared fixed point in the present truncation scheme is no artifact of the
treatment we suggested for the renormalization constant of the 3--gluon loop
$Z_1$. Furthermore, it verifies that this treatment does indeed improve the
ultraviolet behavior, noting that $R(x)$ as obtained for $Z_1 = 1$ does not
approach a momentum independent constant in the ultraviolet on any reasonable
scale. 

\section{Discussion of the Running Coupling}

\begin{figure}[t]
\hskip -.7cm
\parbox{5cm}{
  \epsfig{file=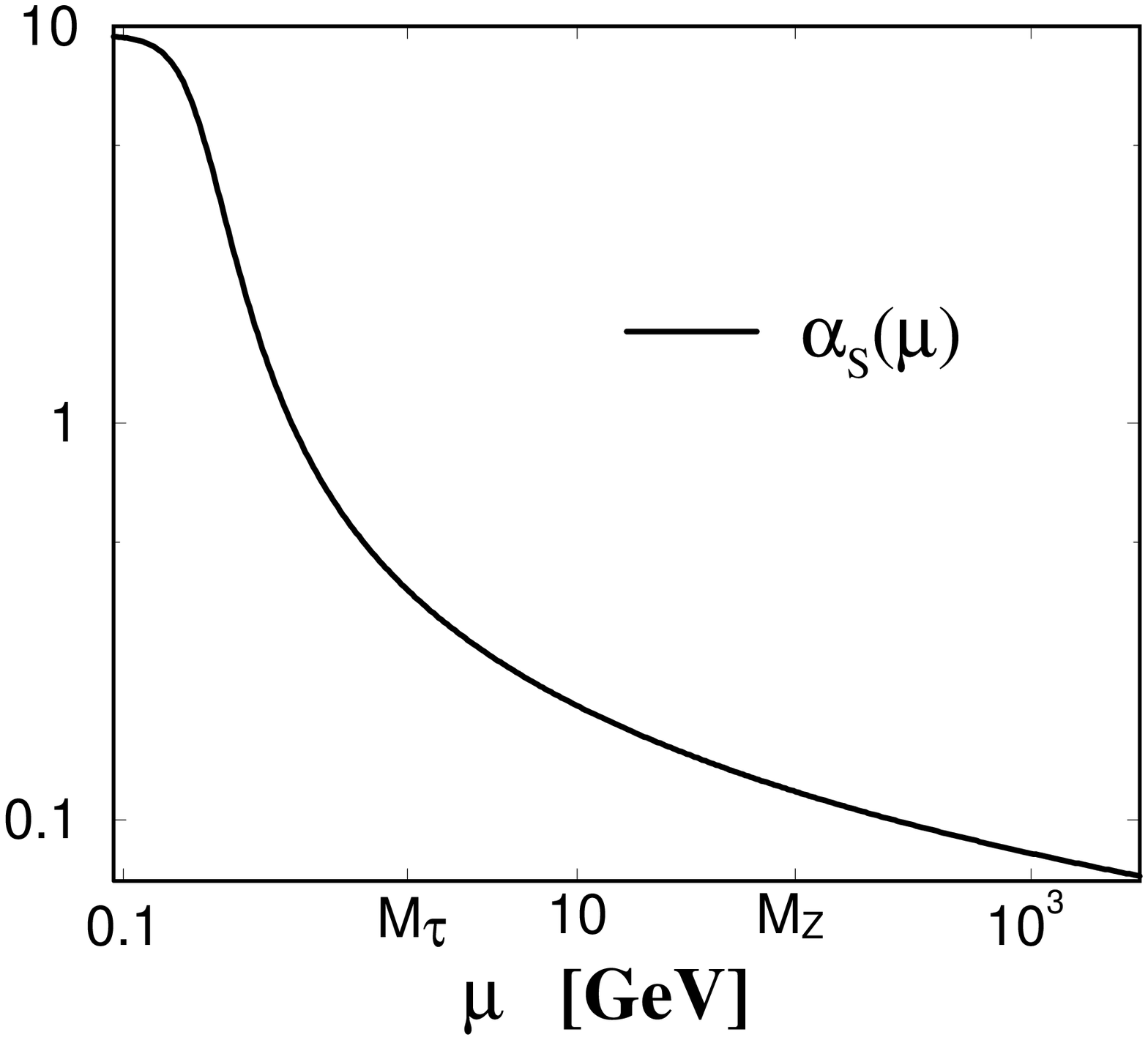,width=7.2cm}}
\hfil
\parbox{5cm}{
  \epsfig{file=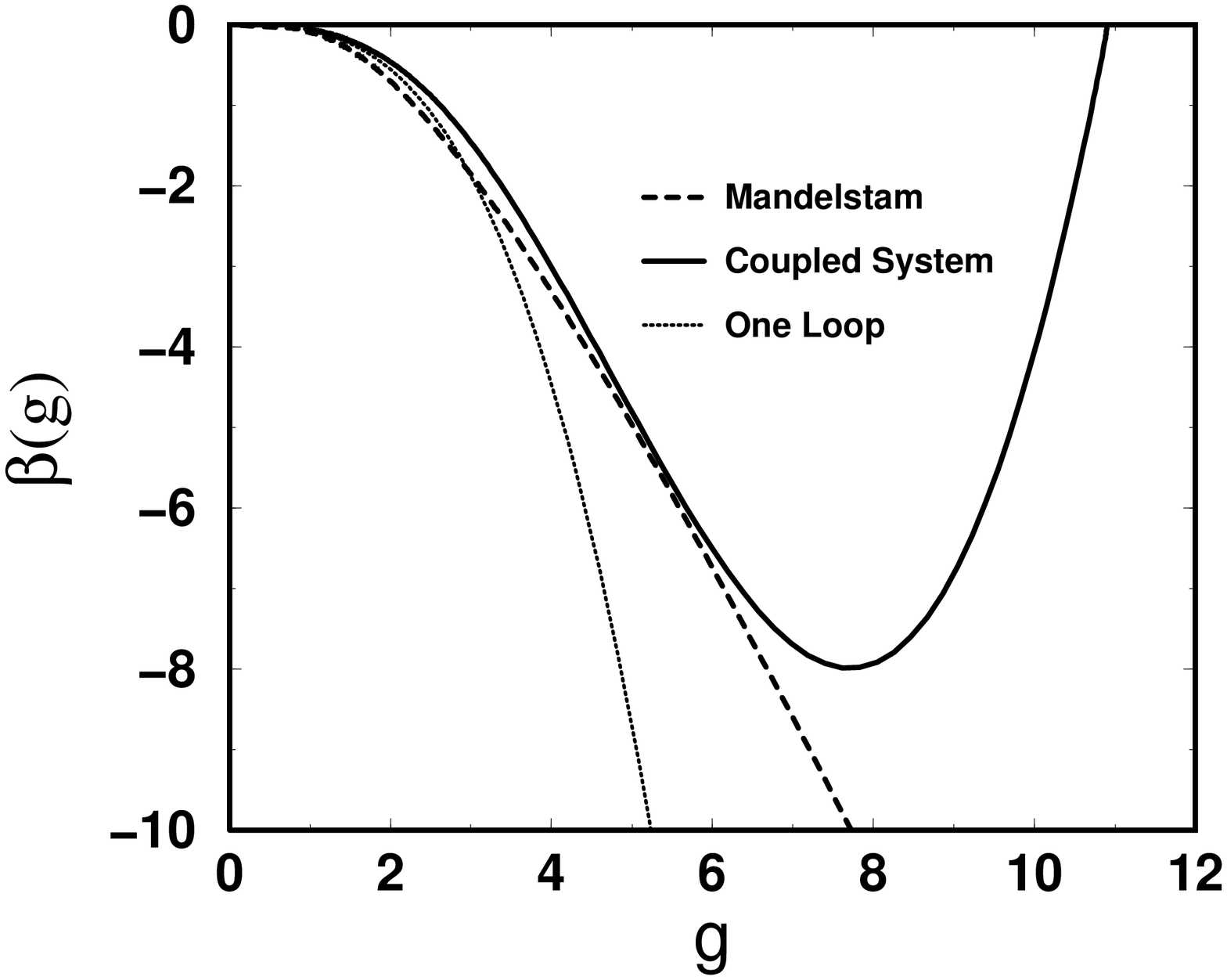,width=7.2cm}}
  \caption{The running coupling $\alpha_S(\mu)$ for $t = 0$ (left), and the
           corresponding $\beta$--function (right) in comparison with its
           leading perturbative form (one--loop) and the $\beta$--function
           as a result of the Mandelstam approximation from
           ref.~\protect\cite{Hau96}.}   
  \label{alpha}
\end{figure}

The running coupling of the present scheme is given in eq. (\ref{rcpl}) by
\begin{equation}
  \alpha_S(\mu) \, = \, \frac{g^2(\mu)}{4\pi}
                \, = \, \frac{1}{4\pi\beta_0} F(s) \; ,
\end{equation}  
where $s = \mu^2/\sigma$ involves the scale $\sigma $ related to the 
infrared boundary condition $b = 1$.  The result for $\alpha_S(\mu) $ is shown
in figure \ref{alpha}. To assign a value to the scale $\sigma$ we use the
phenomenological value of the running coupling $\alpha_S(M_Z) = 0.118$ at the
mass of the $Z$--boson , $M_Z = 91.2$ GeV \cite{PDG96}. For the $t = 0$
solution we obtain this way $M^2_Z/\sigma \simeq 70000 $ and therefore, 
$\sigma \simeq (350 \hbox{MeV})^2$. From the comparison of the solutions for
the renormalization functions $Z$ and $G$ to their leading logarithmic form at
high momenta, this corresponds to a perturbative scale $\Lambda_{\hbox{\tiny
QCD}}$ in $250 \sim 300$ MeV ( for $\lambda $ in $ 1.5 \sim 2$). This is not
a very significant result, however, since the value for $\Lambda_{\hbox{\tiny
QCD}} $  is dependent on the number of quarks $N_f$ and the order of the
perturbative expansion. In particular, in the present framework we expect it
to change when quarks are included. The determination of the momentum scale
in our scale invariant calculation from the dimensionless function $F(s)$ (by
$\alpha_S(M_Z) \stackrel{!}{=} 0.118$) is independent of the value for the
scale parameter $\lambda$ and thus of the value for perturbative scale
$\Lambda_{\hbox{\tiny QCD}} = \sqrt{\sigma/\lambda}$. Equally 
independent of this value is the ratio of the $Z$-- to the $\tau$--mass,
$M_Z/M_\tau \simeq 51.5$, and we obtain from this ratio the value
$\alpha_S(M_\tau) = 0.38$ for the running coupling at the mass of the
$\tau$--lepton. Even though this is in compelling agreement with the
experimental value \cite{PDG96}, we deliberately do not want to claim it to
be a theoretical prediction. We regard this result as nothing but a
consistency check of the present calculation which happens to yield 
better numbers than the present level of truncations and approximations might
suggest.  

The right panel of figure \ref{alpha} shows the corresponding
$\beta$--function with its two fixed points in the ultraviolet at $g = 0$ and
in the infrared at $g = g_c \simeq 10.9$. For comparison we plotted the
leading perturbative form for $g \to 0$ (and $N_f = 0$), $\beta(g) \to -
\beta_0 \, g^3  = - 11/(16\pi^2) \, g^3$, and the infrared singular
non--perturbative result obtained from the analogous subtraction scheme in
the Mandelstam approximation according to ref.~\cite{Hau96}. The solution to
the coupled system of gluon and ghost Dyson--Schwinger equations yields
better agreement with the leading perturbative form at small $g$, since due
to neglecting ghosts, the leading coefficient was obtained to be $\beta_0 =
14/(16 \pi^2)$ in Mandelstam approximation \cite{Hau96}. While this can be
regarded as small quantitative discrepancy, the significant difference
between the present result and the Mandelstam approximation for $g \ge 7$
once more demonstrates the importance of ghosts in Landau gauge, in
particular, in the infrared.  

\section{A Comparison to Lattice Results}

It is interesting to compare our solutions to lattice results available for the
gluon propagator \cite{Ber94,Mar95} and for the ghost propagator \cite{Sum96}
using lattice versions to implement the Landau gauge condition supplemented
by different procedures to eliminate Gribov copies (at least approximately). 
Recently, for the pure $SU(2)$ lattice gauge theory, the influence of such 
copies of gauge equivalent configurations present in the conventional Landau
gauge, has been systematically investigated for gluons and ghosts in
\cite{Cuc97}. 

\begin{figure}[t]
  \centering{ \epsfig{file=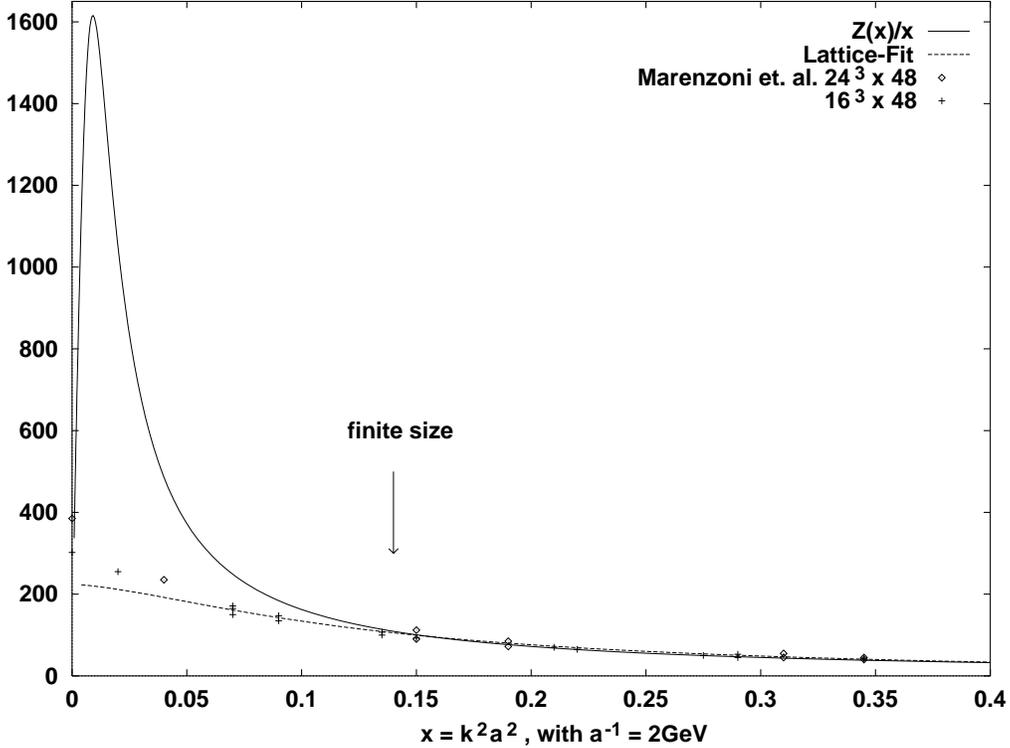,width=\linewidth} }
  \vskip 3mm
  \caption{The numerical result for the gluon propagator from Dyson--Schwinger
           equations (solid line) compared to lattice data from fig.~3 in
           \protect\cite{Mar95}.}
  \label{Zlat}
\end{figure}

In figure \ref{Zlat} we compare our solution for the gluon propagator to
the data from ref.~\cite{Mar95}. The momentum scale in our results, fixed
from $\alpha_S(M_Z)$, is not used as a free parameter. In order to account
for the units used in ref.~\cite{Mar95}, we plotted the gluon propagator,
normalized according to $Z(x = 1) \approx 11.3$, as a function of the
momentum $x = k^2 a^2$ in units of the inverse lattice spacing, using
$a^{-1} = 2$ GeV corresponding to the value $\beta = 6.0$ for $SU(3)$ used
in \cite{Mar95}. According to the authors of \cite{Mar95}, the arrow
indicates a bound below which finite size effects become considerable. As
can be seen in figure~\ref{Zlat}, the essential features of our solution
are beyond the scope of present lattice calculations. Due to different
normalizations, lattice sizes and values of the lattice coupling $\beta$,
it is not quite obvious that presently available lattice data for the
gluon propagator in Landau gauge from the different groups \cite{Ber94}
and \cite{Mar95} is indeed consistent. Since we prefer to adjust the units
of our calculations rather than those of the lattice data, we apologize
for choosing a particular set of data (from figure 3 in ref.~\cite{Mar95})
and refer the reader to the literature, in order to asses the consistency
of the different lattice results.

\begin{figure}[t]
  \centering{ \epsfig{file=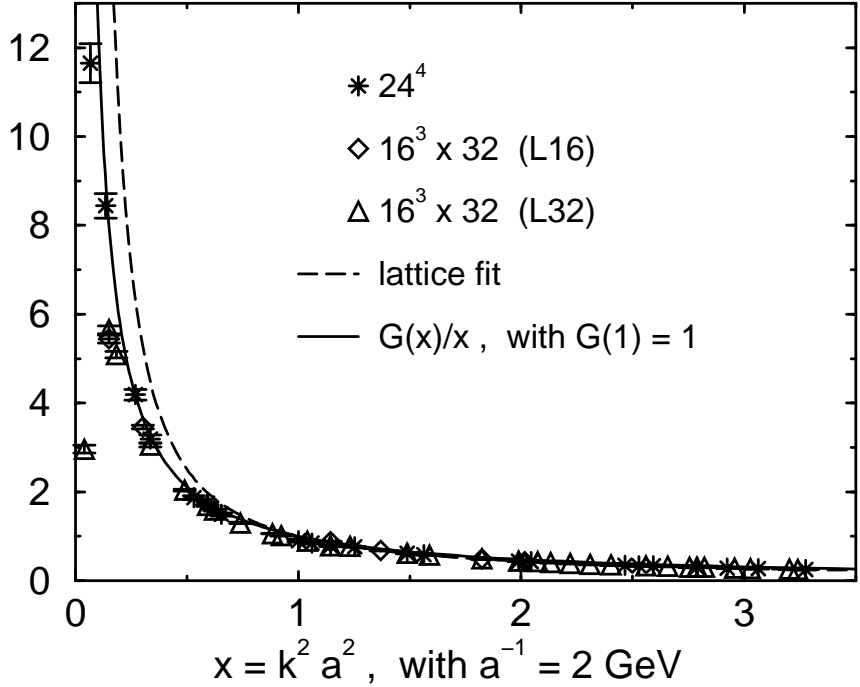,width=0.8\linewidth} }
  \vskip .5cm
  \caption{The numerical result for the ghost propagator from Dyson--Schwinger
           equations (solid line) compared to data from fig.~1 in
           ref.\ \protect\cite{Sum96} for $24^4$ and $16^3\times 32$
           lattices, and a fit of the form $c/x+d/x^2$ (with $c=0.744 , \; 
           d=0.256 $) as obtained in ref.\ \protect\cite{Sum96} for  
           $x \ge 2$ (dashed line).}  
  \label{Glat}
\end{figure}

In figure \ref{Glat} we compared our infrared enhanced ghost propagator 
with normalization such that $G(x=1) = 1$ to the
results of ref.~\cite{Sum96}. We choose to display the lattice results for the
symmetric $24^4$ and the non--symmetric $16^3 \times 32$ lattices for $SU(3)$ 
at $\beta = 6.0$ from figure~1 in ref.~\cite{Sum96} up to $x \approx
3$. Identical results modulo finite size effects were obtained for an $8^4$
lattice (see ref.~\cite{Sum96}). Again, $x = k^2 a^2 $ with $a^{-1} = 2$ GeV
and the momentum scale in our results, fixed from the $Z$--mass as described
in the last section, is {\it not} adjusted.    

The data extracted from the long direction of the $16^3
\times 32$ lattice might indicate the existence of a finite maximum in the
ghost propagator at very low momenta. The fact that the two lowest data
points in this case lie significantly below their neighbors of the $24^4$
lattice was interpreted by the authors of ref.~\cite{Sum96} as a genuine
signal rather than a finite size effect. The reason for this being that, on
the $8^4$ lattice, an enhancement due to finite size was observed for the
lowest point in contrast to the shift downwards of the two points from the 32
direction of the non--symmetric lattice. No such maximum was observed on any
of the smaller lattices.  Our results do not confirm the existence of such a
maximum in the ghost propagator but coincide nicely with all those data
points of the differently sized lattices that lie on a universal curve.
In addition, the $24^4$ and $16^3 \times 32$ lattices are of roughly the same
size as the unsymmetric $16^3 \times 48$ and  $24^3 \times 48$ lattices used
for the gluon propagator in ref. \cite{Mar95}. Considering their
investigation of finite size effects due to deviations between different
components of the gluon propagator at small momenta, one might be led to 
question the significance of the maximum in the ghost propagator
observed for one particular set of data in ref.~\cite{Sum96} at momenta too
low to yield finite size independent results for the gluon propagator, {\it
c.f.}, ref.~\cite{Mar95} on even larger lattices.\footnote{Note the different
$x$--ranges used in figures \protect\ref{Zlat} and \protect\ref{Glat} and the
position of the arrow indicating finite size effects to become considerable.} 
 
It is quite amazing to observe that our numerical solution fits the lattice
data at low momenta ($x \le 1$) significantly better than the fit to an
infrared singular form with integer exponents, $D_G(k^2) = c/k^2  + d/k^4$,
as given in ref.~\cite{Sum96}. Clearly, low momenta ($x<2$) were not included
in this fit, but the authors conclude that no reasonable fit of such a form 
is possible if the lower momentum data is to be included. Therefore, apart
from the question about a possible maximum at the very lowest momenta, the 
lattice calculation seems to confirm the existence of an infrared
enhanced ghost propagator of the form $D_G \sim 1/(k^2)^{1+\kappa}$ with
non--integer exponent $0 < \kappa < 1$. The same qualitative conclusion has
in fact been obtained in a more recent lattice calculation of the ghost
propagator in $SU(2)$~\cite{Cuc97}, where its infrared dominant part was
fitted best by $D_G \sim 1/(k^2)^{1+\kappa}$ for an exponent $\kappa $ of
roughly $ 0.3$ (for $\beta = 2.7$). This also is in qualitative agreement
with the $SU(2)$ calculations of ref.~\cite{Sum96}, again, with the exception
of one data point for the lowest possible lattice momentum.   

Furthermore, in refs.~\cite{Sum96,Cuc97} the Landau gauge condition was
supplemented by algorithms to select gauge field configurations from the
fundamental modular region which is to eliminate systematic errors that might
occur due to the presence of Gribov copies.\footnote{An investigation of the
effectiveness of the stochastic overrelaxation method to achieve this is
given in ref.~\protect\cite{Cuc97}} Thus, the good agreement of our result
with the lattice calculations suggests that the existence of such copies of
gauge configurations might have little effect on the solutions to Landau
gauge Dyson--Schwinger equations. This could also explain the similarity of
our solutions to the qualitative behavior obtained by Zwanziger for gluon and
ghost propagators from implications of complete gauge fixings \cite{Zwa94}.

\section{Confined Gluons} 
 
Since we calculated Euclidean Green's functions, or Schwinger functions, for
gluon and ghost correlations, we resort to the Osterwalder--\-Schrader
reconstruction theorem which states that a G\aa rding--Wightman relativistic
quantum field theory can be constructed from a set of Schwinger functions if
those Euclidean correlation functions obey certain conditions, the
Osterwalder--\-Schrader axioms \cite{QFTbooks}. In particular, the axiom of
{\sl reflection positivity} for Euclidean Green's functions is a direct result
of the positive definiteness of the norm in the Hilbert space of the
corresponding G\aa rding--Wightman quantum field theory. This condition
involves arbitrary partial sums of n--point functions and is hardly provable
in its general form. The special case of a single propagator $G(x-y)$ reads,
{\it i.e.}, the lowest partial sum,
\begin{equation}
  \int d^4\!x\,  d^4\!y  \;  \bar f(-x_0,\hbox{\bf x}) \,  G(x-y)\,
  f(y_0,\hbox{\bf y}) \, \ge \, 0 
  \label{OS2}
\end{equation}
where $f \in {\cal S}_+(\RR^4)$ is a complex valued test (Schwartz) function
with support in $\{(x_0,\hbox{\bf x}) : x_0 > 0 \}$. This special case of
reflection positivity can be shown to be a necessary and sufficient condition
for the existence of a Lehmann representation of the Wightman distribution (the
Minkowski space propagator) corresponding to $G(x-y)$ \cite{Fritz}. Therefore,
the construction of a counter example to this condition (by a suitable choice
of $f$) is sufficient to prove that the Euclidean correlation function cannot
represent physical particle correlations which can be interpreted as a
manifestation of confinement. 

\begin{figure}[t]
  \centering{\
        \epsfig{file=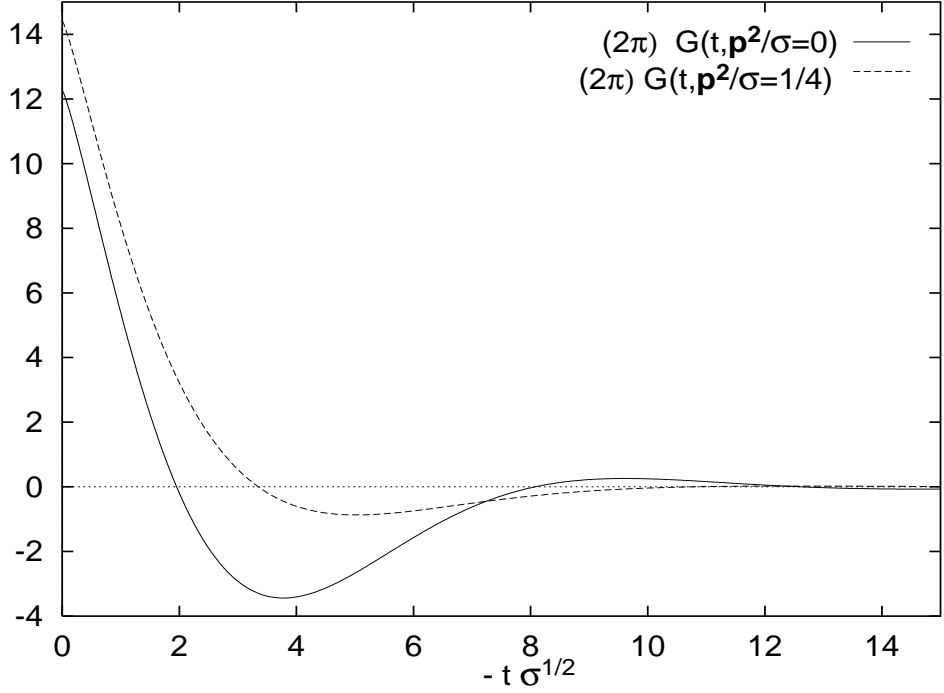,height=0.6\linewidth,width=0.7\linewidth}}
  \caption{The one--dimensional Fourier transform of the gluon propagator,
$G(t,\hbox{\bf p}^2)$.} 
  \label{gluon_ft}
\end{figure}

Heuristically, we can rewrite the condition (\ref{OS2}) after three dimensional
Fourier transformation as 
\begin{equation}
  \int_0^\infty \,   dt' \,  dt \; \bar f(t') \,  G(- (t'+t), \hbox{\bf p}) \,
  f(t) \, \ge \, 0 \; ,
\end{equation}
where $G(x_0,\hbox{\bf p}) := \int d^3\!x \, G(x_0,\hbox{\bf x}) \, e^{i
\,\hbox{\small\bf px}}$, and a separated momentum dependence of the analogous
Fourier transform of the test function  $f$ has to provide for a suitable
smearing around the three--momentum {\bf p}. 

In figure \ref{gluon_ft} we plotted this Fourier transform of essentially the
trace of the gluon propagator, 
\begin{equation}
  G(t,\hbox{\bf p}^2) :=  \int \frac{dp_0}{2\pi}
    \frac{Z(p_0^2 + \hbox{\bf p}^2)}{p_0^2 + \hbox{\bf p}^2} \; e^{i p_0 t}
   \; , \label{eq:gluon_FT}
\end{equation}
for $\hbox{\bf p}^2 = 0$ and $\hbox{\bf p}^2 = \sigma /4$. Choosing a
sufficiently small $\hbox{\bf p}$ and a real function $f(t)$ which peaks
strongly around a point $t_0/2$ for which $G(-t_0,\hbox{\bf p}^2) < 0 $, it
is now very easy to construct a counter example to condition (\ref{OS2})
for the gluon propagator. We interpret this result as confined gluon
correlations described by the Euclidean Green's function $D_{\mu\nu}$. 
The ghost propagator in Euclidean momentum space, being negative for positive
renormalization functions $G(k^2) > 0$ which is the case at tree--level
already, trivially violates reflection positivity. This is of course due to
the wrong spin--statistic of the ghost fields.

Note that the existence of zeros in the Fourier transform (\ref{eq:gluon_FT})
for sufficiently small $\hbox{\bf p}^2$ implies that the gluon propagator in
coordinate space cannot have a well defined inverse for all $(t,\hbox{\bf x})$.
While the corresponding singularities in $G^{-1}(t,\hbox{\bf x})$ will cause
problems in a Hubbard--Stratonovich transformation to bosonize effective
non--local quark theories as the Global Color Model, see ref.~\cite{Tan97},
the same comment we gave at the end of section 4 applies here again. The
infrared structure of the vertex functions, renormalization group invariance
and the existence of zeros in the gluon propagator alone, all lead to the
same conclusion: The combination of ghosts and gluons, $g^2 G^2(k^2)
Z(k^2)/k^2 = \alpha_S(k^2)/k^2 $ is the physically important quantity that
determines the interactions of quarks in Landau gauge (and the
current--current interaction of the Global Color Model). We have verified
explicitly that the Fourier transform of $\alpha_S(k^2)/k^2$
is free of such zeros and positive. 

\section{Summary and Conclusions}

We presented a selfconsistent solution for the propagators of gluons and ghosts
within the Dyson--Schwinger equation approach to QCD in Landau gauge without
quarks. A direct comparison of this solution to the gluon propagator from the
Mandelstam approximation to its Dyson--Schwinger equation shows the
tremendous importance of non--perturbative ghost contributions which are
neglected in the Mandelstam approximation.

We introduced a systematic truncation scheme to close the infinite tower of
Dyson--Schwinger equations at the level of 3--point vertex functions. This
was possible by constructing these vertex functions from their respective
Slavnov--Taylor identities. Such constructions, when contributions from
irreducible 4--point scattering kernels are neglected in the Slavnov--Taylor
identities, allow to express all 3--point vertex functions in terms of the
scalar functions which parameterize the propagators of gluons, ghosts and
quarks. This procedure is consistent with the truncation of the
Dyson--Schwinger equations in which irreducible 4--point correlations are
also neglected. 

In this truncation scheme, at present without quarks, it is found that the
leading infrared behavior of the gluon propagator is dominated completely by
the ghost loop in the gluon DSE clearly demonstrating the importance of
ghosts. In particular, an analytic discussion of the solutions to the closed
set of equations for gluons and ghosts reveals that the infrared behavior of
their 2--point functions is determined by an irrational exponent:
The gluon propagator vanishes in the infrared as $D(k) \sim
(k^2)^{2\kappa-1}$, and the ghost propagator is infrared enhanced (as compared
to a free massless particle pole), $D_G(k)  \sim (1/k^2)^{\kappa+1}$, where
$\kappa =(61-\sqrt{1897})/19 \approx 0.92$. 

The non--perturbative definition of the running coupling $\alpha_S(\mu^2)$ of
our renormalization scheme is given by the renormalization group invariant
product of the ghost and gluon renormalization functions $G$ and $Z$, {\it
i.e.}, $ 4\pi\alpha_S(\mu^2) = g^2 \, G^2(\mu^2)Z(\mu^2)$. Their
infrared behavior implies the existence of an infrared stable fixed point at
a critical coupling, $\alpha_c \simeq 9.5$.

Both, the gluon and the ghost propagator compare favorably with recent results
from lattice gauge simulations. Since these studies implemented various
algorithms to avoid possible Gribov copies, in particular, those of the ghost
propagator, the compelling agreement with our solution for the ghost
propagator indicates that the Gribov problem might have little influence on  
the solutions to Dyson--Schwinger equations.
 
We have furthermore demonstrated that the gluon propagator obtained here 
violates Osterwalder--Schrader reflection positivity, and hence has no
Lehmann representation, which we interpret as a signal for confinement.
To study quarks and, most importantly, their confinement, the present
framework will be extended to include the quark propagator selfconsistently
in future. 

We have given a variety of arguments in section 2 as to why our present
results should not be sensitive to undetermined transverse terms in
the 3--gluon vertex. However, the necessity of an additional assumption on the
renormalization constant of the 3--gluon vertex to restore the ultraviolet
behavior of the solutions, besides being due to neglecting 4--gluon
correlations which appear at order $g^4$, might indicate that the possible
presence of some additional purely transverse terms in the vertex should be
taken into account also, in future. 

The conclusions of the present work rely on the fact that there
are no such undetermined transverse terms in the ghost--gluon vertex of the
present truncation scheme. Therefore, in contrast to the ultraviolet
(perturbative) behavior, the infrared behavior of the propagators, being
entirely due to infrared dominant ghost contributions, is not affected by the
details of the 3--gluon vertex. As for the gluon loop, we were able to
proceed in close analogy to the Mandelstam approximation while accounting
for ghost contributions. This allows a direct comparison and shows the
influence of ghosts in Landau gauge clearly.

Nevertheless, the construction of transverse terms by means of constraints
arising from multiplicative renormalizability in the case of the vertices in
QCD will be an important future project. Such a construction, available at
present for massless quenched QED \cite{Bro91}, will be of particular
importance when quarks are included in the selfconsistent framework.  

It will be furthermore important to compare our present results to calculations
which do not rely on a one--dimensional approximation for the integrals. 
In order to overcome this approximation, it will be
necessary to extend the technique of asymptotic infrared expansions to the
original equations including their full angular dependence. This analytic
technique, developed and applied to the Mandelstam approximation previously
\cite{Atk81,Hau96} and extended to the coupled system of gluon and ghost
Dyson--Schwinger equations in the present paper, proved to be a necessary
prerequisite to finding numerically stable solutions. 

Since confinement might be realized by a quite different mechanism in a
different gauge, and in order to asses the possibility of gauge artifacts in
our results, it will be important to study different gauges in parallel.
Such a study of a coupled system of equations, analogous to the
present one, which determines the gluon propagator in the ghost--free axial
gauge while accounting for its complete tensor structure, is currently in
progress \cite{Alk--}.

\bigskip
\bigskip

\centerline{\Large\bf Acknowledgments}
\smallskip 

We thank F.~Coester, F.~Lenz, M.~R.~Pennington, H.~Reinhardt and M.~Schaden
for helpful discussions. AH and RA gratefully acknowledge the hospitality of
the Physics Division at Argonne during a visit in the early stages of this
work.  This work was supported by the DFG under contract Al 279/3-1, by the
Graduiertenkolleg T\"ubingen (DFG Mu705/3), and the US Department of Energy,
Nuclear Physics Division, contract number\ W-31-109-ENG-38.    

We are indebted to A.~Cucchieri for helping us become aware of a typo in
Eqs. (83), (85) and (90) in previous versions of our manuscript.

\appendix{}
\section*{Appendix}

\section{The Identity for the Ghost--Gluon Vertex}
Without irreducible ghost--ghost correlations, {\it c.f.}, eq. (\ref{redgh}),
the Slavnov--Taylor identity~(\ref{ghSTI}) decomposes into the irreducible
ghost--gluon vertex function, $G_\mu^{abc}(x,y,z)$, and gluon and ghost
propagators as follows, 
\label{app:A}
\begin{equation}
  \frac{1}{\xi}\int d^4\!u \, d^4\!v \, d^4\!w \;
    \partial^x_\mu  D_{\mu\nu}(x-u)  \, G_\nu^{acb}(u,v,w) \,
     D_G(z-v) D_G(w-y)
\end{equation}
\begin{displaymath}
\hskip 1cm  - \frac{1}{\xi}\int d^4\!u \, d^4\!v \, d^4\!w \;
        \partial^y_\mu  D_{\mu\nu}(y-u) \,
    G_\nu^{bca}(u,v,w) \,  D_G(z-v) D_G(w-x)
\end{displaymath}
\begin{displaymath}
  \hskip 3cm = \, g f^{abc} \, D_G(z-x) D_G(z-y) \;.
\end{displaymath}
From the Slavnov--Taylor identity for the gluon propagator, 
\begin{displaymath}
   \partial^x_\mu  D_{\mu\nu}(x-u)  \,=\, \xi \int \frac{d^4k}{(2\pi)^4} \,
 \frac{ik^\nu}{k^2}  \, e^{i k(x-u) } \; ,
\end{displaymath}
and with the ghost propagator 
\begin{displaymath}
D_G(x-y) \,
=\, \int \frac{d^4q}{(2\pi)^4} \, D_G(q)  \, e^{iq(x-y)}
\end{displaymath} 
we obtain after Fourier transformation,
\begin{displaymath}
\frac{ik_\rho G_\rho^{abc}(k,q,p)}{k^2} \,D_G(p) \, D_G(q) 
\, -   \frac{ ip_\rho  G_\rho^{abc}(-p,q,-k)}{p^2} \, D_G(-k) \, D_G(q) 
\end{displaymath}
\begin{equation}
\hskip 3cm   =  \, g f^{abc}\, (2\pi)^4 \delta^4(k+q-p)  
 \,  D_G(-k) \, D_G(p). 
  \label{ghWTIt}
\end{equation}
Here, $G_\mu^{abc}(k,q,p)$ denotes the ghost--gluon vertex in momentum space
which corresponds to the conventions used in section 2 up to the total
momentum conserving $\delta$--function,
\begin{displaymath}
 G_\mu^{abc}(k,q,p) \, =\, \int d^4x \, d^4y \, d^4z \;  G_\mu^{abc}(x,y,z) \;
e^{-ikx} \, e^{-iqy}\,  e^{ipz} 
\end{displaymath}
\begin{displaymath}
\hskip 3cm = \, (2\pi)^4 \,\delta^4(k+q-p)  \,G_\mu^{abc}(p,q) \; .
\end{displaymath}
Using this in eq. (\ref{ghWTIt}) one obtains eq. (\ref{ghWTI}) for
the ghost--gluon vertex function. 
 
\section{3--Gluon Vertex with Tree--Level Ghost Coupling}
In this appendix we demonstrate that the Slavnov--Taylor identity for the
3--gluon vertex in presence of ghosts in Landau gauge has no solution
assuming a tree--level form for the ghost--gluon coupling. Neglecting
irreducible contributions from the ghost--gluon scattering kernel, {\it
i.e.}, setting $\widetilde G_{\mu\nu}(p,q)  = \delta_{\mu\nu}$, the
Slavnov--Taylor identity for the 3--gluon vertex function in Landau gauge
simplifies to the following equation,
\label{app:B}
\begin{equation}
  i k_\rho \Gamma_{\mu\nu\rho} (p,q,k) \, =\, G(k^2) \, \left\{ \,
  {\cal P}_{\mu\nu} (q) \, \frac{q^2}{Z(q^2)}
  \, - \, {\cal P}_{\mu\nu} (p) \,
  \frac{p^2}{Z(p^2)} \right\}  \; .
  \label{glSTItlgh} 
\end{equation} 
Following the general Ansatz of ref.~\cite{Bal80} for those parts of the
3--gluon vertex which can be constrained by its Slavnov--Taylor identity,
we write,
\begin{eqnarray}
 \Gamma_{\mu\nu\rho}(p,q,k) 
       &=&  - A(p^2,q^2;k^2)\,  \delta_{\mu\nu}\,  i(p-q)_\rho\,  
          - \,  B(p^2,q^2;k^2)\,  \delta_{\mu\nu} i(p+q)_\rho  \nonumber\\
&& \hskip -2.5cm - \, C(p^2,q^2;k^2)  ( \delta_{\mu\nu} pq \, -\,  p_\nu
q_\mu) \, i(p-q)_\rho\, \, + \,\frac{1}{3} \, S(p^2,q^2,k^2) \, i(p_\rho q_\mu
k_\nu + p_\nu q_\rho k_\mu )  \nonumber \\
&& \hskip 2cm    + \; \hbox{cyclic permutations} \; .
  \label{an3gv}
\end{eqnarray}
The most general tensor structure contains these 10 terms plus 4 purely
transverse terms involving 2 more scalar functions \cite{Bal80}. The latter
cannot be constrained by Slavnov--Taylor identities and are generally
omitted. The scalar functions appearing in (\ref{an3gv}) have the following
symmetry properties,
\begin{displaymath}
  A(x,y;z) = A(y,x;z) \; , \quad B(x,y;z) = - B(y,x;z) \; , \quad
  C(x,y;z) = C(y,x;z) \; ,
\end{displaymath}
and $S(x,y,z)$ is totally antisymmetric with respect to any two
arguments. With this Ansatz in (\ref{glSTItlgh}) we obtain the following
equations from comparing the coefficients of the independent tensors on both
sides:
\begin{eqnarray}
\delta_{\mu\nu} \, : \hskip .6cm &&  \Bigl( A(x,y;z) - B(x,y;z)
\Bigr) y \,   - \, \Bigl( A(x,y;z) +  B(x,y;z) \Bigr) x \nonumber\\
&&  \hskip 3cm = \, \Biggl( \frac{y}{Z(y)} - \frac{x}{Z(x)} \Biggr) \, G(z) 
\label{A1}\\ 
\hbox{and}\hskip .6cm  && -2B(x,y;z) + C(x,y;z) \,(x-y) \, = \, 0\; , 
\label{A2}\\
q_\mu q_\nu \, : \hskip .6cm && -2 A(z,y;x) + A(x,z;y) - B(x,z;y) \nonumber\\
&& \hskip 2cm -\frac{1}{2}
(z + x - y) S(x,y,z)  \, = \, - \frac{G(z)}{Z(y)} \; ,
\label{A3}\\
p_\mu p_\nu \, : \hskip .6cm && \hskip 1cm \hbox{same as above for
$q_\mu q_\nu$ with $ x\leftrightarrow y $ } \; , \nonumber \\
p_\mu q_\nu \, : \hskip .6cm && -A(z,y;x) - B(z,y;x) + A(x,z;y) - B(x,z;y) \,
= \, 0  \; ,
\label{A4} \\
q_\mu p_\nu \, ;\hskip .6cm &&   - 2 B(x,y;z) - 2 A(z,y;x) + 2
A(x,z;y) - z S(x,y,z) \, = \, 0   \; .
\label{A5}
\end{eqnarray}
The first observation here is that with (\ref{A5}) in (\ref{A4}) we find that
\begin{equation}
  S(x,y,z) \, =\, 0  \; .
\end{equation}
We could than in principle use (\ref{A5}) to eliminate $B$ in (\ref{A3}) thus
obtaining an equation for $A$. Eq.~(\ref{A5}) would then give us the
corresponding function $B$ which would fix $C$ from (\ref{A2}). 
However, eliminating $B$ from (\ref{A3}) yields,
\begin{equation} 
  A(x,y;z) - A(x,z;y) + A(z,y;x) \, = \, \frac{G(z)}{Z(y)} \; . \label{WS}
\end{equation}  
From the necessary symmetry of $A(x,y;z)$ it follows that the l.h.s.\ in
(\ref{WS}) is symmetric under exchange of $x$ and $z$ whereas the r.h.s.\ 
is not. It is therefore not possible to find a solution $A$ for this
equation with the correct symmetry.

This can be cured in two ways. The first is to allow for a non--trivial
ghost--gluon scattering kernel of the form given in eq.~(\ref{glghkern})
in section 2 which would not affect the ghost--gluon vertex since the
correction to the kernel is transverse in the momentum of the incoming
ghost. With the Ansatz (\ref{glghkern}) we can redo the above analysis
obtaining modifications to the different terms on the r.h.s.\ of
eqs.~(\ref{A1}) to (\ref{A3}). In particular, eq.~(\ref{A1}) is replaced
by
\begin{eqnarray}
 \Bigl( A(x,y;z) - B(x,y;z)
\Bigr) y &-& \Bigl( A(x,y;z) +  B(x,y;z) \Bigr) x 
\label{A1n}\\ 
&& \hskip -2.2cm = \, \Biggl(
\frac{y}{Z(y)} \Bigl(1 + y\, a(z,y;x) \Bigr) - \frac{x}{Z(x)} \Bigl( 1 + x
\, a(z,x;y) \Bigr)  \Biggr) \, G(z)  \; ,  \nonumber
\end{eqnarray}
eq. (\ref{A2}) by,
\begin{equation}
-2B(x,y;z) + C(x,y;z) \,(x-y) \, = \, \Biggl( \frac{y \, a(z,y;x)}{Z(y)} -
\frac{x \, a(z,x;y)}{Z(x)} \Biggr) \, G(z)  , 
\label{A2n}
\end{equation}
and eq. (\ref{A3n}) by
\begin{equation}
-2 A(z,y;x) + A(x,z;y) - B(x,z;y)\, = \, - \frac{G(z)}{Z(y)} \, \Bigl(1 + y\,
a(z,y;x)  
\Bigr)   \; .
\label{A3n}
\end{equation}
Eqs.~(\ref{A4}) and (\ref{A5}) remain unchanged and, in particular, $S =
0$, which we used in (\ref{A3n}) already. Defining 
\begin{equation}
  f(z,x;y) := \frac{G(z)}{Z(y)}\, \Bigl( 1 + y \, a(z,y;z) \Bigr) \; ,
\end{equation}
we realize that the program outlined above to obtain the functions $A,B$ and $C$
can give a solution if $f(x,y;z) = f(y,x;z)$. The equation to obtain $A$
which replaces (\ref{WS}) is now,
\begin{equation} 
A(x,y;z) - A(x,z;y) + A(z,y;x) \, = \, f(z,x;y) \; . \label{WSrep}
\end{equation}  
Note that using (\ref{A5}) here again, this is yields,
\begin{equation} 
A(x,y;z) - B(x,y;z) \, = \, f(z,x;y) \; , \label{A1ok}
\end{equation}  
showing that a solution to (\ref{WSrep}) for $A$ with $B$ obtained from
(\ref{A5}) also satisfies (\ref{A1n}). With $C$ from (\ref{A2n}) we find that
all necessary equations (\ref{A1n},\ref{A2n},\ref{A3n}) and
(\ref{A4},\ref{A5}) are solved (with $S=0$). So far, we did not specify the
unknown function $a(x,y;z)$ resulting from a non--trivial ghost--gluon
scattering kernel. The necessary symmetry of $f(x,y;z)$  provides certainly
not enough information to fix $a(x,y;z)$ completely. However, since we do not
want to modify the scaling properties of the vertex artificially, it may seem
reasonable to solve the system of equations with an Ansatz for $f(x,y;z)$ as
the most general sum of terms representing ratios $G/Z$ with all possible
arguments $x,y$ and $z$ obeying the necessary symmetry of $f$ in its first
two arguments. This restriction, which yields terms of the same
structure as the r.h.s. of eq. (\ref{WS}) now in (\ref{WSrep}), allows to
determine the function $a(x,y;z)$ up to an unknown constant $\Delta$,
\begin{displaymath}
  a(x,y;z) = \frac{1}{y} \left( \frac{G(z)}{G(x)} - \frac{G(y)}{G(x)} \right)
  \, - \, \frac{\Delta}{2y} \Bigg\{ \frac{G(y)}{G(x)} \Biggl( \frac{Z(y)}{Z(z)} +
  \frac{Z(y)}{Z(x)}  - 2 \Biggr)
\end{displaymath}
\begin{equation} 
  \hskip 3cm + \frac{G(z)}{G(x)} \Biggl( 1 - \frac{Z(y)}{Z(x)}
  \Biggr) + 1 - \frac{Z(y)}{Z(z)} \Bigg\} 
  \; .
\end{equation} 
The solutions for $A, B$ and $C$ follow as outlined above, in this case,
\begin{eqnarray}
A(x,y;z) &=& \frac{1}{2} \left( \frac{G(z)}{Z(x)} +  \frac{G(z)}{Z(y)}
\right) + \frac{1-\Delta}{2} \left( \frac{G(x)}{Z(y)} +  \frac{G(y)}{Z(x)}
\right)  \nonumber\\
&& \hskip 1cm -\frac{1-\Delta}{2} \left( \frac{G(x)}{Z(x)} +
\frac{G(y)}{Z(y)}  \right) \; ,\\
B(x,y;z) &=& \frac{1}{2} \left( \frac{G(y)}{Z(x)} -  \frac{G(x)}{Z(y)}
\right) + \frac{1-\Delta}{2} \left( \frac{G(z)}{Z(x)} -  \frac{G(z)}{Z(y)}
\right) \nonumber \\
&& \hskip 1cm 
-\frac{1-\Delta}{2} \left( \frac{G(x)}{Z(x)} - \frac{G(y)}{Z(y)}\right) -
\frac{\Delta}{2}  \left( \frac{G(x)}{Z(z)} - \frac{G(y)}{Z(z)} \right) \; ,\\
C(x,y;z) &=& \frac{1}{x-y} \left( \frac{G(z)}{Z(x)} - \frac{G(z)}{Z(y)}
\right) \; .
\end{eqnarray}
This solution to the 3--gluon Slavnov--Taylor identity corresponds to a
specific form for the ghost--gluon scattering kernel, in the simplest case,
for $\Delta = 0$, the dependence on ratios of gluon renormalization
functions $Z$ disappears and it is given by
\begin{equation}
  \widetilde G_{\mu\nu} (p,q) = \delta_{\mu\nu} + \frac{1}{q^2}
  \frac{G(k^2) - G(q^2)}{G(p^2)}
  \Bigl( \delta_{\mu\nu} pq \, - \, p_\nu q_\mu \Bigr) \; . 
\end{equation} 
The second way to account for the presence of ghosts in the Slavnov--Taylor
identity in Landau gauge is to consider a possible dressing of the
ghost--gluon vertex function as described in section 2. For the present
truncation scheme this led to a vertex as given in eq. (\ref{fvs}) from the
new Slavnov--Taylor identity (\ref{ghSTI}). The structure of the
corresponding ghost--gluon scattering kernel as far as it is relevant to the
3--gluon Slavnov--Taylor identity, however, is then simply 
\begin{equation} 
  \widetilde G_{\mu\nu}(p,q) \, =\, \frac{G(k^2)}{G(q^2)} \,
  \delta_{\mu\nu}
  \; , 
\end{equation}
which, used in the Slavnov--Taylor identity (\ref{glSTI}), led to
eq. (\ref{glWTI}). The solution to this equation, as given in eqs.
(\ref{3gv}/\ref{3gva}) in section 2, can be obtained from an entirely
analogous procedure as outlined in this appendix, setting, 
\begin{equation}
f(x,y;z) \, = \, \frac{G(x) G(y)}{G(z) Z(z)}\; ,
\end{equation}
as it corresponds to the {\sl Abelian}--like identity (\ref{glWTI}). This
results in the solution $A(x,y;z) = A_+(x,y;z)$, $B(x,y;z) = A_-(x,y;z) $ and
$C(x,y;z) = 2A_-(x,y;z)/(x-y)$ with the notations of section 2. The latter
way to implement ghost contributions in the construction of the 3--gluon
vertex function from its Slavnov--Taylor identity is considerably simpler
than the alternative as described in this appendix (based on a tree--level
ghost--gluon vertex), and it is furthermore much more consistent with the
assumption underlying the truncation scheme of section 2.

\section{Functions in the 3--Gluon Loop}
Below we give the functions $N_1$ and $N_2$ which appear in the 3--gluon loop
of the gluon DSE using the solution (\ref{3gv}/\ref{3gva}) for the 3--gluon
vertex function and contracting the gluon DSE with the projector ${\cal
R}_{\mu\nu} (k)$ of eq.~(\ref{proR}) (see sec.~4), 
\label{app:C} 
\begin{eqnarray}
N_1(x,y;z) &=&    {{29\,x}\over 4} + {{{x^2}}\over {4\,y}} + {{29\,y}\over 4} + 
   {{{y^2}}\over {4\,x}} + {{\left( 9\,{x^2} + 50\,x\,y + 9\,{y^2} \right) \,
       z}\over {4\,x\,y}} \\ 
&&\hskip -1cm -\, {{9\,\left( x + y \right) \,{z^2}}\over 
     {4\,x\,y}} - {{{z^3}}\over {4\,x\,y}} + 
   {{24\,{x^2} - 10\,x\,y + {y^2}}\over {2\,\left(z -x \right) }} + 
   {{ 24\,{y^2} - 10\,x\,y +{x^2}}\over {2\,\left(z -y \right) }} \nonumber
\end{eqnarray}
and
\begin{eqnarray}
N_2(x,y;z) &=&   {{5\,{x^3} + 41\,{x^2}\,y + 5\,x\,{y^2} - 3\,{y^3}}\over 
     {4\,x\,\left(y -x \right) }} + 
   {{{x^2} - 10\,x\,y + 24\,{y^2}}\over {2\,\left( y - z \right) }} \\
&& \hskip -2cm + \, 
   {{{x^3} + 9\,{x^2}\,y - 9\,x\,{y^2} - {y^3}}\over {x\,z}} + 
   {{\left( 2\,{x^2} + 11\,x\,y - 3\,{y^2} \right) \,z}\over 
     {2\,x\,\left( x - y \right) }} + 
   {{\left( x + y \right) \,{z^2}}\over {4\,x\,\left(y -x \right) }}
\; .\nonumber 
\end{eqnarray}

%
%
\newpage


\begin{thebibliography}{99}
\bibitem{Fritz} We are indebted to F.~Coester for pointing this out to us.
\end{thebibliography}
\end{document}